\documentclass[12pt,hyper,a4paper]{JHEP3}
\usepackage{cite,graphicx,amsmath,amssymb,epsfig,slashed}

\newcommand{\mbf}[1]{\mbox{\boldmath $#1$}}
\newcommand{\bk}{\mbf{k}}

\newcommand{\bp}{\mbf{p}}

\newcommand{\bzero}{\mbf{0}}
\newcommand{\disc}{\textrm{disc}}
\newcommand{\tr}{\operatorname{tr}}

\newcommand{\cK}{{\cal K}}
\newcommand{\cG}{{\cal G}}
\newcommand{\cH}{{\cal H}}
\newcommand{\cV}{{\cal V}}

\newcommand{\sK}{\slashed{\cal K}}

\newcommand{\osK}[3]{\overset{#1}{{}_{#2}\slashed{\cal K}_{#3}}}
\newcommand{\ocs}[4]{\overset{#2}{{}_{#3}\slashed{\cal #1}_{#4}}}
\newcommand{\os}[4]{\overset{#2}{{}_{#3}\slashed{#1}_{#4}}}
\newcommand{\ov}[3]{\overset{#2}{{#1}_{#3}}}
\newcommand{\oK}[1]{\overset{#1}{\cal K}}
\newcommand{\oc}[3]{\overset{#2}{{\cal #1}_{#3}}}
\newcommand{\op}[4]{\overset{#2}{{}_{#3}{#1}_{#4}}}
\newcommand{\oZ}[3]{\overset{#1}{{}_{#2}Z_{#3}}}
\newcommand{\obZ}[3]{\overset{#1}{\langle {}_{#2}Z_{#3} \rangle}}

\newcommand{\obs}[4]{\overset{#2}{\langle {}_{#3}\slashed{\cal #1}_{#4} \rangle}}
\newcommand{\oo}[1]{\overset{#1}{\omega}}

\def\o{\omega}

\title{\bf Inclusive 1-jet Production Cross Section\\ at Small $x$ in QCD:
Multiple Interactions} 

\author{J.Bartels$^a$, M.Salvadore$^a$ and G.P.Vacca$^b$ \\
${}^a$ II Inst. f. Theor. Physik, Univ. of Hamburg,\\
Luruper Chaussee 149, 22761 Hamburg, Germany\\
E-mail: bartels@mail.desy.de, michele.salvadore@gmail.com\\

${}^b$ INFN - Sezione di Bologna, Dip. di Fisica,\\
Via Irnerio 46, Bologna, Italy.\\
E-mail: vacca@bo.infn.it}

\preprint{DESY-08-016}


\abstract{We study corrections due to two Pomeron exchanges 
to the inclusive 1-jet production cross section in the Regge 
limit of perturbative QCD for a finite number of colors.
By considering deep inelastic scattering 
on a weakly bound two-nucleon system, we carefully follow the logic of the 
AGK cutting rules and show, for the single inclusive cross
section, that, due to the reggeization of the gluon, modifications 
of the AGK cutting rules appear. As our main result, we investigate and 
calculate the jet production vertex in the presence of a two-Pomeron 
cut correction. Compared to previous studies, we find a novel structure  
of the jet vertex which has not been considered before. We discuss a few 
implications of this new piece.}
\begin{document}
\section{Introduction}
High gluon densities and saturation in high energy QCD have attracted
much interest in recent years. Experimental evidence has been discussed
in connection with both HERA and RHIC data, and with the advent of the LHC
there will be interest in signals for high densities also in proton-proton
collisions. In this context inclusive jet production plays a central role:
whereas for moderate values of longitudinal momenta, $x_1$ and $x_2$,
the cross sections for inclusive jet production will be described by
collinear factorization and leading twist parton densities,
the forward region may require substantial corrections.
The LHC will allow, close to the forward direction of one of the protons,
a very asymmetric configuration of jet or Drell-Yan production,
for example $x_1 \ll x_2$. This leads,
for not too high momenta of jets, to very small values of $x_1$, and may
require multiple exchanges between the produced jet and
proton '1' (Fig.1). In more physical terms, the produced jet
may originate from a configuration where the density of gluons from
proton '1' is high.
\begin{center}
\epsfig{file=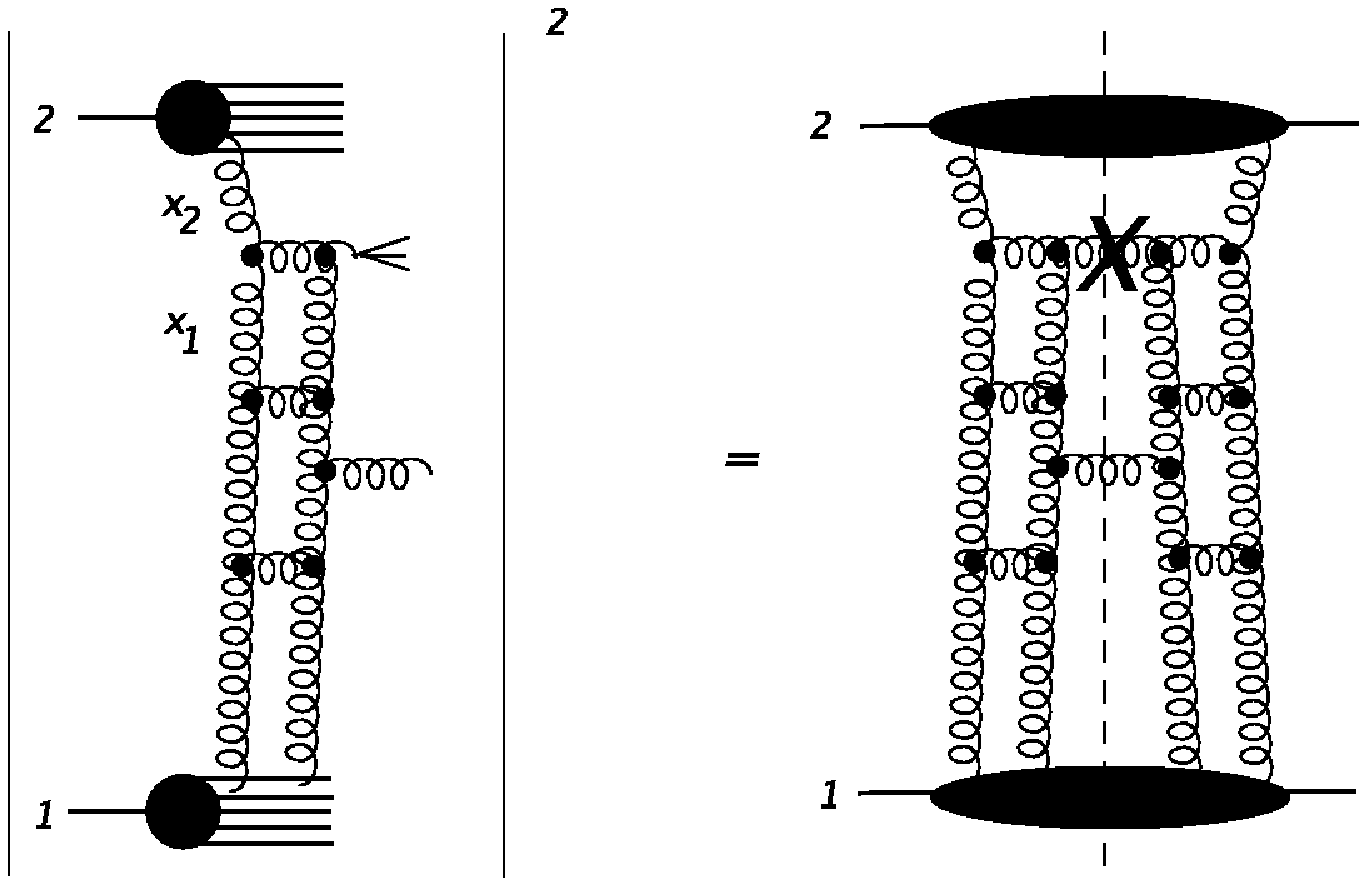,width=10cm,height=5.8cm}\\
Fig.1: forward jet in $pp$ collisions
\end{center}
It is therefore important to provide, from the theoretical side,
cross section formulae which allow to incorporate multiple exchanges
between the produced jet and the proton. Within the collinear approximation,
these corrections belong to higher twist and are suppressed by powers of the
jet transverse momentum. However, at small $x$, resummation of powers of
$\ln 1/x$ are expected to partly compensate such a suppression. It may
therefore be more suitable to start within the BFKL approach. It is also this
approach which, in deep inelastic electron-proton or electron-nucleus
scattering, provides the framework for the discussion of high gluon
densities and saturation.

In this paper we make an attempt to address, within the  
BFKL framework in momentum space, for finite $N_c$, the issue of 
multiple interaction in inclusive jet production. As a theoretical framework 
we use the scattering of a virtual photon on a nuclear target consisting 
of two (different) nucleons (Fig.2): the energy discontinuity of this 
process consists of different classes of final states, and within these 
final states we fix one gluon which generates the jet. In particular, we 
search for the jet vertex illustrated in Fig.1a-c, where below the jet vertex  
we have to sum over all possible cuttings. It will turn out that the 
vertex is more complicated than suggested by Fig.1.   

The single inclusive jet cross section, mostly in the large-$N_c$ limit approximation, 
has been studied before in ~\cite{B2000,KT,KL,BGV,B2005,B2006}. 
Whereas the first study ~\cite{B2000} had explicitly been based upon the AGK 
~\cite{Abramovsky:1973fm} cutting rules (see also ~\cite{agk-qcd1,agk-qcd2} for a
QCD analysis), it was then in ~\cite{KT} observed that the 
emission of the jet inside the triple Pomeron vertex might lead to 
deviations from the AGK rules. Results of ~\cite{KT} have been supported 
in ~\cite{KL,BGV}. In ~\cite{B2005} a new investigation
was reported, more detailed than ~\cite{B2000} but still based upon 
assumptions, which lead to the discovery of 
new contributions to the effective production vertex. An improved and 
more accurate investigation was given more recently in ~\cite{B2006}.
Whereas the calculations 
reported in ~\cite{KT,KL,BGV} have been done in configuration space, the  
studies in ~\cite{B2000,B2005,B2006} were done in momentum space, and their method 
is similar to the one used in this paper. Nevertheless, our results, 
which - in contrast to ~\cite{B2000,B2005,B2006} - are valid for 
an arbitrary number of colors, are in partial conflict with those of 
~\cite{B2006}.       

\section{The strategy}
We consider deep inelastic scattering on a nucleus consisting of two
weakly bound nucleons (Fig.2).
\begin{center}
\epsfig{file=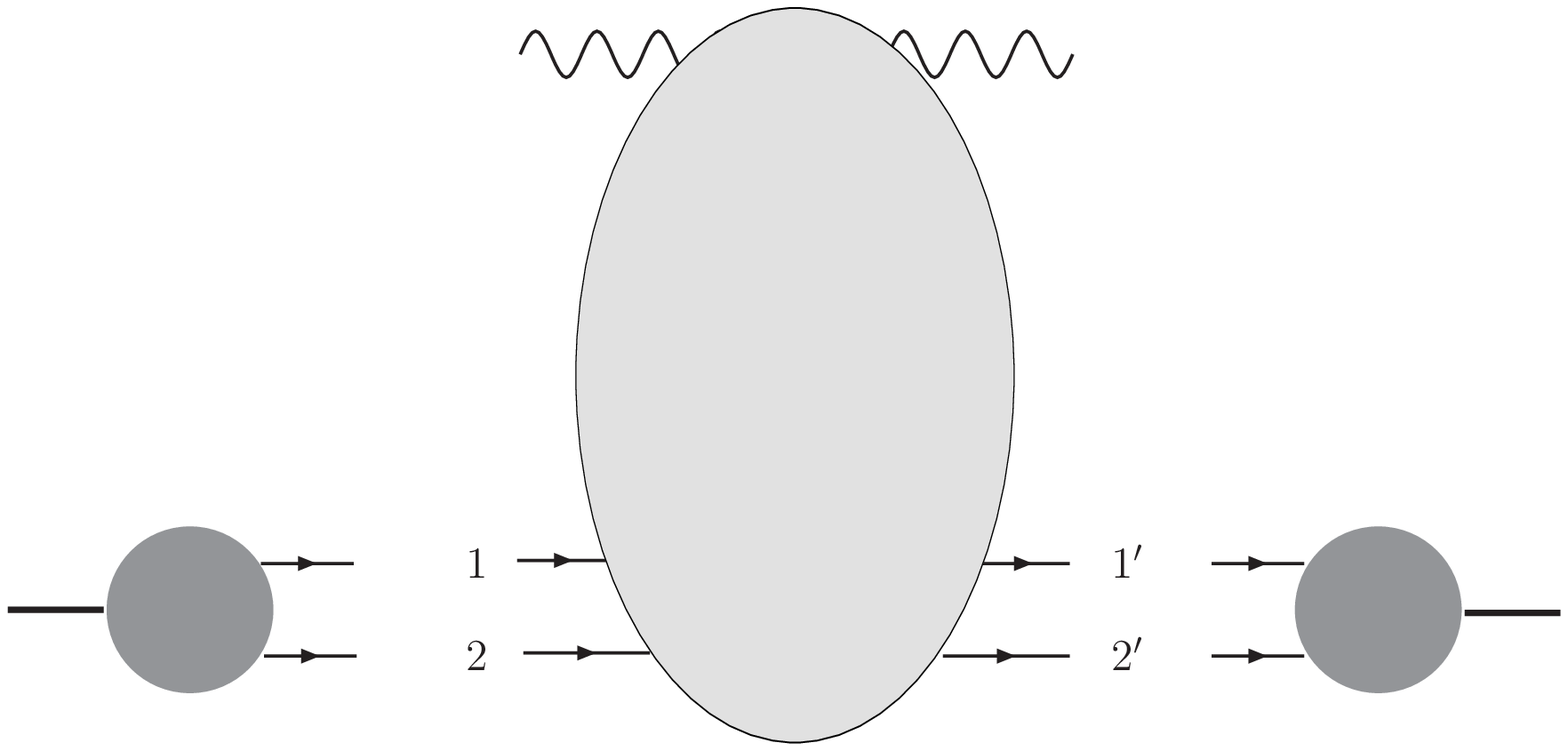,width=10cm,height=4.5cm}\\
Fig.2: Scattering of a virtual photon on a weakly bound nucleus
\end{center}
The total cross section is obtained 
from the elastic scattering amplitude, $T_{\gamma^*(pn) \to \gamma ^* (pn)}$:
\begin{equation}
\sigma^{tot}_{\gamma^*(pn) \to \gamma ^* (pn)} = \frac{1}{S} {\rm Im}\, 
T_{\gamma^*(pn) \to \gamma ^* (pn)}.
\end{equation} 
where $S$ denotes the total energy of the scattering process.
Before we consider the inclusive cross section we find it useful to 
recapitulate the computation of the total cross section.   
The kinematics is illustrated in Fig.3: 
\begin{center}
\epsfig{file=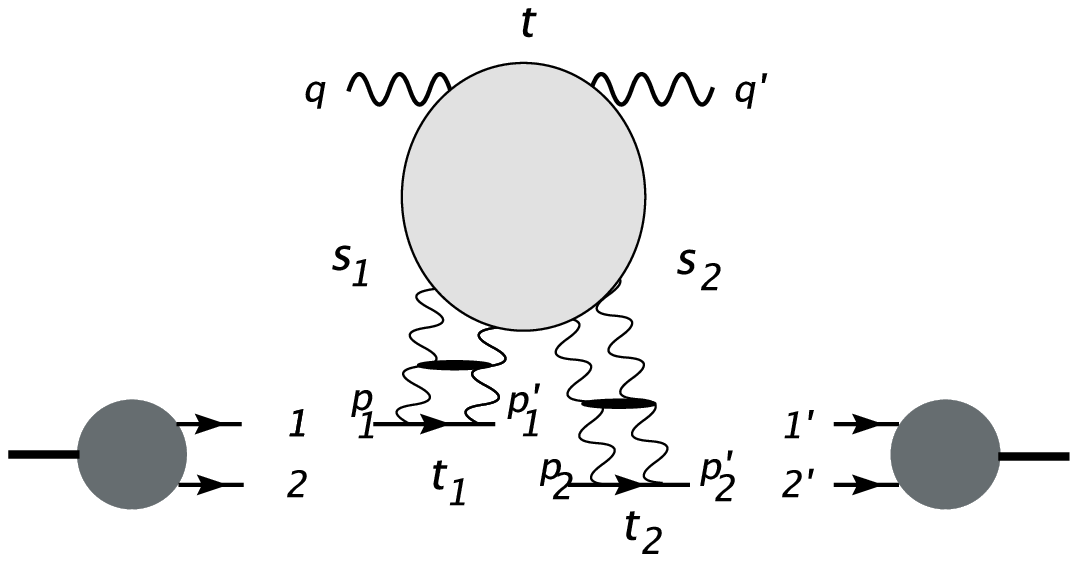, width=9cm,height=5cm}\\
Fig 3: kinematics of the $3\to3$ process.
\end{center} 
the energy variables $s_1=(q+p_1)^2$,
$s_2=(q+p'_2)^2$, $M^2 = (q+p_1 - p'_1)^2$, $S=(q+p_1+p_2)^2$ are assumed to 
be much larger than the momentum transfer variables $t=(q-q')^2$, 
$t_1=(p_1-p'_1)^2$, $t_2=(p_2-p'_2)^2$. We will distinguish between $s_1$ 
and $s_2$, but at the end we set $s_1= s_2 =s \gg M^2$ and $t=0$. 
Throughout this paper we use Sudakov variables 
with the lightlike reference vectors $q'$ and $p$, such that $s=2p'q = 2pq$, 
$S= 4pq =2s$, $q=q'-xp$ with $x=2pq/Q^2$ and $M^2= x_P \,s$. 
Neglecting the nucleon masses we have 
\begin{equation}
p_1 = p_2 =p,\,\,p'_1 = p(1-x_P) +p_{1 \perp},\,\ p'_2= p(1+x_P) + 
p_{2 \perp}. 
\end{equation}
Internal momenta are then written as 
\begin{equation}
k_i=\alpha_i q'+ \beta_i p + k_{i\,\perp} 
\end{equation}
with 
$k^2_{i\,\perp} = - \bk_i^2$.
The fact that the two nucleons are in a weakly coupled bound state implies 
that we will allow the two nucleons to have small losses of longitudinal 
and transverse momenta, i.e. we will integrate over $x_P$ and 
$p_{1 \perp} = - p_{2 \perp}= k_{\perp}$. 

\subsection{The total cross section}
For the total cross section we will be interested in the 
imaginary part of the amplitude  
$T_{\gamma^*(pn) \to \gamma ^* (pn)}(s_1, s_2, M^2; t_1,t_2,t)$, integrated 
over $x_P$ and $p_{1 \perp}= -p_{2 \perp}= k_{\perp}$.
Following the discussion in ~\cite{Abramovsky:1973fm}, this imaginary part consists 
of the three contributions illustrated in Fig.4. They are often referred to 
as 'diffractive cut' (Fig.4a), 'single cut' (Fig.4b), and 'double cut' 
(Fig.4c).
\begin{center}
\epsfig{file=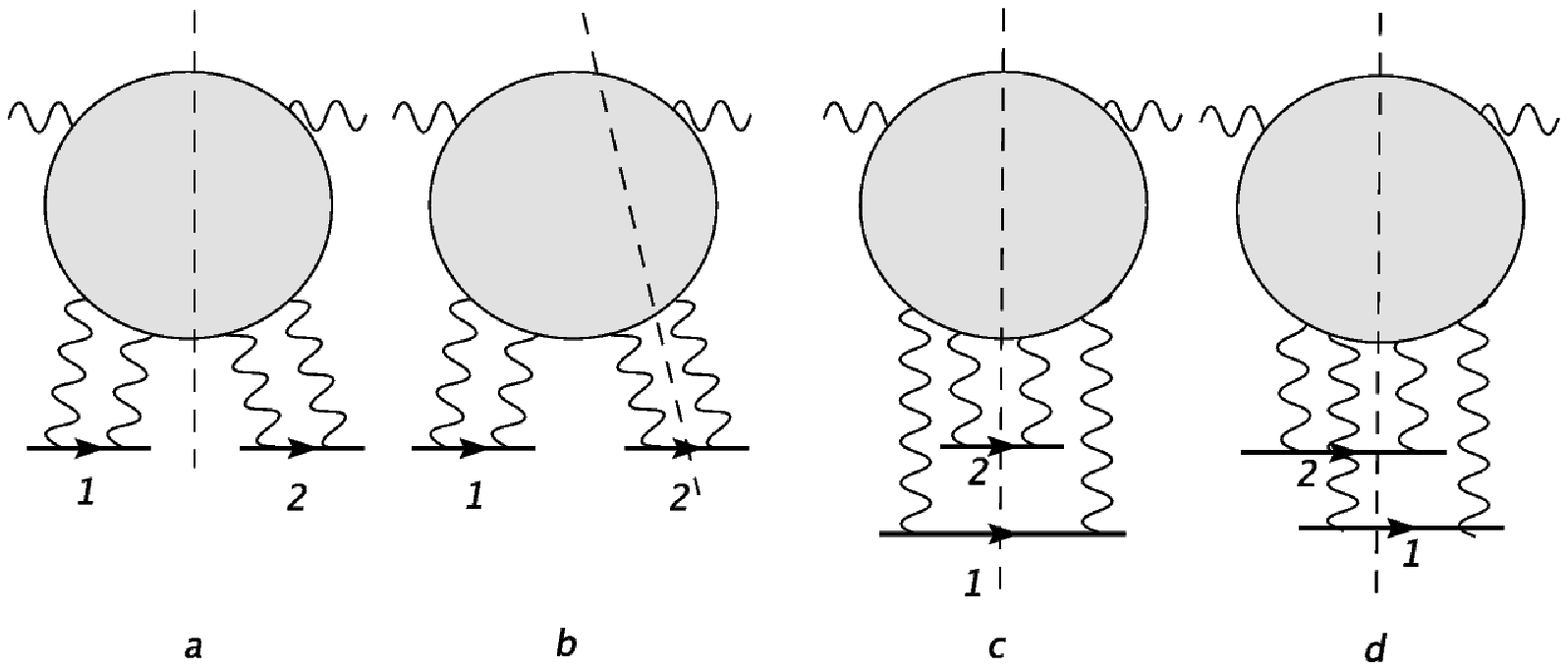, width=12cm,height=5cm}\\
Fig 4: different energy cuts:\\
(a) diffractive cut, (b) single cut, (c) double cut.
\end{center} 
The total cross section is obtained from the sum of these terms, where 
we have to observe that, in Fig.4a, we have to add the configuration where 
the nucleons $1$ and $2$ are interchanged. Similarly, in Fig.4b 
the cut line can pass through nucleon $1$ or $2$, and for both configurations 
we also have to add their complex conjugates. Finally, in  
Fig.4c we show two of the four configurations; the remaining ones are 
obtained by interchanging nucleons $1$ and $2$. 

Let us analyse these contributions in more detail. In all cases we encounter 
subamplitudes, $A_4(\bk_1,\beta_1, \bk_2,\beta_2, \bk_3,\beta_3,\bk_4,\beta_4)$ (Fig.5),
which differ from each other by the way in which the $\beta$ integrals are
done.

Gluons $1$,...$4$ are labelled from the left to the right.
In Fig.4a, gluons $1$ and $2$ will couple to nucleon $1$, and gluons 
$3$ and $4$ to nucleon $2$; in Fig.4c1 gluons $1$ and $4$ are attached to 
nucleon $1$, and gluons $2$ and $3$ to nucleon $2$.
The variables $\beta_i$ (with $\sum \beta_i =0$) denote
the $\beta$ components of the gluons, which are integrated, 
and which can be interpreted as (dimensionless) energy variables of the subamplitudes. For the 
moment we will ignore the color indices of the gluons.

\begin{center}
\epsfig{file=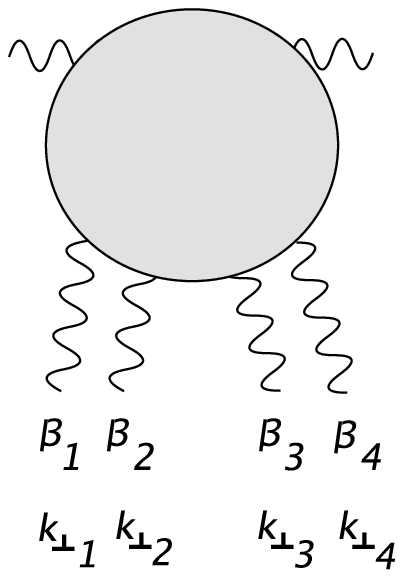,width=3cm,height=4cm}\\
Fig 5: a subamplitude of Fig.4
\end{center} 

To begin with Fig.4a, we introduce $x_P = \beta_1 + \beta_2 =
- \beta_3 - \beta_4$ and consider, as integration variables, 
$x_P$, $\beta_1$, and $\beta_3$. With $M^2 = s x_P$, the discontinuity  
in Fig.4a indicates that, for the diffractive cut, we are integrating 
across the discontinuity in $x_P$. Obviously, $M$ denotes 
the invariant mass of the process: photon + (gluon $1$ + gluon $2$) $\to $
photon + (gluon $3$ + gluon $4$). 
\begin{center}
\epsfig{file=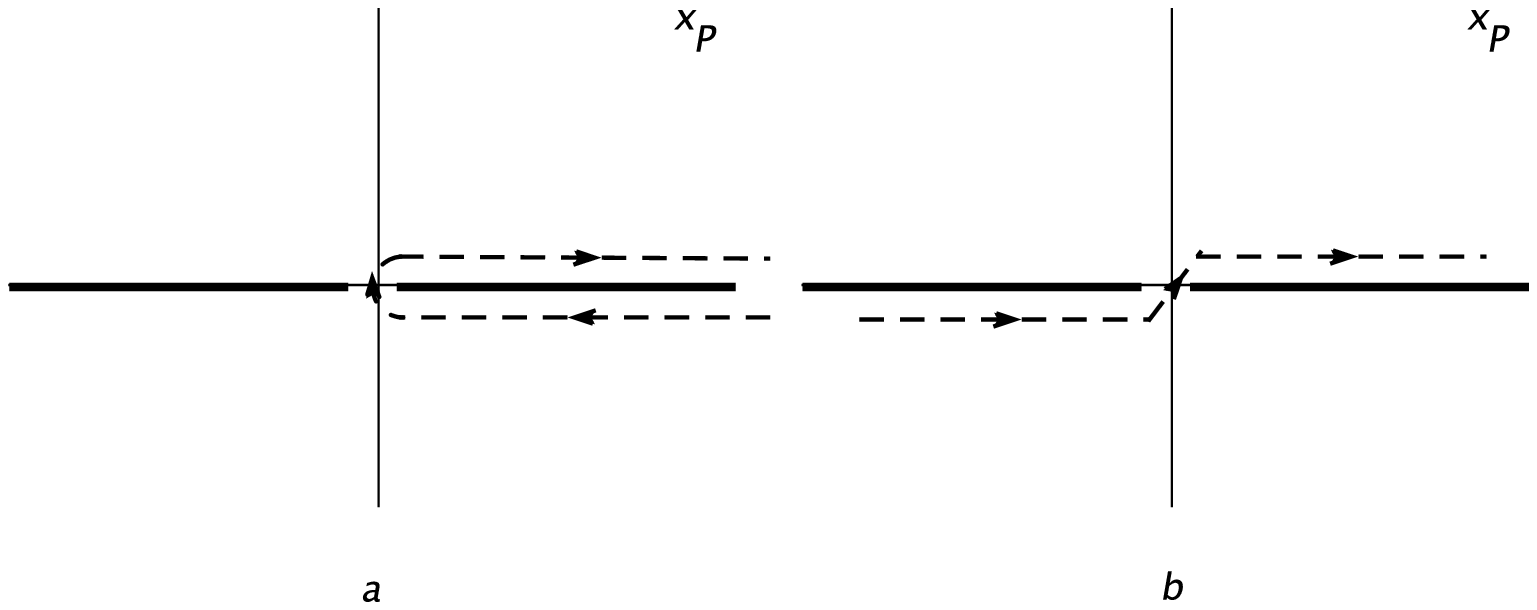,width=9cm,height=4cm}\\
Fig 6:  integration contours
\end{center}
The remaining $\beta$ variables, $\beta_1$ and $\beta_3$ are inside the 
ladders to the right and to the left hand side of the 
energy cut, and the integration contours are taken along the real axis.
In Fig.4b, it is the variable $\beta_1$ in which we take the discontinuity,
and the contour goes around the right hand cut, whereas the other 
$\beta$ variables (including $x_P$) run along the real axis. Here $s \beta_1$ 
denotes the squared energy of the subprocess: photon + gluon $1$ $\to$ photon + 
(gluon $2$ + gluon $3$ + gluon $4$). 
Finally, in Fig.4c1 we introduce $\beta= \beta_1 + \beta_3$ as the 
subenergy variable in which the discontinuity is taken, and its integration 
goes around the right hand cut. At this stage, the subamplitudes of Figs.4a - c
are different from each other. 

Provided that in all three cases the 
subamplitude $A_4$, taken as a function of the three independent $\beta$ variables, 
falls off sufficiently fast for large $|\beta|$, one can redraw the 
contour containing the discontinuity along the real axis (Fig.6b). 
Furthermore, we need the amplitudes to be symmetric under permutations 
of the gluons. If all these conditions are satisfied, all three cases 
can be reduced to one and the same integral, 
where all three $\beta$-integrations run along the real axis:
\begin{equation}
N_4 (\bk_1,\bk_2,\bk_3,\bk_4)= \int d\beta_1 \int d\beta_2 \int d\beta_3 
A_4(\bk_1,\beta_1, \bk_2,\beta_2, \bk_3,\beta_3,\bk_4,\beta_4),
\end{equation} 
and the function $N_4$ is symmetric under permutations of the gluons. 
Alternatively, $N_4$ can be written as a triple discontinuity integral and 
is, therefore, a real-valued function. This is what is required for the 
AGK rules to be valid.        

Applying this discussion to Figs.4a-c, it is then clear that 
all three different cuts, after integration over $x_P$, will have the same 
expression for the subamplitude, $N_4$, and they differ from each other 
only by the phases for the ladders below.  This allows, 
to write the sum of all three terms in the simple form:   
$$
2 {\rm Im} T = 2 S \sigma_{tot} = S \int dx_P \int d\mu(\{\bk\}) \,\,{\rm Im} T_{\gamma^*(pn)\to \gamma^*(pn)}= 
$$
\begin{align}
= \int \frac{d\omega}{2 \pi i} e^{\omega Y} 
\int \frac{d\omega_1}{2 \pi i} \int \frac{d\omega_2}{2 \pi i} \int \frac{d\mu(\{\bk\})}{(2 \pi)^6} 
2 \pi i \delta(\omega - \omega_1 -\omega_2) N_4 \nonumber \\
\cdot \Big[\xi(\omega_1) \xi(\omega_2)^* + \xi(\omega_2) \xi(\omega_1)^* \nonumber \\
+ 2 {\rm Im} \xi(\omega_1) \left( i \xi(\omega_2)+ (i\xi(\omega_2))^* \right) +
 2 {\rm Im} \xi(\omega_2) \left( i \xi(\omega_1)+ (i\xi(\omega_1))^* \right) \nonumber \\
+ 4 {\rm Im} \xi(\omega_1 {\rm Im} \xi(\omega_2)\Big] \otimes \Phi_{1,2}(\{\bk\}),
\label{sumofcuts}
\end{align}
where $y=\ln S$, and $\bk$ is the momentum transfer of nucleon $1$ and we have
considered the kinematics corresponding to the measure
$d\mu(\{\bk\})=d^2\bk_1 d^2\bk_2 d^2\bk_3d^2\bk_4 \delta^{(2)}(\bk_1+\bk_2)\delta^{(2)}(\bk_3+\bk_4)$.
The  signature factors have the form 
\begin{equation} 
\xi(\omega) = \frac{1 - e^{-i \pi \omega}}{\sin \pi \omega},
\end{equation}
and $\Phi_{1,2}(\{\bk\})$ contains the two nucleon form 
factors and the deuteron wave function (including the integration 
over the $\alpha$-variables). In the large-$N_c$ limit, the subamplitude 
$N_4$ will contain the product of two Pomeron propagators, 
$G_2(\bk_1,\omega_1)$ and $G_2(\bk_3,\omega_2)$, which couple to nucleons 
$1$ and $2$, resp. Eq.(\ref{sumofcuts}) can also be written 
in the more familiar form:
$$
2 {\rm Im} T = 2 S \sigma_{tot} = S \int dx_P \int d^2 \bk \,\,{\rm Im} T_{\gamma^*(pn)\to \gamma^*(pn)}= 
$$
\begin{align}
= \int \frac{d\omega}{2 \pi i} e^{Y \omega} 
\int \frac{d\omega_1}{2 \pi i} \int \frac{d\omega_2}{2 \pi i} \int \frac{d\mu(\{\bk\})}{(2 \pi)^6} 
2 \pi i \delta(\omega - \omega_1 -\omega_2) N_4 \nonumber \\
\cdot 2 {\rm Im} \Big[ (-i)(i\xi(\omega_1)) (i\xi(\omega_2)) \Big] \otimes \Phi_{1,2}(\{\bk\}),
\label{AGKform}
\end{align}
in agreement with the AGK argument.

In a practical calculation of the total cross section, we  
compute the three different cut contributions in Fig.4, term by term. 
In each term we have to calculate   
production amplitudes on both sides of the cutting line, 
evaluate the unitarity integrals, and then sum over the intermediate states. 
As a result, we should find that the subamplitudes in all three cases, 
in fact, are equal and symmetric under permutations: 
otherwise the assumptions stated above would prove to be incorrect. 

As to the calculation of the production amplitudes, 
we can show that they also can be derived from discontinuities 
in their own energy variables. Beginning with Fig.4a and concentrating on the 
phases, we have, on the lhs of the cutting line, the production amplitude illustrated in Fig.7a:
\begin{center}
\epsfig{file=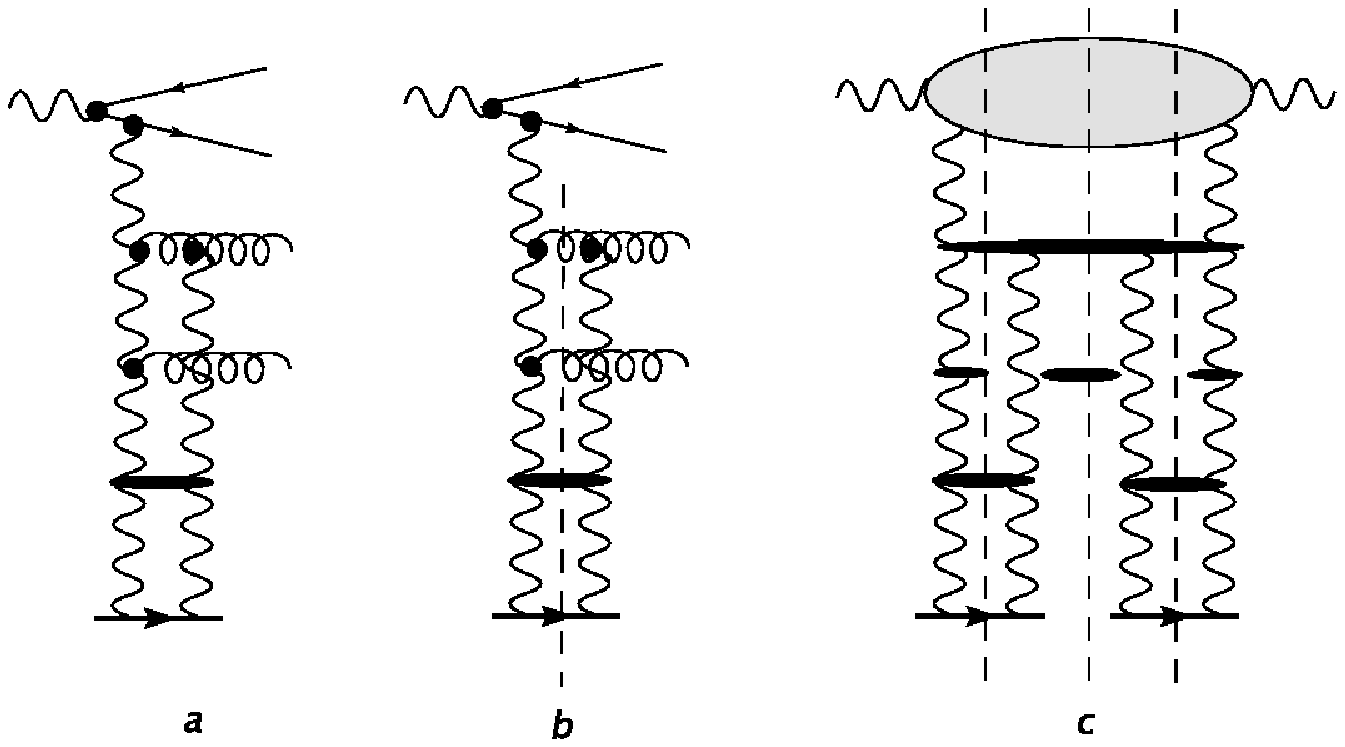, width=9cm,height=5cm}\\
Fig 7: construction of Fig.4a:\\
(a) production amplitude; (b) energy discontinuity of the production amplitude; 
(c) triple energy discontinuity 
\end{center}
In the leading logarithmic approximation, it is proportional to its 
energy discontinuity, shown in Fig.6b. Symbolically:
\begin{equation}
T = \xi(\omega_1) \disc T.
\end{equation} 
The same argument applies to the rhs of the cutting line in Fig.4a 
(with $\xi(\omega_1) \to \xi(\omega_2)^*$).
Together, the single discontinuity from which we have started can be expressed 
in terms of the triple discontinuity (Fig.7c). The 
analytic expression for this contribution, therefore, is the same as for the 
first term in (\ref{sumofcuts}), 
with $N_4$ being replaced by the triple discontinuity illustrated in Fig.8:
\begin{center}
\epsfig{file=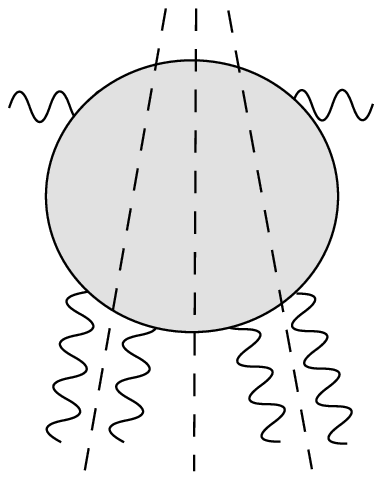, width=4cm,height=4cm}\\
Fig 8: multiple energy discontinuity\\
\end{center}
In this triple discontinuity, all the $\beta$ integrals are closed on the
rhs, analogous to Fig.6a. Because of the assumptions for the large-$\beta$
behavior, we can redraw the 
contours as in Fig.6b, and the triple discontinuity coincides with $N_4$.

For the next term, Fig.4b, the situation is similar, although a bit more complicated.
We illustrate the situation in Fig.9:
\begin{center}
\epsfig{file=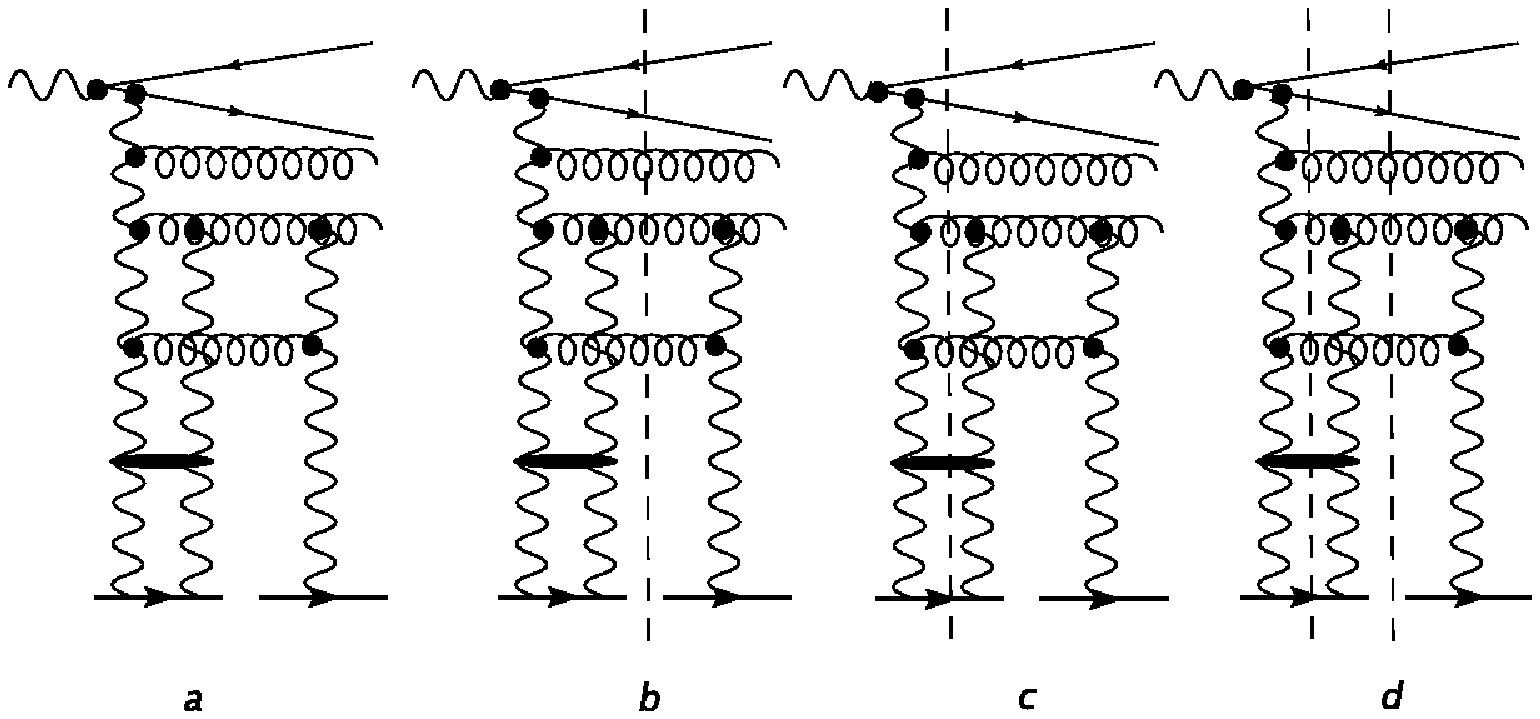,width=12cm,height=5cm}\\
Fig 9: multiple energy discontinuity\\
\end{center}
Fig.9a illustrates the production amplitude we need to find. It has the phase structure
$(-i) (i) (i\xi(\omega_1))$,
and we can find it from its energy discontinuity, provided the amplitude above satisfies the assumptions 
stated before (good behavior for large $\beta$, and symmetry under the exchange of gluons). Following 
the AGK arguments, the energy discontinuity consists of the two parts shown in Figs.9b and c.
The first one has the phase $\xi(\omega_1) +\xi(\omega_1)^*$, whereas the second one vanishes 
(after adding the analogous contribution with the the cut ladder on the rhs and the single 
exchange on the lhs). As a result, the production amplitude is proportional to its 
triple discontinuity in Fig.9d, and, returning to Fig.4b, we obtain the second term of (\ref{sumofcuts}), with 
$N_4$ being replaced by the triple discontinuity of Fig.8. But as we have already said, this triple 
discontinuity equals $N_4$. 

Finally the cuts in Fig.4c1 and c2. In Fig.10a we illustrate the production 
amplitude 
to the left of the discontinuity line in Fig.4c1. It contains a further cut (Fig.10b),
and it is proportional to this discontinuity.    
\begin{center}
\epsfig{file=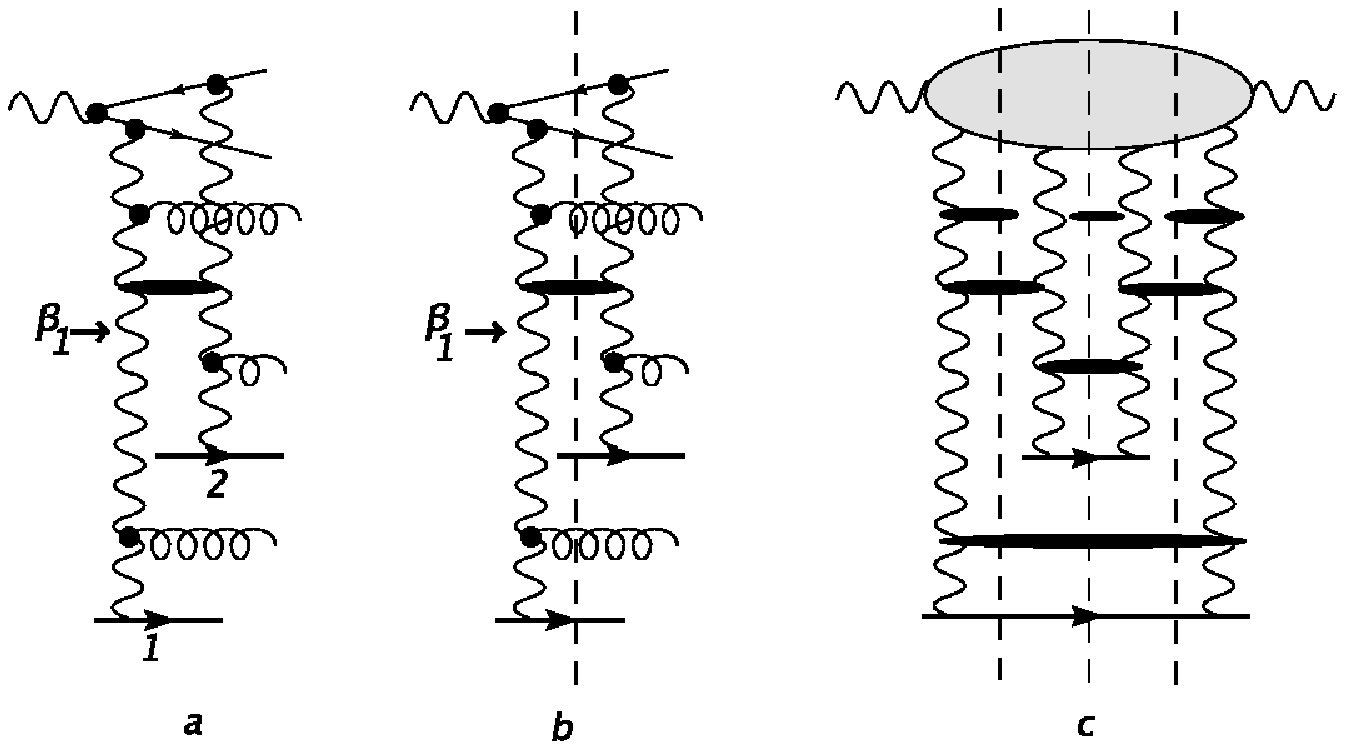,width=12cm,height=5cm}\\
Fig 10: multiple energy discontinuity\\
\end{center}
As a function of $\beta_1$, Fig.10b presents the right hand cut. An additional term (corresponding to 
Fig.4c2) where nucleon $1$ and $2$ at the lower end are interchanged, provides the left hand cut.
When inserting these production amplitudes into Fig.4c1 and 4c2 and performing the $\beta_1$ integrals, we close 
the contour on the rhs and include Fig.10b.  
As a result we find that the contribution Fig.4c1 is proportional to the 
triple discontinuity, and the phases can be read off from (\ref{sumofcuts}).     

\subsection{The single-gluon inclusive cross section}
So far all our discussion has been for the total cross section. 
Turning to the single gluon inclusive cross section,
we find it convenient to assign the rapidity value $y=0$ to the virtual 
photon, and $y=Y$ to the nucleons (i.e. in our figures, 
we draw the rapidity axis downwards, starting from $y=0$ 
at the upper photon and ending with $Y$ at the target). 
The rapidity of the inclusive jet will be denoted by $y_1$ with 
$0 < y_1 < Y$, and its transverse momentum by $\bp$. 
 
For the calculation we follow the same procedure, i.e. we compute the 
discontinuities in Fig.4. But in all the three energy cuts in 
Figs.4a - c, we now fix, in the sum over the intermediate states, 
for one gluon the values of rapidity and transverse momentum
$y_1$ and $\bp$, resp. This leads to the inclusive cross section
illustrated in Figs.11a - c.
\begin{center}
\epsfig{file=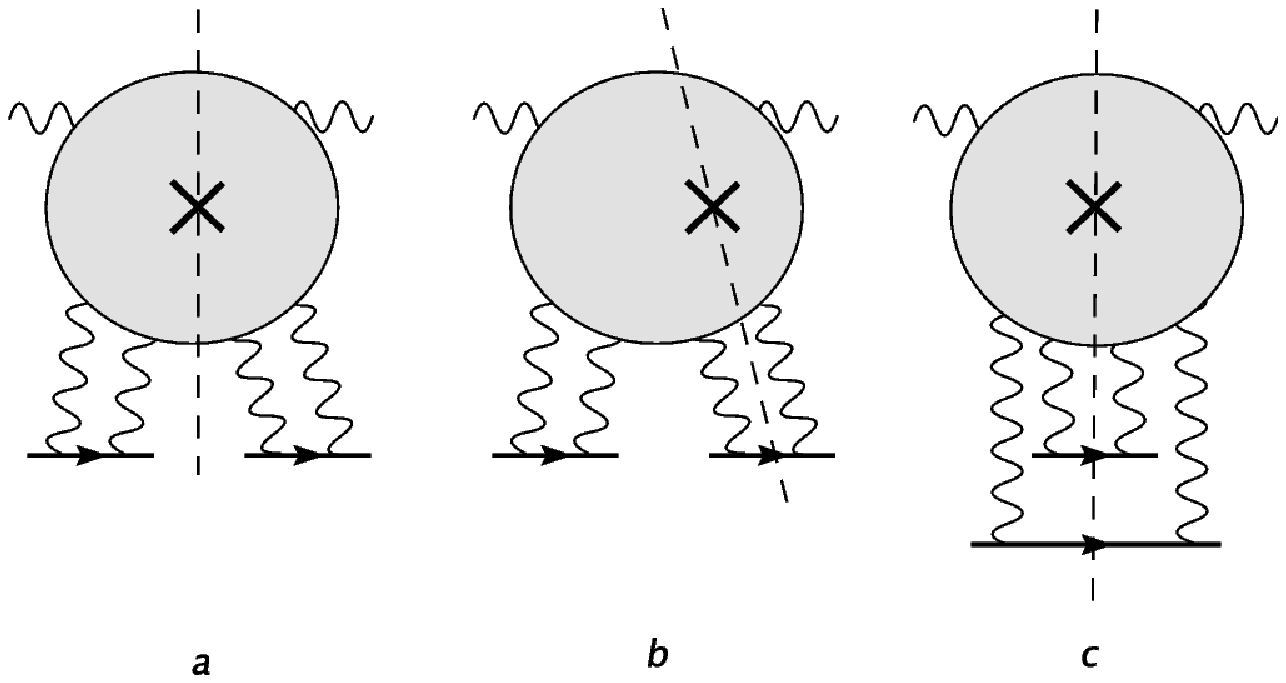, width=8cm,height=4cm}\\
Fig 11: different energy cuts:\\
(a) diffractive cut, (b) single cut, (c) double cut.
\end{center}  
The crosses mark the fixed final state gluon inside the unitarity sum. 
We then, again, have to compute the production amplitudes on both sides of the cutting 
line and sum over the intermediate states (keeping now the one final state gluon 
fixed). Since, before doing the summation over the intermediate states, 
the production amplitudes are the same as for the total cross section, 
we can proceed as outlined above, and we can make use of the results 
described before. In particular, we have the same phases factors.

However, unlike the case of the total cross section described before, we can no longer 
expect that the subamplitudes which appear in the three different cuttings are equal to 
each other. On general grounds we have to expect that, when going from a fully
inclusive total cross section to a slightly less inclusive quantity, 
we loose part of the coherence and of the cancellations. In our particular case,
the equality of the subamplitudes in Figs.11a - c will, in fact, be lost.  
Our calculations described below will confirm this. Depending upon where, 
inside the grey blob the fixed gluon is produced, there exist some 
contributions where the equality still exists (and the AGK rules are valid), 
and others where it is not the case. The inclusive cross section 
has, therefore, to be written in the following form:
\begin{eqnarray}
_{}&&\hspace{4cm}\frac{d\sigma}{dy d^2 \bp} = \nonumber\\
= &&\!\!\!\!\int \frac{d\omega'}{2 \pi i} e^{y_1 \,\omega'}  \!\!\!
 \int \frac{d\omega}{2 \pi i} e^{(Y- y_1) \,\omega} 
\!\!\!\int \!\frac{d\omega_1}{2 \pi i} \!\int \!\frac{d\omega_2}{2 \pi i} 
\!\int \!\frac{d \mu(\{\bk\})}{(2 \pi)^6} 
2 \pi i \delta(\omega - \omega_1 -\omega_2) \hspace{3cm}\nonumber \\
&&\cdot \Big[N_4^c (1,2|3,4;\bp) \xi(\omega_1) \xi(\omega_2)^* + 
N_4^c (3,4|1,2;\bp) \xi(\omega_2) \xi(\omega_1)^*+ \hspace{3cm} \nonumber \\
&&\hspace{0.5cm}+ 2  {\rm Im} \xi(\omega_1)\,\, \left( N_4^c (1|2,3,4;\bp) \,i \xi(\omega_2) 
\,\,+ \,\,c.c.\,\,\right) \,\,+\hspace{3cm} \nonumber \\ 
&&\hspace{0.5cm}+  2 {\rm Im} \xi(\omega_2) \,\, \left( N_4^c(1,2,3|4;\bp) \,i
\xi(\omega_1) \,\,+\,\, 
c.c.\,\,\right) \,\,+\hspace{3cm} \nonumber \\
&&\hspace{1cm}+ 4 N_4^c(1,3|2,4;\bp) {\rm Im} \xi(\omega_1) {\rm Im} \xi(\omega_2)\Big]\cdot  
\Phi_{1,2}(\{\bk\}),
\label{sumofcuts2}
\end{eqnarray}

where the argument structure of the $N_4$ indicates where the cutting line containing 
the produced jet enters the subamplitude: for example, in $N_4^c(1,2|3,4;\bp)$,
the line runs between gluon $3$ and $4$. 
We find that, in general, the amplitudes $N_4^c$ are different for different 
positions of the cutting line. Here and in the following we suppress the dependence 
of the $N_4^c$ upon the variables $\omega'$, $\omega$, $\omega_1$, and  
$\omega_2$. 

Having given this general description of how to compute the total 
cross section and the single jet inclusive cross section we now turn to 
QCD calculations. We first return to the total cross section, for which we can make 
use of earlier results and review the main results (a few more 
details will be given in the following section). The discussion of the 
inclusive case - which represents the main result of this 
paper - will be presented in the following section.  
We first need to address the question of the 
large-$\beta$ behavior of the subamplitudes in QCD. Here the gluon reggeization 
plays an important role. If we compute, in pQCD, the subamplitude 
illustrated in Fig.8, which, in total, is in color singlet state
there will be pieces in which, at the lower end, subsystems 
are in antisymmetric color octet states: 
\begin{center}
\epsfig{file=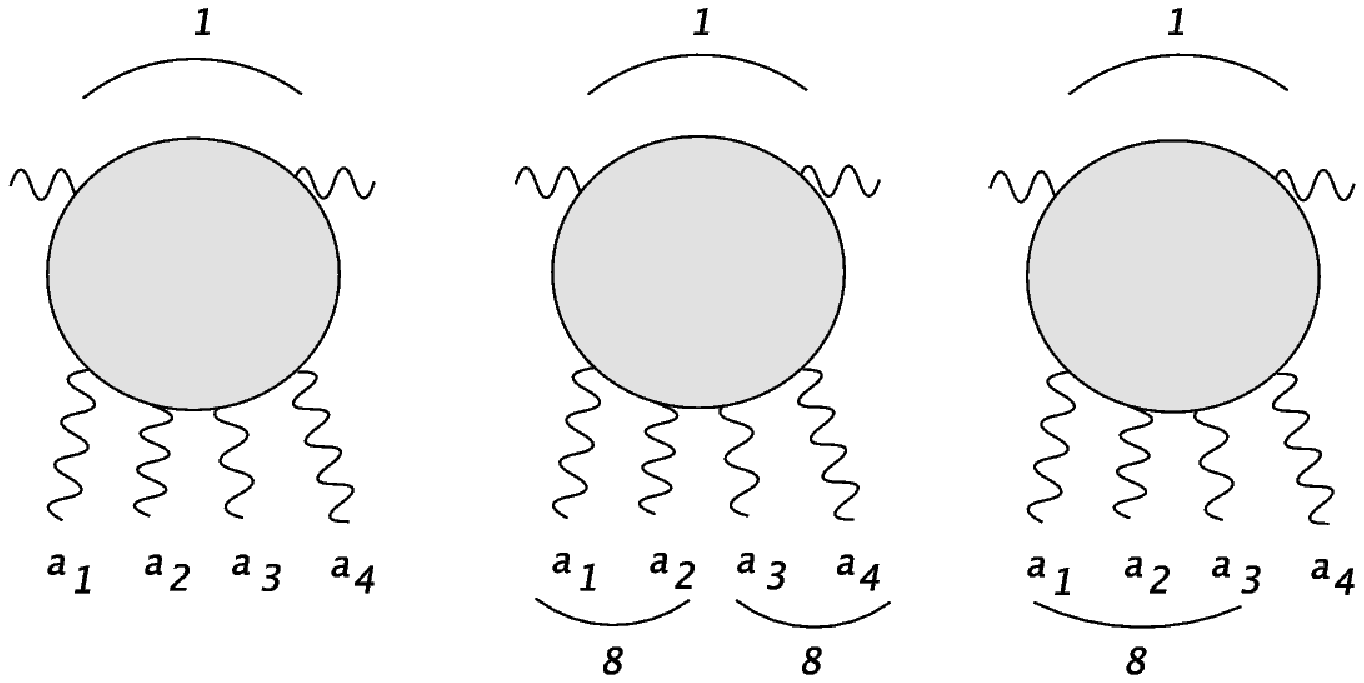, width=9cm,height=5cm}\\
Fig 12: color configurations of $t$-channel gluons\\
(a) general tensor structure; (b) two color octet pairs; 
(c) one triplet of gluons in a color octet state 
\end{center} 

We expect that these pieces belong to the reggeization of the gluon: they 
do not satisfy the naive Ward identies, i.e. they do not vanish when the 
transverse momentum, $k_{\perp}$, of one of the gluons goes to zero. 
Connected with the
lack of the Ward identities, we expect that also their high 
energy behavior does not satisfy the requirements listed above, i.e. 
the large-$\beta$ behavior does not allow to redraw the contour as indicated 
in Fig.6. Consequently, when computing the different pieces illustrated in 
Fig.4, we will attempt to first isolate and remove the potentially dangerous reggeizing 
pieces, and retain only those such for which Ward identities and large-$\beta$ behavior  
are in accordance with what has been postulated before. This goal is achieved by 
the decomposition described in ~\cite{Bartels:1994jj} where, for the subamplitueds 
$D_4$ the reggeizing pieces, $D_4^R$, have been separated from the remaining 
part, $D_4^I$. The latter ones, in fact, satisfy the Ward identities and are
fully symmetric under permutations of the outgoing gluons, whereas the former ones do
not. Consequently we expect (although this has not fully been proven yet) that
also their large $\beta$ behavior is 'good'.        

After these general remarks it is fairly straightforward to follow the procedure 
outlined above and to obtain the 
different cut contributions in Fig.4. As an example, consider Fig.4a. On the 
lhs of the discontinuity line we need the set of production amplitudes 
illustrated in Fig.7a. 
Restricting ourselves to the 
(generalized) leading-log approximation and to even signature in the 
lower $t$-channel, they contain only single energy discontinuities (Fig.7b), 
from which one easily reconstructs the full production amplitudes (by simply 
multiplying by $i\pi$). Taking the square of these production amplitudes,
we see that the triple discontinuity is sufficient to obtain the 
energy discontinity in Fig.4a (Fig.7c). As seen in Fig.7c, there is a 'last' 
interaction between the two lower ladders: the sum of all diagrams above 
this last interaction (including the last rung) coincides with the amplitude
$D_4$ analyzed in ~\cite{Bartels:1994jj}. Using the separation $D_4= D_4^R +
D_4^I$ which has been described in detail, and retaining only $D_4^I$,
we arrive at the QCD result for the subamplitude in  Fig.4a, $N_4$.
It is important to stress that this amplitude is completely symmetric under the exchange of 
any two gluons below.
In an analogous way one computes the other cut-contributions in Figs.4b and c.     
When adding Figs4a, b, and c and making use of the symmetry (under permutations) of $D_4^I$: . 
we can combine all contributions Fig.4a - c in the way outlined above.
The symmetry of the $D_4^I$ under permutation of the lower gluons, together with 
the fulfillment of the Ward identities can be viewed as strong hint that 
also the large-$\beta$ behavior satifies the requirements discussed before.
This then allows, in particular, to draw all three integration contours 
of Fig.6 along the real axis: this property is required by the AGK rules.    
  
In the following section we turn to the inclusive cross section and 
compute the discontinuities shown in Figs.11a-c. We repeat the same steps 
as those for the total cross section, until we reach the analogue of 
$D_4$. In particular,we\\ 
(1) start from the triple discontinuites,\\  
(2) repeat the decomposition into 'reggeizing'and 'irreducible' 
pieces, filtering out those terms which do not satisfy the Ward identities 
and, hence, threaten to have a bad large-$\beta$-behavior.
This decomposition is different from the one carried out in
~\cite{Bartels:1994jj} for the total cross section,
and it represents the main achievement of this paper.\\
(3) The remaining terms (the analogue of $D_4^I$) have to computed for
each term in Fig.11. We shall find that they satisfy the symmetry requirements
and Ward identitites,
which, however, are less restrictive than in the case of the total cross section.

As a result, in the inclusive case the different subamplitudes in
Figs.11a-c, $N_4^c$, are no longer identical.
Our final cross section, therefore, will be written as in (\ref{sumofcuts2}):
it consist of several pieces which cannot be combined
in a simple way.

\section{The cut amplitudes $N_4^c$ in QCD}
\subsection{Review of the total cross section}
We begin with a brief review of the amplitudes $N_4$ which enter the total  
cross section. As we have stated above, we start from triple discontinuities 
and, in a second step, decompose them into reggeizing pieces (which do not 
satisfy the Ward identities and have a 'bad large $\beta$' 
behavior), and a remainder with 'good properties'. In the notation of     
\cite{Bartels:1994jj}, they are denoted by $D_4^R$ and $D_4^I$, resp.
In the context of this short review, we also introduce a compact 
notation that will be used throughout the paper.

The triple discontiunuity, $D_4$, is illustrated in Fig.13.
\begin{center}
\epsfig{file=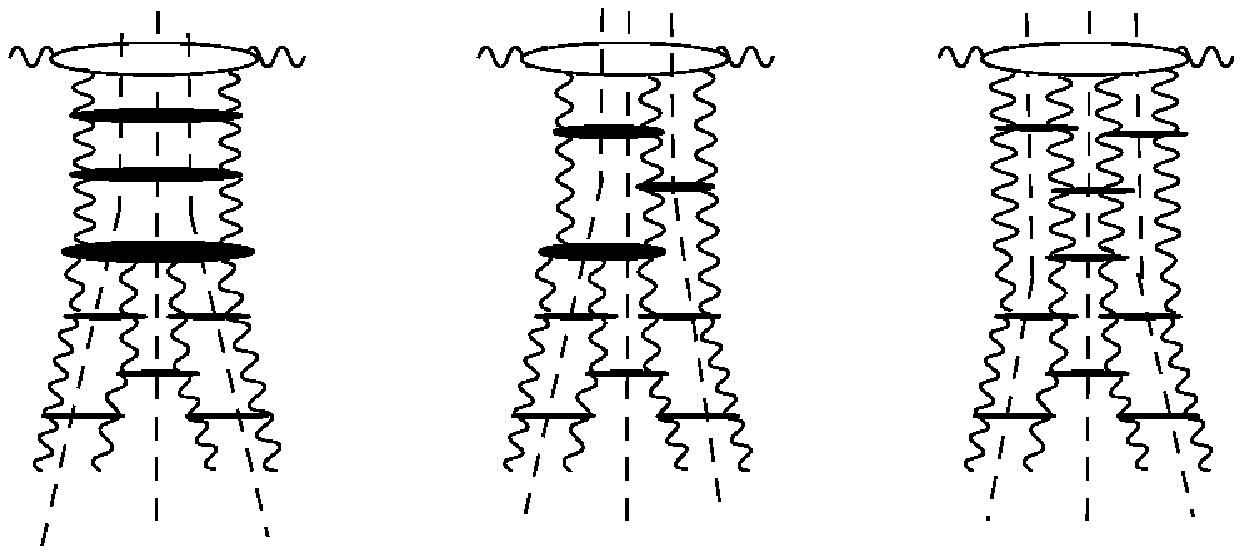,width=10cm,height=5cm}\\
Fig.13: illustration of the triple discontinuity
\end{center}
Here we have removed the couplings to the nucleons at the lower end:
for Fig.4a, we attach gluon $1$ and $2$ to nucleon $1$ and gluon 
$3$ and $4$ to nucleon $2$, for Fig.4c we attach gluon $1$ and $4$ to 
nucleon $1$ and so on. However, provided the triple discontinuity (Fig.8) 
is symmetric under the exchange 
of the lower gluons, all terms in Fig.4 are obtained from the 
same triple discontinuity, and the order in which the gluons are attached 
to the two nucleons does not matter. The summmation of all diagrams shown 
in Fig.13 is done in terms of integral equations. We introduce amplitudes
$D_2$ (associated to the BFKL
evolution~\cite{Kuraev:1976ge,Kuraev:1977fs,Balitsky:1978ic})
and $D_3$ which, together with $D_4$, satisfy a set of coupled integral equations (Fig.14).   
\begin{center}
\epsfig{file=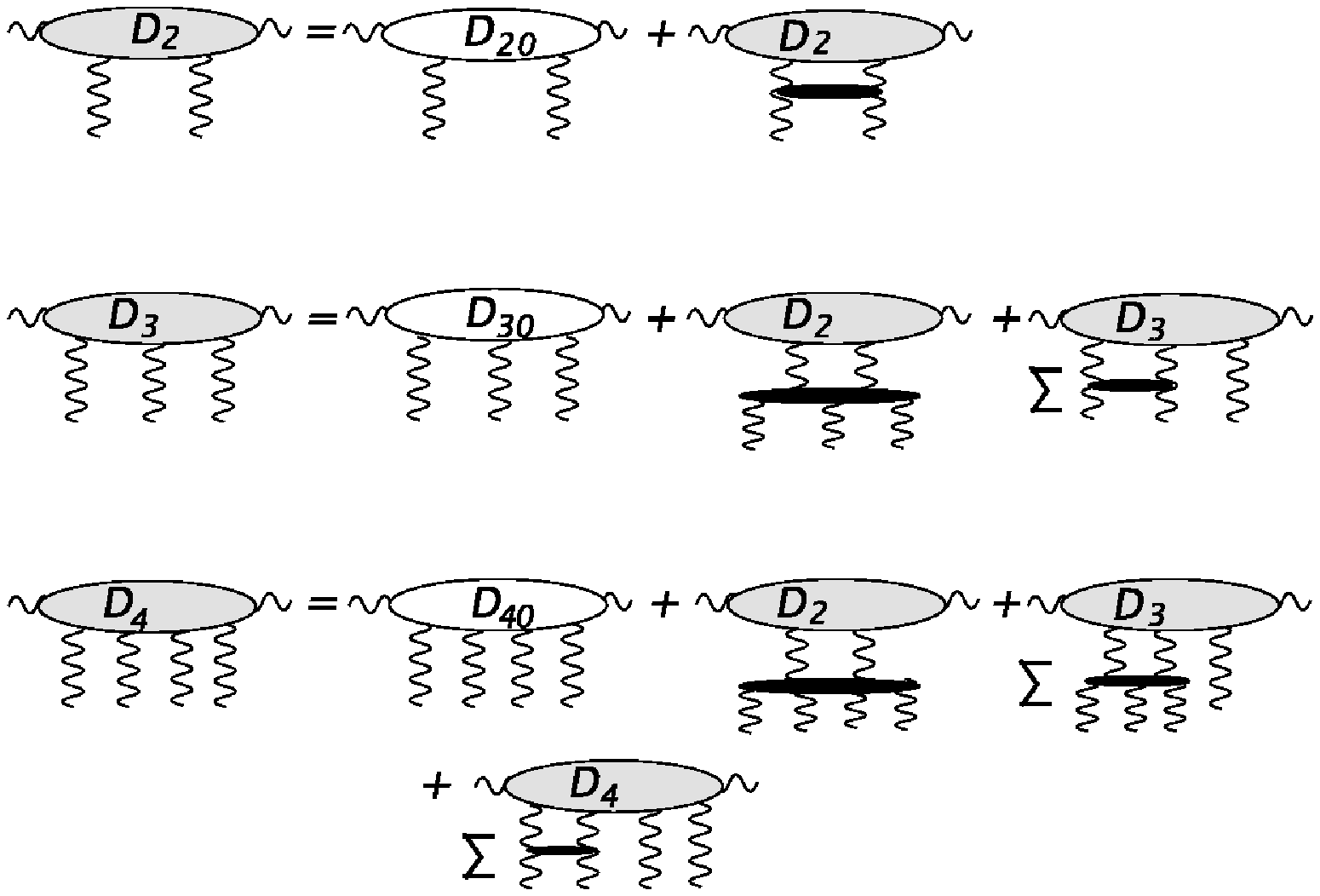,width=11cm,height=7cm}\\
Fig.14: integral equations for $D_2$, $D_3$, and $D_4$.
\end{center}

Writing these equation as evolution equations in rapidity $y=\ln s$, we 
find: 
\begin{subequations}
  \label{eq:DnReqInc}
  \begin{align}
    \label{eq:D2ReqInc}
    (\partial_y-\cH_2) D_2 &= \delta(y) D_{2;0} \\[10pt]
    \label{eq:D3ReqInc}
    (\partial_y-\cH_3) D_3 &= \delta(y) D_{3;0} +
    \cK_3 D_2 \\[10pt]
    \label{eq:D4ReqInc}
    (\partial_y-\cH_4) D_4 &= \delta(y) D_{4;0} + \nonumber \\
    &+\cK_4 D_2 +
    \oK{123}_3 ~\ov{D}{\cdot \cdot 4}{3}+
    \oK{124}_3 ~\ov{D}{\cdot 3 \cdot}{3}+
    \oK{234}_3 ~\ov{D}{1 \cdot \cdot}{3}+
    \oK{134}_3 ~\ov{D}{\cdot 2 \cdot}{3}
  \end{align}
\end{subequations}
with the boundary conditions $D_n(y)=0$ for $y<0$.
The notation used in these equations should be clarified by writing an 
explicit example:
\begin{equation}
  \label{eq:K3-D3-example}
  \begin{split}
  \oc{K}{123}{3} \;& \ov{D}{\cdot \cdot 4}{3} =
  \begin{minipage}{2cm}
    \includegraphics[width=2cm]{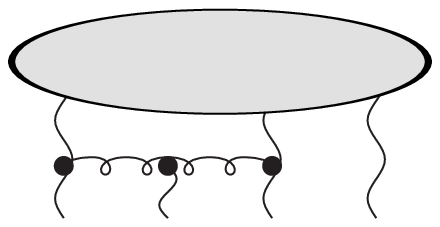}
  \end{minipage}
  = \\
  & = \cK_3(1,2,3;1', 2')
  \otimes D_3(1', 2', 4) \, ,
  \end{split}
\end{equation}
where the convolution '$\otimes$' denotes an integral in the
transverse momentum space and includes propagators. In our notation, 
the convolution acts on the primed variables.
The two dots above $D_3$ denote those gluon variables on which
$\cK_3$ acts: $D_3$ is a function of three gluon variables, and
$\cK_3$ acts just on those which are marked by the dots. In our 
example, this are the gluons $1$ and $2$, while the third gluon, $4$, 
remains a spectator.

We now list the operators $\cK_n$ and $\cH_n$ appearing in
(\ref{eq:DnReqInc}a-c). The former \cite{Bartels:1978fc,Bartels:1980pe} 
are integral kernels which describe the transition from $2$ to $n$ 
reggeized gluons in the $t$-channel. The latter are the BKP hamiltonians
\cite{Bartels:1978fc,Bartels:1980pe,Kwiecinski:1980wb,Jaroszewicz:1980mq},
which generalize the BFKL hamiltonian $\cH_2$,
and describe the interaction of a fixed number $n$ of reggeized gluons;
we will denote their Green's function $\cG_n$. All these objects are
integral operators acting in the transverse momentum and color spaces.
The integral kernels $K_n$ of $\cK_n$ are:
\begin{equation}
  \label{eq:Knkernel}
  \begin{split}
    K_n&(\bk_1,\bk_2,...,\bk_n;\bk'_1,\bk'_2)=\\
    &\frac{g^n}{(2 \pi)^3} \bigg( \bk_{12...n}^2
    -\frac{\bk_{12...n-1}^2\bk'^2_2}{(\bk'_2-\bk_n)^2}
    -\frac{\bk_{23...n}^2\bk'^2_1}{(\bk'_1-\bk_1)^2}
    +\frac{\bk'^2_1\bk'^2_2\bk_{2...n-1}^2}{(\bk'_1-\bk_1)^2(\bk'_2-\bk_n)^2}
    \bigg)\, ,
  \end{split}
\end{equation}
and the action of $\cK_n$ on a two point function $\phi(\bk_1,\bk_2)$
is given by
\begin{equation}
  \label{eq:Knphi}
  \begin{split}
    \cK_n \phi&(\bk_1,...,\bk_n) =
    \begin{minipage}{2cm}
      \includegraphics[width=2cm]{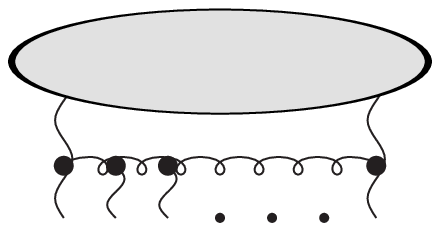}
    \end{minipage} =\\&
    =\int \!\! \frac{d^2\bk'_1d^2\bk'_2}{\bk'^2_1\bk'^2_2}
    \delta^{(2)}(\bk_{1...n}-\bk'_{12}) \,
    K_n(\bk_1,...,\bk_n;\bk'_1,\bk'_2)
    \phi(\bk'_1,\bk'_2) \, .
  \end{split}
\end{equation}
We have introduced the notation $\bk_{ijk\ldots}=\bk_i+\bk_j+\bk_k+\ldots$
for the sum of transverse momenta.
The Lipatov kernel $K_2$ is can be obtained from $K_3(\bk_1,\bzero,\bk_2)$,
where the last term in \eqref{eq:Knkernel} vanishes.
In the color space these integral operators are multiplied 
by color tensors originating from the gluon vertices:
\begin{equation}
  \label{eq:KnphiColor}
  f^{a'_1 a_1 b_1} f^{b_1 a_2 b_2} ... f^{b_{n-1} a_n a'_2}
  \cK_n \phi^{a'_1 a'_2} \, .
\end{equation}

The virtual corrections are encoded in the gluon Regge trajectory function 
$\o$, whose action on a function $\phi$ is multiplicative in momentum space:
\begin{equation}
  \label{eq:OmegaOp}
  \oo{i} \, \phi(\bk_1,\bk_2) =
  -\frac{N_c}{2} \o(\bk_i) \, \phi(\bk_1,\bk_2) \quad i=1,2 \, ,
\end{equation}
with the function $\o(\bk)$ being \footnote{A regularization
of the IR divergences is understood.}
\begin{equation}
  \label{eq:omega}
  \o (\bk) =
  \frac{g^2}{(2\pi)^3}
  \int\!\! d^2\bk' \frac{\bk^2}{\bk'^2(\bk-\bk')^2} \, .
\end{equation}

The BKP hamiltonians $\cH_n$ are defined as
\begin{equation}
  \label{eq:Hn}
  \cH_n = \sum_{i=1}^{n} \oo{i} +
  \sum_{1 \le i < j \le n}
  \vec{t}_i \cdot \vec{t}_j \; \overset{ij}{\cK_2} \, ,
\end{equation}
where we have introduced the $SU(N_c)$ generators in the adjoint
representation $\vec{t}_i = {}^t (t^1_{a_i a'_i},...,t^{N_c^2-1}_{a_i a'_i})$
with $t^b_{a_i a'_i} = i f^{a_i b a'_i}$.
The BKP Green's functions $\cG_n$ satisfy the equations
\begin{equation}
  \label{eq:Gn}
  (\partial_y - \cH_n) \; \cG_n(y) = \delta (y) \, ,
\end{equation}
with the formal solutions
\begin{equation}
  \label{eq:GnFormExp}
  \cG_n(y) = \Theta(y) e^{y \cH_n} \, .
\end{equation}

The action of $\cH_2$ on a color singlet function
$\phi^{a_1 a_2} = \delta^{a_1 a_2} \phi$ gives the BFKL hamiltonian:
\begin{equation}
  \label{eq:BFKLhamiltonian}
  \cH_2 \phi^{~a_1 a_2} =
  \delta^{a_1 a_2} \Big(
  \oo{1} + \oo{2} - N_c \cK_2
  \Big) \phi \, .
\end{equation}
When acting, in a color octect state, on a function which depends only on the 
sum of transverse momentum of the two gluons:
$\psi^{a_1 a_2} = f^{a_1 a_2 b} \tilde{\psi}^b (\bk_{12})$,
the hamitonian leads to the \emph{bootstrap equation}:
\begin{equation}
  \label{eq:bootstrap}
  \cH_2 \psi^{~a_1 a_2} =
  -\frac{N_c}{2} \o(\bk_{12}) \psi^{a_1 a_2} \, .
\end{equation}

Finally, the initial conditions $D_{n;0}$ are the lowest order
\emph{impact factors} for the coupling of $n$ reggeized gluons
to the external photon at rapidity $y=0$.
These couplings are given by a simple quark loop.

Eq. \eqref{eq:D2ReqInc} is just the BFKL equation
\cite{Kuraev:1976ge,Kuraev:1977fs,Balitsky:1978ic},
starting from the initial condition $D_{2;0}$.
Its solution, formally given by
\begin{equation}
  \label{eq:BFKLsol}
  D_2(y) = \cG_2 (y) D_{2;0} \, ,
\end{equation}
can be solved explicitly, thanks to the invariance of the BFKL
equation under M\"obius transformation \cite{Lipatov:1985uk,Bartels:2005ji}. It 
satisfies the Ward identity, i.e. it vanishes as one of the  
gluons carries zero momentum, and it is symmetric under the exchange of 
the two gluons. This property is crucial to have the possibility to obtain a
dual description, the dipole picture~\cite{Mueller:1993rr,Nikolaev:1994uu},
as has been discussed in~\cite{Bartels:2004ef}.

Green's functions for a higher number $n$ of reggeized gluons have been widely
studied: the case $n=3$ is associated to the Odderon exchange and is a completely
integrable problem~\cite{integrab}; the solutions have been found~\cite{JW,BLV} and physical
amplitudes constructed\cite{BBCV,Bartels:2003zu}. For $n\ge 4$ the kernels lead to an integrable
problem only in the planar limit~\cite{integrab,dv-lip2,dkkm,Vacca:2000bk}
whereas even the estimate of non planar corrections is an extremely difficult
problem~\cite{Lotter:1996vk,Iafelice:2007dc}.
Let us note that the integrability found in this framework is the first example of integrable
structures present in gauge theories and now such symmetries are deeply investigated in the
framework of the AdS/CFT correspondence between $N=4$ SYM theories and
superstring sigma models.  

Let us continue to discuss the results for the case discussed, wherein the
number of reggeizing gluons in the $t$-channel may change.
For the amplitudes $D_3$ and $D_4$ it will be necessary to isolate the 
reggeizing pieces. Beginning with $D_3$, the particular form of $D_{3;0}$ in
eq. \eqref{eq:D3ReqInc} allows to write the solution in the following form:
\begin{equation}
  \label{eq:D3sol}
  \begin{split}
    D_3 &= \frac{g}{2} f^{a_1 a_2 a_3} \bigg(
    \ov{D}{(12)3}{2} -
    \ov{D}{(13)2}{2} +
    \ov{D}{1(23)}{2}
    \bigg)   = \\
    & = \frac{1}{2} \Bigg(
    \begin{minipage}{1.5cm}
      \includegraphics[width=1.5cm]{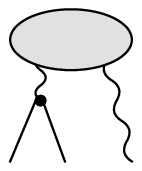}
    \end{minipage}
    +
    \begin{minipage}{1.5cm}
      \includegraphics[width=1.5cm]{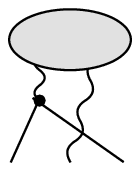}
    \end{minipage}
    +
    \begin{minipage}{1.5cm}
      \includegraphics[width=1.5cm]{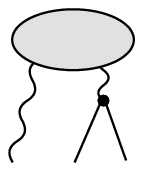}
    \end{minipage}
    \Bigg) \, .
  \end{split}
\end{equation}
Here we have introduced the notation
$\ov{D}{(12)3}{2} = D_2(\bk_{12}, \bk_3)$. $D_3$ is said to be
``reggeized'', in the sense
that a real three gluon state never appears until the last step
of the evolution, when the three gluon state is reached through
a local splitting of one of the reggeized gluons.
It is easy to see that $D_3$, as a function of its three gluon momenta 
and color labels, (i) does not satify the Ward identities (i.e. it does 
not vanish as $k_2$ goes to zero); (ii) individual terms are not symmetric 
under permutations of the gluons.

$D_4$ is more involved, and it contains both a reggeized part $D_4^R$
and an irreducible one $D_4^I$,
\begin{equation}
  \label{eq:D4RplusD4I}
  D_4 = D_4^R + D_4^I \, .
\end{equation}
This decomposition, from a diagrammatic point of view, is nothing but a 
reordering of the sum of diagrams in Fig.13. In the triple discontinuity 
illustrated in Fig.13. each horizontal line (or vertex) denotes an 
on-shell gluon, and each vertical wavy line a reggeized gluon. After the 
rearrangement we end up with the two terms of (\ref{eq:D4RplusD4I}).
The first term, $D_4^R$, is illustrated in Fig.15:
\begin{center}
\epsfig{file=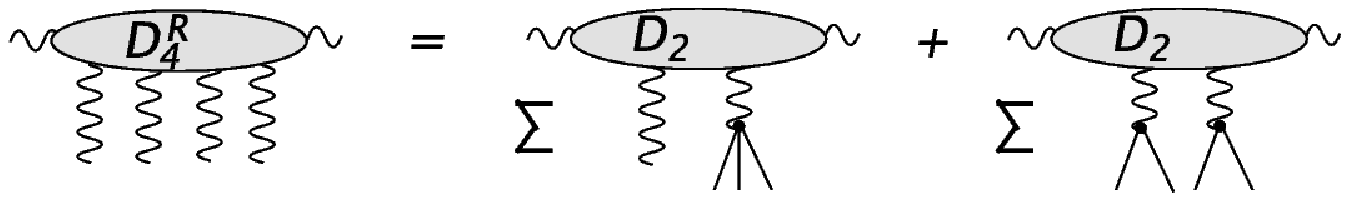,width=10cm,height=2.5cm}\\
Fig.15: illustration of $D_4^R$.
\end{center}
In detail, its structure is inferred from the 
initial condition:
\begin{equation}
  \label{eq:D40}
  \begin{split}
    D_{4;0} =
    &- g^2 d^{a_1 a_2 a_3 a_4} \bigg(
    \ov{D}{(123)4}{2;0} +
    \ov{D}{1(234)}{2;0} -
    \ov{D}{(14)(23)}{2;0} \bigg) + \\
    &- g^2 d^{a_1 a_2 a_4 a_3} \bigg(
    \ov{D}{(124)3}{2;0} +
    \ov{D}{2(134)}{2;0} -
    \ov{D}{(12)(34)}{2;0} -
    \ov{D}{(13)(24)}{2;0} \bigg) \, ,
  \end{split}
\end{equation}
and has the same form:
\begin{equation}
  \label{eq:D4R}
  \begin{split}
    D_4^R =
    &- g^2 d^{a_1 a_2 a_3 a_4} \bigg(
    \ov{D}{(123)4}{2} +
    \ov{D}{1(234)}{2} -
    \ov{D}{(14)(23)}{2} \bigg) + \\
    &- g^2 d^{a_1 a_2 a_4 a_3} \bigg(
    \ov{D}{(124)3}{2} +
    \ov{D}{2(134)}{2} -
    \ov{D}{(12)(34)}{2} -
    \ov{D}{(13)(24)}{2} \bigg) \, ,
  \end{split}
\end{equation}
The remainder, $D_4^I$, is illustrated in Fig.16:\\
\begin{center}
\epsfig{file=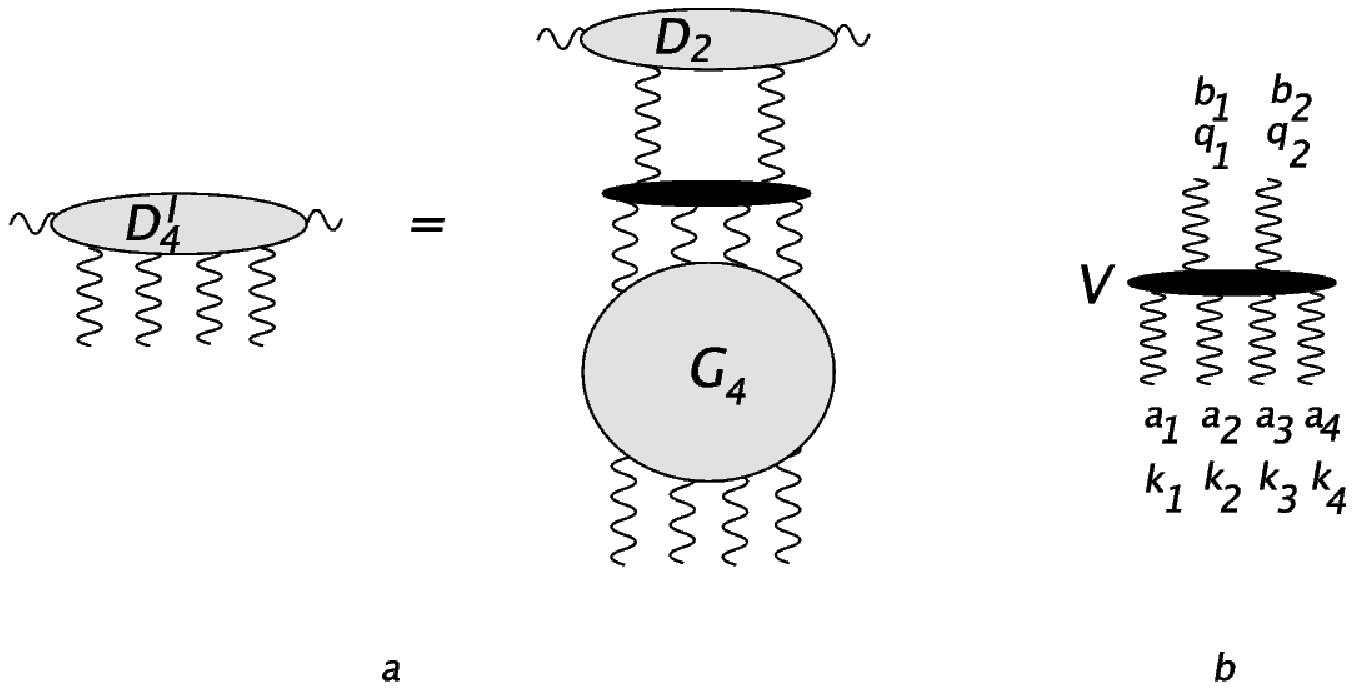,width=12cm,height=6cm}\\
Fig.16: illustration of $D_4^I$.
\end{center}
It has the appealing form:
\begin{equation}
  \label{eq:D4I}
  D_4^I (y) =
  \int_{0}^y \!\! dy'~
  \cG_4(y-y') \, \cV_4 \, D_2(y') \, .
\end{equation}
where the effective $2$-to-$4$ vertex, $\cV_4$, 
when acting on the space of 2-gluon gauge invariant functions,
has remarkable properties:\\
(i) it is infrared safe,\\
(ii) vanishes whenever one of the gluon momenta goes to $0$:
(Ward identities),\\
(iii) is completely symmetric in the 4 gluons and\\
(iv) is M\"obius invariant.\\
The explicit expression for $\cV_4$, first obtained in \cite{Bartels:1994jj},
can be found in appendix \ref{App:V4}. It is these 'good' properties which
support the expectation that the assumptions listed above are, in fact, 
satisfied. Finally we note that the vertex $\cV_4$ in Fig.16 contains 
disconnected (virtual) parts: they are analogous to the 'virtual' pieces 
inside the BFKL kernel which have their origin in the gluon trajectory 
function and do not contribute to $s$-channel gluon production.   

So far we have given attention only to the irreducible pieces,
$D_4^I$, which, because of their 'good' properties, represent the 
building blocks of the two-ladder contributions.  
The reggeizing pieces, $D_4^R$, provide a different class of corrections 
to $T_{\gamma^*(pn) \to \gamma^*(pn)}$. First we remind that these subamplitudes
(Fig.12), when considered as function of the {\it gluon momenta}, do not satisfy 
Ward identities and symmetry properties. However, as function of {\it reggeon
momenta} (e.g., in a piece of the the second term in Fig.15, 
$\ov{D}{(12)(34)}{2;0}$, as function of $\bk_1+\bk_2$), we again have 
the good properties (Ward identities). In this sense, the reggeizing pieces 
$D_4^R$ can be viewed as higher order corrections to $D_2$. Their 
contribution to $T_{\gamma^*(pn) \to \gamma^*(pn)}$ is illustrated in 
Fig.17:
\begin{center}
\epsfig{file=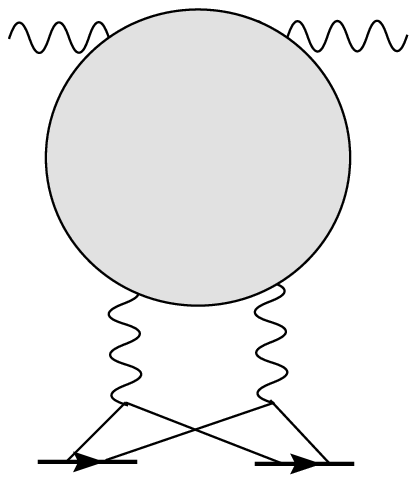,width=3cm,height=3cm}\\
Fig.17: illustration of $D_4^I$.
\end{center}
They contribute to the double cut, and they introduce higher order 
color correlators inside the two-nulceon target. This way of classifying 
corrections due to single, double, triple ... ladder exchanges 
can be viewed as a hierarchy: when generalizing the analysis of 
$D_4$ to $D_6$, the reggeizing pieces of $D_6$ 
contain contributions with four reggeizing gluons which, in the scattering 
of a photon on a nucleus with three gluons, will provide a two-ladder 
correction with higher correlators inside the three nucleon target. 
The analysis of $D_6$ has been started in ~\cite{Bartels:1999aw}.  

\subsection{The single-jet inclusive cross section: integral equations}
After these preparations we now turn to the main part of this paper, 
the calculation of the 1-jet inclusive cross section.
Following the discussion in section 2, we again consider the triple 
discontinuities of Fig.13, keeping in mind that, for the inclusive 
jet cross section, one s-channel gluon 
is kept fixed, both in rapidity and in transverse momentum.  
Depending upon the position of the $s$-cut line (Fig.11) we are considering, 
the gluon with fixed kinematics, in Fig.13, can belong to the left, the 
central, or the right hand cut: we will label these three possibilities by
a subscript $j=1,2,3$, resp. Furthermore, inside the three different 
classes of contributions of Fig.14 the gluon can appear at different places,
inside a transition kernel or inside a rung connecting two $t$-channel 
gluons of a two-gluon, a three-gluon or of a four-gluon state.

Following \cite{MSthesis},
we define the triple discontinuities for single jet production,
$_jZ_n$, where $j$ indicates the position of the 
$s$-channel cut to which the jet belongs \footnote{
  Such a notation is suited for an easy generalization to the case of
  $m$-jet production: $_j^mZ_n$. In \cite{MSthesis} a tecnique based
  on generating functionals has been devised for the computation
  of the evolution equations for couplings with an arbitrary number of
  jets produced.}.
Later on, we will relate $_jZ_4$ to the subamplitudes $N_4^c$. They are 
functions of:
\begin{itemize}
\item
  the rapidity differences $y_1$ between the external photon and
  the emitted jet and the difference $Y-y_1$ between the jet and the reggeized gluons;
\item
  the tranverse momentum $\bp_1$ of the produced jet;
\item
  the tranverse momenta $\bk_i$ of the reggeized gluons;
\item
  the photon virtuality and polarization, encoded in the impact factors.
\end{itemize}
In the following we will omit to write these variables explicitely,
unless it is necessary or we feel that their explicit appearance would 
clarify the meaning of the expressions. 

The summation of all diagrams will be organized in integral equations as 
follows. We concentrate 
on the evolution below the jet, i.e. $y > y_1$. For this evolution we 
define, as initial conditions, $_iZ_{n;0}$, the sum of all diagrams above 
the jet vertex (including the vertex), such that the gluon generating 
the jet is inside the lowest kernel or rung.
It is then easy to see that 
the equations for $n=2,3,4$ read\footnote{Note that
$Z_2 \equiv {}_1Z_2$ since there is only one possible cut.}:
\begin{subequations}
  \label{eq:BSeqs}
  \begin{eqnarray}
    \label{eq:BSeq2}
    &&(\partial_y-\cH_2) \; Z_2 =
    \delta(y-y_1) \; Z_{2;0} \, , \\[10pt]
    \label{eq:BSeq3}
    &&(\partial_y-\cH_3) \; {}_iZ_3 =
    \delta(y-y_1) \; {}_iZ_{3;0} +
    \oK{123}_3 \; \oZ{\cdot \cdot}{}{2} \, , \qquad\qquad i=1,2 \\[10pt]
    \label{eq:BSeq4}
    &&(\partial_y -\cH_4) \; _iZ_4 =
    \delta(y-y_1) \; _iZ_{4;0} \, + \oc{K}{}{4} \,
    \oZ{}{}{2} + \nonumber \\
    && \qquad \qquad + \oc{K}{123}{3} \; \oZ{\cdot \cdot 4}{1}{3} +
    \oc{K}{124}{3} \; \oZ{\cdot 3 \cdot}{1}{3} +
    \oc{K}{234}{3} \; \oZ{1 \cdot \cdot}{2}{3} +
    \oc{K}{134}{3} \; \oZ{\cdot 2 \cdot}{2}{3} \; , \quad i=1,2,3
  \end{eqnarray}
\end{subequations}
They are similar to the equations for the inclusive couplings
$D_n \equiv {}_0Z_n$ in (\ref{eq:DnReqInc}a-c),
the only difference being the initial conditions.


Let us look in more detail at the initial conditions $_iZ_{n;0}$.
As a new ingredient we need to introduce
the cut operators $_j\sK_n$: they are the cut counterpart of
\eqref{eq:Knphi} in which the transverse momentum of the $s$-channel
gluon exchanged between the reggeized gluons $j$ and $j+1$ has been fixed
to $\bp$; we still sum over its color degree of freedom. Its explicit action
is defined as
\begin{equation}
  \label{eq:Kncut}
  \begin{split}
    _j\sK_n &\phi(\bp;\bk_1,...,\bk_n) = 
    \begin{minipage}{2cm}
      \includegraphics[width=2cm]{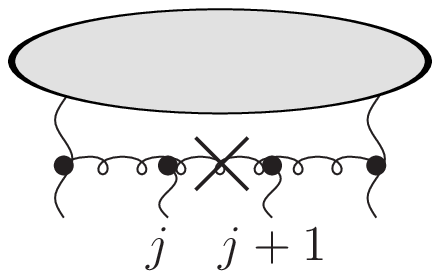}
    \end{minipage} =\\&
    =\frac{K_n(\bk_1,...,\bk_n;\bk_{1...j}+\bp,\bk_{j+1...n}-\bp)}
    {(\bk_{1...j}+\bp)^2(\bk_{j+1...n}-\bp)^2}
    \phi(\bk_{1...j}+\bp,\bk_{j+1...n}-\bp) \, ,
  \end{split}
\end{equation}
and in the color space we have the same tensor as in \eqref{eq:KnphiColor}.
With these cut kernels, the initial conditions appearing in the evolution 
equations (\ref{eq:BSeqs}a-c) are given by the following integral equations
\footnote{Note that $\sK_2 \equiv {}_1\sK_2$.
Here we have omitted to write explicitely the action in the color space
as has been shown in \eqref{eq:KnphiColor},
but it should be understood that they are present.}:
\begin{subequations}
  \label{eq:BSeqsInCond}
  \begin{eqnarray}
    && Z_{2;0} = \sK_2 \, D_2 \, , \\[10pt]
    && _1Z_{3;0} =
    \osK{12}{}{2} \, \ov{D}{\cdot \cdot 3}{3} +
    \osK{13}{}{2} \, \ov{D}{\cdot 2 \cdot}{3} +
    \osK{123}{1}{3} \, \ov{D}{\cdot \cdot}{2} \, , \\[10pt]
    && _2Z_{3;0} =
    \osK{23}{}{2} \, \ov{D}{1 \cdot \cdot}{3} +
    \osK{13}{}{2} \, \ov{D}{\cdot 2 \cdot}{3} +
    \osK{123}{2}{3} \, \ov{D}{\cdot \cdot}{2} \, , \\[10pt]
    && _1Z_{4;0} =
    \osK{12}{}{2} \, \ov{D}{\cdot \cdot 3 4}{4} +
    \osK{13}{}{2} \, \ov{D}{\cdot 2 \cdot 4}{4} +
    \osK{14}{}{2} \, \ov{D}{\cdot 2 3 \cdot}{4} + \nonumber \\ &&\qquad+
    \osK{123}{1}{3} \, \ov{D}{\cdot \cdot 4}{3} +
    \osK{124}{1}{3} \, \ov{D}{\cdot 3 \cdot}{3} +
    \osK{134}{1}{3} \, \ov{D}{\cdot 2 \cdot}{3} +
    \osK{1234}{1}{4} \, \ov{D}{\cdot \cdot}{2} \, , \\[10pt]
    && _2Z_{4;0} =
    \osK{13}{}{2} \, \ov{D}{\cdot 2 \cdot 4}{4} +
    \osK{14}{}{2} \, \ov{D}{\cdot 2 3 \cdot}{4} +
    \osK{23}{}{2} \, \ov{D}{1 \cdot \cdot 4}{4} +
    \osK{24}{}{2} \, \ov{D}{1 \cdot 3 \cdot}{4} + \nonumber \\ &&\qquad+
    \osK{123}{2}{3} \, \ov{D}{\cdot \cdot 4}{3} +
    \osK{124}{2}{3} \, \ov{D}{\cdot 3 \cdot}{3} +
    \osK{234}{1}{3} \, \ov{D}{1 \cdot \cdot}{3} +
    \osK{134}{1}{3} \, \ov{D}{\cdot 2 \cdot}{3} +
    \osK{1234}{2}{4} \, \ov{D}{\cdot \cdot}{2} \, , \\[10pt]
    && _3Z_{4;0} =
    \osK{14}{}{2} \, \ov{D}{ \cdot 2 3 \cdot}{4} +
    \osK{24}{}{2} \, \ov{D}{1 \cdot 3 \cdot}{4} +
    \osK{34}{}{2} \, \ov{D}{1 2 \cdot \cdot}{4} + \nonumber \\ &&\qquad+
    \osK{234}{2}{3} \, \ov{D}{1 \cdot \cdot}{3} +
    \osK{134}{2}{3} \, \ov{D}{\cdot 2 \cdot}{3} +
    \osK{124}{2}{3} \, \ov{D}{\cdot 3 \cdot}{3} +
    \osK{1234}{3}{4} \, \ov{D}{\cdot \cdot}{2} \, .
  \end{eqnarray}
\end{subequations}

The notation is the same as in section 3.1, except for the cut kernel 
$_j\sK_n$: here the subscript on the lhs denotes the position of the 
$s$-channel gluon which generates the jet.
A pictorial representation of one of the equations (\ref{eq:BSeqsInCond})
will illustrate their content:
\begin{equation}
  \label{eq:1Z30}
  _1Z_{3;0} = \sum
  \begin{minipage}{3cm}
    \includegraphics[width=3cm]{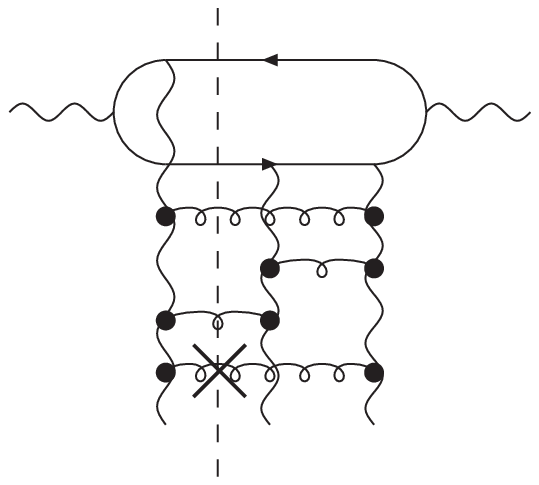}
  \end{minipage} +
  \sum
  \begin{minipage}{3cm}
    \includegraphics[width=3cm]{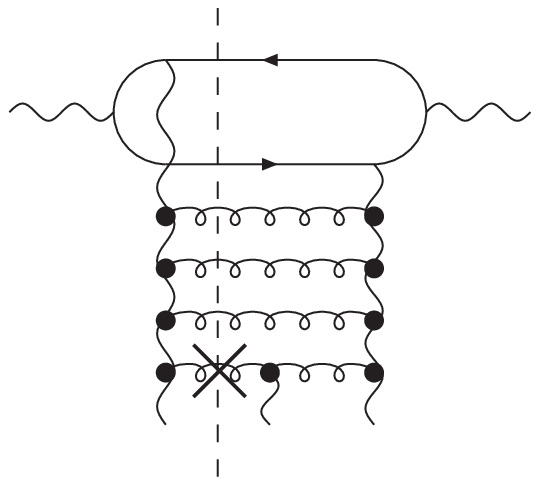}
  \end{minipage} \, .\nonumber
\end{equation}
The amplitude $_1Z_{3;0}$
contains the contributions from all the diagrams where the
jet is produced by the lowest $s$-channel gluon . Above, between the
external photon and the jet, the inclusive functions $D_2$ and $D_3$ appear.
We finally note that the eqs. (\ref{eq:BSeqs}a-c) with the initial conditions
(\ref{eq:BSeqsInCond}a-f) are free from infrared divergences.

\subsection{The single-jet inclusive cross section: reduction}
As the main step of our analysis we now perform the reduction which, 
similar to the case of the total cross section, separates the reggeizing pieces 
with 'bad properties' from those which satisfy Ward identities and 
symmetry requirements. However, once we fix the momenta of the jet, 
we can no longer expect to find the same symmetry properties as in the 
case of the total cross section. For example, in Fig.11a (the diffractive cut),
$N_4^c$ should be symmetric in gluon pair $1$ and $2$, and in the pair $3$ and
$4$, but not in $1$ and $3$ etc., in Fig.11b we expect symmetry in the triplet
$(123)$, and in Fig.11c $N_4^c$ is expected to be symmetric in the pairs
$(13)$ and $(24)$.
In other words, we expect full symmetry on each side of the cutting line but 
not across the cutting line. Nevertheless, we still will find some 
left-right symmetry: when summing over all different cuttings in
(\ref{sumofcuts2}), as for Fig.4 in (\ref{sumofcuts}), we also interchange
the ladders attached to nucleon $1$ and $2$, assuming even signature
in the $t$ channel.
This signature property will show up also in the inclusive cross section.

Following the strategy developed in \cite{Bartels:1994jj} for the total 
cross section, we begin with a careful analysis of the initial conditions, 
which serves as a guideline for the \emph{reggeization pattern}. 
As a result, the amplitudes $_jZ_n$ will be written as a sum of a 
\emph{reggeized part} (a linear combination of solutions with $<n$  
reggeized gluons) and a \emph{irreducible part} 
which satisfies Ward identities and symmetry properties:
\begin{equation}
 {}_iZ_n  = {}_iZ_n^R + {}_iZ_n^I
\end{equation}

\subsubsection*{2 Reggeized gluons:}
The simplest case of two gluons (eq. \eqref{eq:BSeq2}) is trivial: 
there is only one gluon on each side of the cutting line, and no  
reduction is necessary. The solution to the integral equation is 
the evolution of the initial condition
by means of the BFKL Green's function $\cG_2$:
\begin{equation}
  \label{eq:2Rsolution}
  Z_2^{a_1 a_2} = \big(
  \cG_2(y - y_1) \sK_2 (\bp_1) D_2 (y_1)
  \big)^{a_1 a_2}\, .
\end{equation}
More explicitely, since $D_2$ is a color singlet,
$D_2^{a_1 a_2} = \delta^{a_1 a_2} D_2$, we can use the well known
relation $f^{a'_1 a_1 b} f^{b a_2 a'_1} = -N_c \delta^{a_1 a_2}$ and
factorize the color tensor from \eqref{eq:2Rsolution}:
\begin{eqnarray}
  \label{eq:2RsolutionExpColor}
  Z_2^{~a_1 a_2} &=& \delta^{a_1 a_2} Z_2 \nonumber \\
  Z_2 &=& -N_c \cG_2(y - y_1) \sK_2 (\bp_1) D_2 (y_1) \, ,
\end{eqnarray}
where the operators are now those acting just in the transverse momentum space.

\subsubsection*{3 Reggeized gluons:}
The case of three gluons, ${}_iZ_{3;0}$, is already already more involved.
Namely the presence of the jet breaks the coherence in the initial conditions,
which, in the fully inclusive case, leads to the complete reduction
of $D_3$ in terms of $D_2$'s. In the present case this is no longer true.
Imposing the condition that, after subtraction of the reggeizing 
term ${}_iZ_{3}^R$, the irreducible piece ${}_iZ_{3}^I$ has to satisfy 
Ward identities, we find, after some calculations, that we have to form 
even and odd combinations 
\begin{equation}
\label{signature}
{}_iZ_{3}^{\pm}(\bp_1) = \frac{1}{2} 
\Big( {}_iZ_{3}(\bp_1) \pm {}_iZ_{3}(-\bp_1) \Big)
\end{equation}
Keeping in mind that, in order to arrive at the inclusive cross section,
all transverse momenta (except for $\bp_1$) will be integrated, we have 
complete azimuthal symmetry, and the negative signature combination does
not contribute. We note, however, that the appearance of even and odd 
combinations, from a signature point of view, is quite natural: 
in Fig.11a, the jet momentum $\bp_1$ is equal to the momentum transfer 
across the left lower Pomeron (flowing upwards) and across the right 
Pomeron (flowing downwards). When interchanging the nucleons below, we thus 
reverse the direction of the jet momentum. Therefore, the two combinations     
in (\ref{signature}) belong to even and odd symmetry under interchange 
of the lower Pomerons. This distinction will become relevant, for example, 
for 2-jet inclusive cross sections where azimuthal correlations come 
into play. In the following we will always refer to the even combination. 
We use the average symbol:
\begin{equation}
\label{eq:averagedCouplings}
\langle {}_iZ_{n} \rangle(\bp_1) = \frac{1}{2} 
\Big( {}_iZ_{n}(\bp_1) + {}_iZ_{n}(-\bp_1) \Big).
\end{equation}
These signatured combinations satisfy the same set of eqs. (\ref{eq:BSeqs}a-c),
with all the functions being replaced by their symmetrized counterpart.
This includes also the initial conditions (\ref{eq:BSeqsInCond}a-f).

Let us now proceed with the decomposition into reggeized and irreducible 
pieces: 
\begin{equation}
  \label{eq:nRG-RplusI}
  \obZ{}{i}{n} = \obZ{}{i}{n}^R + \obZ{}{i}{n}^I \, ,
\end{equation}
the latter defining new effective production vertices
which should satisfy Ward identities. Imposing this condition,
we find that the reggeized part has the same form as the one appearing in the 
inclusive coupling $D_3$ (see \eqref{eq:D3sol}):
\begin{equation}
  \label{eq:1Z3R}
  \begin{split}
    \obZ{}{1}{3}^R &= \frac{g}{2} f^{a_1 a_2 a_3} \Big(
    \obZ{(12)3}{}{2} - \obZ{(13)2}{}{2} + \obZ{1(23)}{}{2} \Big) = \\
    & = \frac{1}{2} \Bigg(
    \begin{minipage}{1.5cm}
      \includegraphics[width=1.5cm]{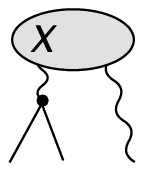}
    \end{minipage}
    +
    \begin{minipage}{1.5cm}
      \includegraphics[width=1.5cm]{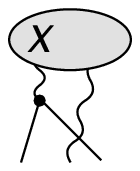}
    \end{minipage}
    +
    \begin{minipage}{1.5cm}
      \includegraphics[width=1.5cm]{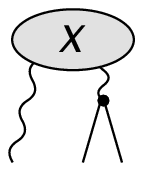}
    \end{minipage}
     \Bigg) \,
  \end{split}
\end{equation}
\begin{equation} 
\label{eq:2Z3R}
\begin{split}
    \obZ{}{2}{3}^R &= \frac{g}{2} f^{a_1 a_2 a_3} \Big(
    \obZ{(12)3}{}{2} - \obZ{2(13)}{}{2} + \obZ{1(23)}{}{2} \Big) = \\
    & = \frac{1}{2} \Bigg(
    \begin{minipage}{1.5cm}
      \includegraphics[width=1.5cm]{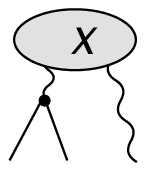}
    \end{minipage}
    +
    \begin{minipage}{1.5cm}
      \includegraphics[width=1.5cm]{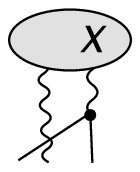}
    \end{minipage}
    +
    \begin{minipage}{1.5cm}
      \includegraphics[width=1.5cm]{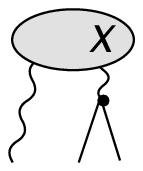}
    \end{minipage}
     \Bigg) \, .
  \end{split}
\end{equation}
Here $\obZ{}{}{2}$ is obtained from \eqref{eq:2RsolutionExpColor} and
\eqref{eq:averagedCouplings}:
\begin{equation}
  \label{eq:Z2justmomentumpart}
  \obZ{}{}{2} = -N_c \; \cG_2 \; \os{\Gamma}{}{}{2} \; D_2 \, .
\end{equation}
$\os{\Gamma}{}{}{2}$ is simply the symmetrized version of
$\sK_2$ in the jet transverse momentum,
\begin{equation}
  \label{eq:Gamma2}
  \os{\Gamma}{}{}{2} = \obs{K}{}{}{2} =
  \frac{1}{2} \big(
  \sK_2 (\bp_1) + \sK_2 (-\bp_1)
  \big) \, .
\end{equation}
On the rhs of (\ref{eq:1Z3R}), (\ref{eq:2Z3R}) the crosses mark the 
positions of the jet. In the first two terms of (\ref{eq:1Z3R}) and in the 
last two terms of (\ref{eq:2Z3R}), one of the 
reggeized gluons is cut. As an example, Fig.18 illustrates the inner 
structure of the first term:
\begin{center}
\epsfig{file=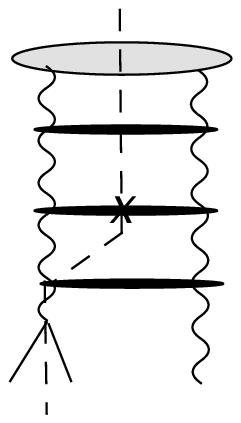,width=2cm,height=3cm}\\
Fig.18: discontinuity inside a cut gluon
\end{center}
On the rhs of (\ref{eq:1Z3R}), the sum of the first two terms is 
\emph{symmetric} 
under the exchange of gluon $2$ and $3$ (momenta and color), 
the third one is \emph{antisymmetric}. An analogous remark applies to (\ref{eq:2Z3R}).  

The remaining irreducible part contains new
effective production vertices $\os{\Gamma}{}{i}{3}$:
\begin{subequations}
  \label{eq:12ZA3I}
\begin{eqnarray}
  \label{eq:1ZA3I}
  \obZ{}{1}{3}^I &=&
  - N_c f^{a_1 a_2 a_3} \cG_3 \; \os{\Gamma}{}{1}{3} \; D_2 =
  \begin{minipage}{1.5cm}
    \includegraphics[width=1.5cm]{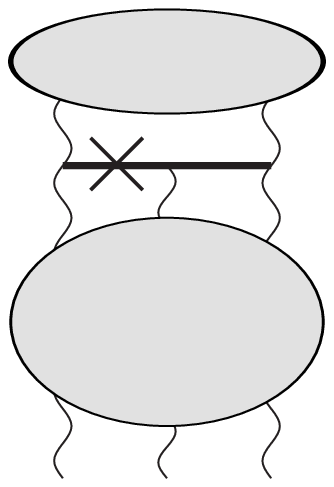}
  \end{minipage}
  \, , \\
  \label{eq:2ZA3I}
  \obZ{}{2}{3}^I &=&
  - N_c f^{a_1 a_2 a_3} \cG_3 \; \os{\Gamma}{}{2}{3} \; D_2 =
  \begin{minipage}{1.5cm}
    \includegraphics[width=1.5cm]{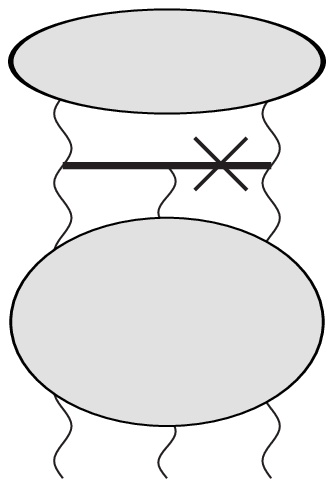}
  \end{minipage}
  \, ,
\end{eqnarray}
\end{subequations}
where the the cross marks the position of the produced gluon
inside the effective production vertices $\os{\Gamma}{}{i}{3}$.
The detailed analytic expression of the vertex is presented in Appendix \ref{App:iGamman},
eqs.(\ref{eq:Gamma3}), (\ref{eq:1Gamma3}), (\ref{eq:2Gamma3}).
It is important to point out that the $\os{\Gamma}{}{i}{3}$ ($i=1,2$), 
when acting 
on a gauge invariant impact factor $\phi$ with $\phi(\bk_j=\bzero)=0$
($j=1,2$), satisfy the required Ward identities:
\begin{equation}
  \label{eq:sJ3WardId}
  \left(\os{\Gamma}{}{i}{3}\right)  \phi (\bk_j=\bzero) = 0, \qquad j=1,2,3 \, ,
\end{equation}
Moreover, due to the symmetry properties of $\os{\Gamma}{}{i}{3}$
\begin{eqnarray}
  \label{eq:iGamma3SymProp}
  \os{\Gamma}{123}{1}{3} &=& - \os{\Gamma}{132}{1}{3} \, , \nonumber \\
  \os{\Gamma}{123}{2}{3} &=& - \os{\Gamma}{213}{2}{3} \, ,
\end{eqnarray}
and of the color tensor
$f^{a_1 a_2 a_3} = - f^{a_1 a_3 a_2} = -f^{a_2 a_1 a_3}$,
the amplitudes $\obZ{}{i}{3}^I$ are \emph{symmetric} under the exchange of
the two reggeized gluons on the same side of the cut (both color and momentum).

\subsubsection*{4 Reggeized gluons:}
For four reggeized gluons, $\obZ{}{i}{4}$, it is again the initial conditions 
which suggest the reggeization pattern.
Following the analysis of the total cross section, it is convenient to 
separate the reggeizing part into two pieces,
\begin{equation}
  \label{eq:4Req4R1plus4R2}
  \obZ{}{i}{4}^R = \obZ{}{i}{4}^{R1} + \obZ{}{i}{4}^{R2} \, .
\end{equation}
The $R1$ component is the same for any position of the $s$-channel cut,
$i=1,2,3$, and it coincides with the expression obtained in
\cite{Bartels:1994jj} for the reggeized part of the inclusive coupling
$D_4$,
\begin{equation}
  \label{eq:4R1}
  \begin{split}
    \obZ{}{i}{4}^{R1} &=
    - g^2 d^{a_1 a_2 a_3 a_4} \bigg(
    \obZ{(123)4}{}{2} +
    \obZ{1(234)}{}{2} -
    \obZ{(14)(23)}{}{2} \bigg) + \\
    &- g^2 d^{a_1 a_2 a_4 a_3} \bigg(
    \obZ{(124)3}{}{2} +
    \obZ{(134)2}{}{2} -
    \obZ{(12)(34)}{}{2} -
    \obZ{(13)(24)}{}{2} \bigg) 
\end{split}
\end{equation}
Here we have introduced another compact notation, e.g.
$\obZ{(123)4}{}{2} = \obZ{}{}{2}(\bk_{123},\bk_4)$ and
$\obZ{(12)(34)}{}{2} = \obZ{}{}{2}(\bk_{12},\bk_{34})$.
For the case $i=2$ (where the cut runs between reggeon $2$ and $3$) 
we illustrate this equation as follows. 
\begin{eqnarray}
\obZ{}{2}{4}^{R1} &=&  
  \label{Z4R1}
\begin{minipage}{1.5cm}
    \includegraphics[width=1.5cm]{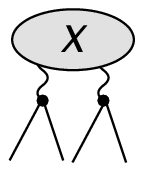}
  \end{minipage} + 
  \begin{minipage}{1.5cm}
    \includegraphics[width=1.5cm]{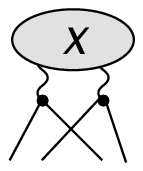}
   \end{minipage} +
\begin{minipage}{1.5cm}
    \includegraphics[width=1.5cm]{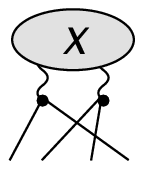}
   \end{minipage}  \nonumber \\ & +&
 \begin{minipage}{1.5cm}
    \includegraphics[width=1.5cm]{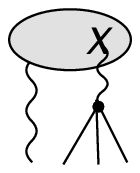}
   \end{minipage} +
\begin{minipage}{1.5cm}
    \includegraphics[width=1.5cm]{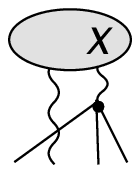}
   \end{minipage} +
 \begin{minipage}{1.5cm}
    \includegraphics[width=1.5cm]{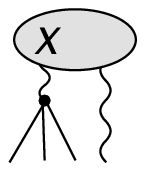}
   \end{minipage} +
  \begin{minipage}{1.5cm}
    \includegraphics[width=1.5cm]{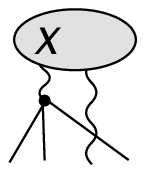}
   \end{minipage}\,.
\end{eqnarray}
The interpretation is analogous to the discussion after (\ref{eq:Gamma2}).
In the first diagram in the first line the cut runs between the reggeons.
All diagrams on the second line contain a cut reggeon; if we open any 
of these diagrams we find structures like those of Fig.18. The second and third 
diagrams of the first line have both reggeons cut. 

The $R2$ component is different for each
cut and is expressed in term of the vertex $\os{\Gamma}{}{}{3}$
defined in \eqref{eq:Gamma3},
\begin{subequations}
  \label{eq:4R2-123}
\begin{eqnarray}
  \obZ{}{1}{4}^{R2} &=&
  ~~gN_c d^{a_1 a_2 a_3 a_4} \bigg(
  \oc{G}{1(23)4}{3} - \oc{G}{14(23)}{3} \bigg) \os{\Gamma}{}{}{3} \;
  D_2 + \nonumber \\
  \label{eq:4R2-1}
  &&+gN_c d^{a_1 a_2 a_4 a_3} \bigg(
  \oc{G}{1(24)3}{3} - \oc{G}{13(24)}{3} \bigg) \os{\Gamma}{}{}{3} \;
  D_2 + \\
  &&+gN_c d^{a_1 a_3 a_4 a_2} \bigg(
  \oc{G}{1(34)2}{3} - \oc{G}{12(34)}{3} \bigg) \os{\Gamma}{}{}{3} \;
  D_2 \nonumber \\
  &=&\begin{minipage}{1.5cm}
    \includegraphics[width=9cm]{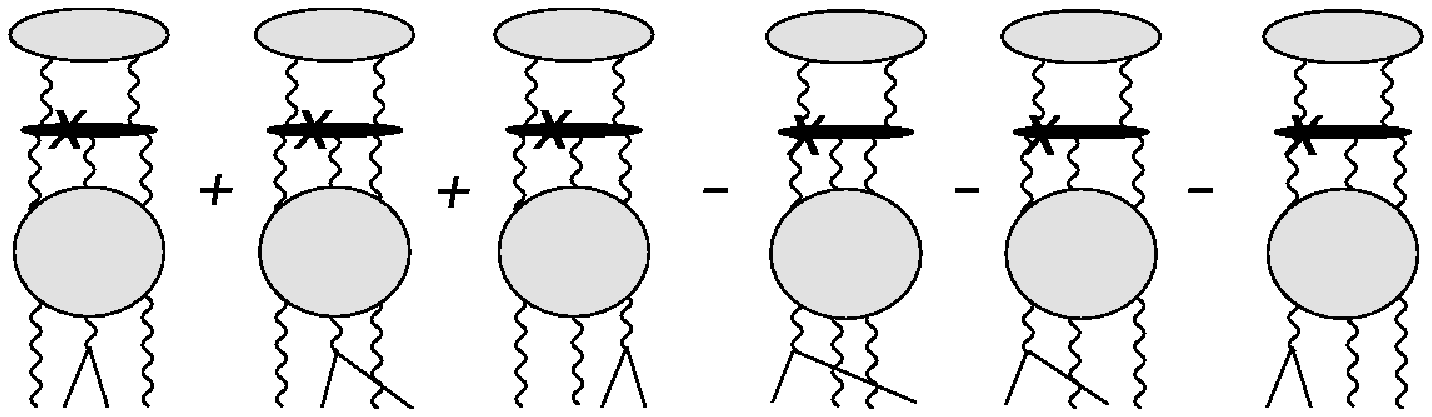}
   \end{minipage}
 , \nonumber \\[10pt]
  \obZ{}{2}{4}^{R2} &=&
  ~~gN_c d^{a_1 a_2 a_3 a_4} \bigg(
  \oc{G}{1(23)4}{3} + \oc{G}{2(14)3}{3} -
  \oc{G}{(12)43}{3} - \oc{G}{21(34)}{3} \bigg) \os{\Gamma}{}{}{3} \;
  D_2 + \\
  \label{eq:4R2-2}
  &&+gN_c d^{a_1 a_2 a_4 a_3} \bigg(
  \oc{G}{1(24)3}{3} + \oc{G}{2(13)4}{3} -
  \oc{G}{(12)34}{3} - \oc{G}{12(34)}{3} \bigg) \os{\Gamma}{}{}{3} \;
  D_2 \nonumber \\
  &=&\begin{minipage}{1.5cm}
    \includegraphics[width=11cm]{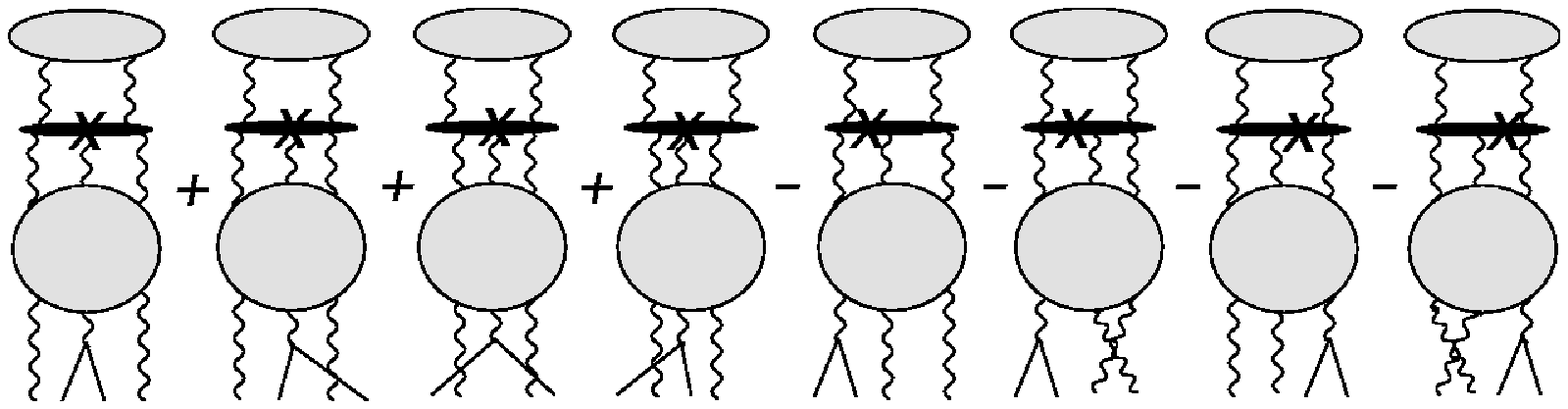}
   \end{minipage}
  \nonumber \\[10pt]
  \obZ{}{3}{4}^{R2} &=&
  ~~gN_c d^{a_1 a_2 a_3 a_4} \bigg(
  \oc{G}{1(23)4}{3} - \oc{G}{(23)14}{3} \bigg) \os{\Gamma}{}{}{3} \;
  D_2 + \nonumber \\
  \label{eq:4R2-3}
  &&+gN_c d^{a_2 a_1 a_3 a_4} \bigg(
  \oc{G}{2(13)4}{3} - \oc{G}{(13)24}{3} \bigg) \os{\Gamma}{}{}{3} \;
  D_2 + \\
  &&+gN_c d^{a_3 a_1 a_2 a_4} \bigg(
  \oc{G}{3(12)4}{3} - \oc{G}{(12)34}{3} \bigg) \os{\Gamma}{}{}{3} \;
  D_2 \,  \nonumber
\end{eqnarray}
\end{subequations}
(in the last equations, the diagrams are analogous to those of 
the first equation, eq.(\ref{eq:4R2-3})). Let us note that
eqs. (\ref{eq:4R2-1}) and (\ref{eq:4R2-3}) can be easily written in terms of  
$\os{\Gamma}{}{1}{3}$ and $\os{\Gamma}{}{2}{3}$  making use of
the relations (\ref{eq:1Gamma3}), (\ref{eq:2Gamma3}) and then one may recognize
a form with a sum of three terms $\obZ{}{1}{3}^I$
and $\obZ{}{2}{3}^I$ respectively, with a gluon splitting at rapidity $Y$. 

The irreducible part of $\obZ{}{i}{4}$ consists of four pieces:
\begin{equation}
  \label{eq:4IeqI1pI2pI3pI4}
  \obZ{}{i}{4} =
  \obZ{}{i}{4}^{I1} + \obZ{}{i}{4}^{I2} +
  \obZ{}{i}{4}^{I3} + \obZ{}{i}{4}^{I4} \, .
\end{equation}
In the first term the jet emission is above 
the effective vertex $\oc{V}{}{4}$, inside the BFKL ladder. 
Here all values of $i$ lead to the same expression, 
i.e. the contribution is independent of the position of the cut, 
\begin{equation}
  \label{eq:iZ4I1}
  \obZ{}{i}{4}^{I1} = \int_{y_1}^{y} \!\!\! dy' \;
  \cG_4(y - y') \; \oc{V}{}{4} \; \langle Z_2 \rangle (y') =
  \begin{minipage}{2cm}
    \includegraphics[width=2cm]{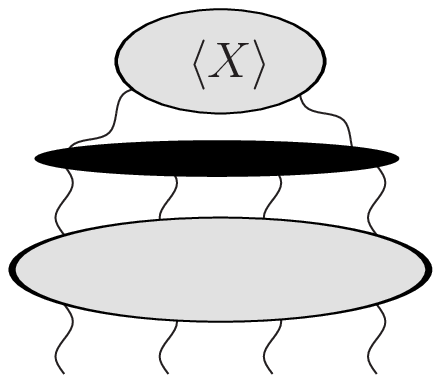}
  \end{minipage} \, .
\end{equation}
The appearance of the same vertex $\oc{V}{}{4}$ below the emission of
the jet is a remarkable result of our analysis: within our approach 
it is absolutely not trivial, since \emph{a priori} one might expect the emission of the jet to
break the reggeization pattern leading to $\oc{V}{}{4}$.

Let us stress that the $2\to4$ vertex is fully symmetric under the exchange of 
any pair of gluons, and it satisfies the Ward identities in all four gluon 
lines. This property implies that also the first and the second term 
in eq.(\ref{eq:4IeqI1pI2pI3pI4}) satisfy the Ward identities, and they have 
the required symmetry features on both sides of the cut.  

The second term can be illustrated by the following figure:
\begin{equation}
  \label{eq:Z4I2-pic}
  \obZ{}{i}{4}^{I2} = \sum
  \begin{minipage}{2cm}
    \includegraphics[width=2cm]{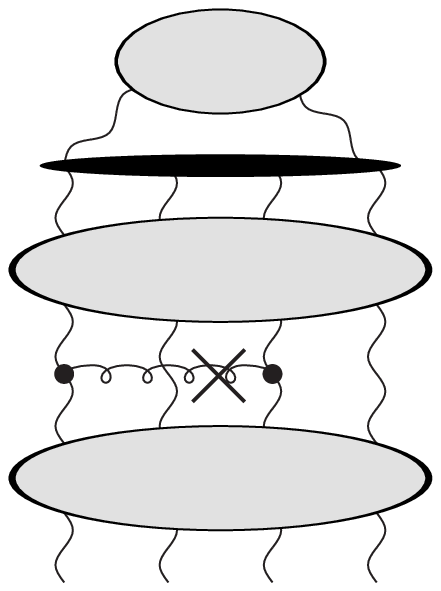}.
  \end{minipage} \, . \nonumber
\end{equation}
The jet is emitted below the $2 \to 4$ vertex, inside the four gluon state, 
and the label $i$ singles out the participating rungs. For example, 
for $i=1$ the possible rungs are between 
gluon $1$ and $2$, between $1$ and $3$, or between $1$ and $4$.
Above the emission we have the same structure, $D_4^I$, as the total cross 
section.
In particular, it contains, again, the $2\to4$ effective vertex $\oc{V}{}{4}$ of
\cite{Bartels:1994jj}. Writing as usual
\begin{eqnarray}
  \label{eq:D4RpD4I}
  D_4 &=& D_4^R + D_4^I \, \nonumber \\
  D_4^I &=& \int_{y_0}^{y_1} \!\!\! dy' \;
  \oc{G}{}{4}(y_1,y') \; \oc{V}{}{4} \; D_2(y') \, ,
\end{eqnarray}
we have
\begin{subequations}
\label{eq:iZ4I2}
\begin{eqnarray}
\label{eq:1Z4I2-1}
  \obZ{}{1}{4}^{I2} &=&
  \cG_4 \; \bigg(
  \os{\Gamma}{12}{}{2} + \os{\Gamma}{13}{}{2} + \os{\Gamma}{14}{}{2}
  \bigg) \; D_4^I \, , \\
\label{eq:2Z4I2-2}
  \obZ{}{2}{4}^{I2} &=&
  \cG_4 \; \bigg(
  \os{\Gamma}{13}{}{2} + \os{\Gamma}{14}{}{2} + \os{\Gamma}{23}{}{2} + \os{\Gamma}{24}{}{2}
  \bigg) \; D_4^I \, , \\
\label{eq:1Z4I2-3}
  \obZ{}{3}{4}^{I2} &=&
  \cG_4 \; \bigg(
  \os{\Gamma}{14}{}{2} + \os{\Gamma}{24}{}{2} + \os{\Gamma}{34}{}{2}
  \bigg) \; D_4^I \, ,
\end{eqnarray}
\end{subequations}

The third group of terms contains new effective production vertices
$\ocs{V}{}{i}{4}$:
\begin{equation}
  \label{eq:iZ4I3}
  \obZ{}{2}{4}^{I3} =
  \cG_4 \; \ocs{V}{}{2}{4} \; D_2 =
  \begin{minipage}{2cm}
    \includegraphics[width=2cm]{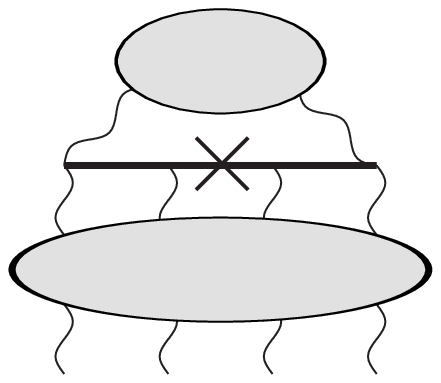}
  \end{minipage} \, ,
\end{equation}
with analogous expressions for $\obZ{}{1}{4}^{I3}$ and $\obZ{}{3}{4}^{I3}$.
The produced jet is inside the $2\to4$ transition vertex, and  
the new production vertices are conveniently expressed in terms of
new cut operators $\os{\Gamma}{}{i}{4}$ defined in Appendix
\ref{App:iGamman}, eq.(\ref{eq:iGamma4}):
\begin{subequations}
  \label{eq:iV4}
\begin{eqnarray}
  \label{eq:1V4}
  \ocs{V}{}{1}{4} &=& ~~\delta^{a_1 a_2} \delta^{a_3 a_4}
  \; \os{\Gamma}{1234}{1}{4} +
  \delta^{a_1 a_3} \delta^{a_2 a_4}
  \; \os{\Gamma}{1324}{1}{4} +
  \delta^{a_1 a_4} \delta^{a_2 a_3}
  \; \os{\Gamma}{1423}{1}{4} \, , \\
  \label{eq:2V4}
  \ocs{V}{}{2}{4} &=& ~~
  \delta^{a_1 a_2} \delta^{a_3 a_4}
  \; \os{\Gamma}{1234}{2A}{4} +
  \delta^{a_1 a_3} \delta^{a_2 a_4}
  \; \os{\Gamma}{1234}{2B}{4} +
  \delta^{a_1 a_4} \delta^{a_2 a_3}
  \; \os{\Gamma}{1243}{2B}{4} \, , \\
  \label{eq:3V4}
  \ocs{V}{}{3}{4} &=& ~~\delta^{a_1 a_2} \delta^{a_3 a_4}
  \; \os{\Gamma}{1234}{3}{4} +
  \delta^{a_1 a_3} \delta^{a_2 a_4}
  \; \os{\Gamma}{1324}{3}{4} +
  \delta^{a_2 a_3} \delta^{a_1 a_4}
  \; \os{\Gamma}{2314}{3}{4} \, .
\end{eqnarray}
\end{subequations}
One can show that the operators $\os{\Gamma}{}{i}{4}$ satify Ward identities.
This then also holds for the vertices (\ref{eq:iV4}a-c). 
Moreover, due to the symmetry properties of $\os{\Gamma}{}{i}{4}$,
\begin{eqnarray}
  \label{eq:iGammanSymProp}
  \os{\Gamma}{1234}{1}{4} &=& \os{\Gamma}{1243}{1}{4} \, , \nonumber \\[10pt]
  \os{\Gamma}{1234}{2A}{4} &=& \os{\Gamma}{1243}{2A}{4} =
  \os{\Gamma}{2134}{2A}{4} = \os{\Gamma}{2143}{2A}{4} \, , \nonumber \\[10pt]
  \os{\Gamma}{1234}{2B}{4} &=& \os{\Gamma}{2143}{2B}{4} \, , \\[10pt]
  \os{\Gamma}{1234}{3}{4} &=& \os{\Gamma}{2134}{3}{4} \, , \nonumber
\end{eqnarray}
$\ocs{V}{}{i}{4}$ are \emph{symmetric} under the exchange of any two gluons
(color and momentum) on each side of the $s$-channel cut.

The fourth group of terms part is novel and has no counterpart in the 
total cross section. Then transition from two gluons to four gluons now 
proceeds in two steps, and the produced jet is inside the $2 \to 3$ vertex. 
For the cut line on the lhs ($i=1$), it has the form
\begin{equation}
  \label{eq:1Z3I4}
  \obZ{}{1}{4}^{I4} =
  N_c \int_{y_1}^{y} \!\!\! dy' \;
  \cG_4(y - y') \; {}_1\oc{W}{}{4} \; \cG_3(y' - y_1) \; \os{\Gamma}{}{}{3}
  \; D_2 (y_1) =
  \begin{minipage}{2cm}
    \includegraphics[width=2cm]{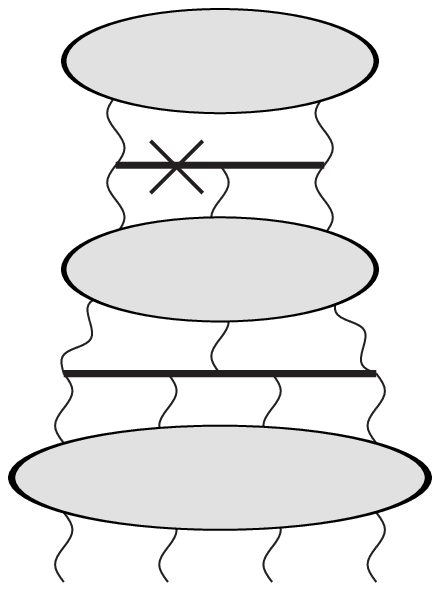}
  \end{minipage} \, .
\end{equation}
For the cut on the rhs $(i=3)$ we have an analogous expression, whereas 
the central cut $(i=2)$ receives two contributions:
\begin{equation}
  \label{eq:2Z3I4}
  \obZ{}{2}{4}^{I4} =
  N_c \int_{y_1}^{y} \!\!\! dy' \;
  \cG_4(y - y') \; {}_2\oc{W}{}{4} \; \cG_3(y' - y_1) \; \os{\Gamma}{}{}{3}
  \; D_2 (y_1) =
  \begin{minipage}{2cm}
    \includegraphics[width=2cm]{pics/1Z4I4.eps}
  \end{minipage}+
  \begin{minipage}{2cm}
    \includegraphics[width=2cm]{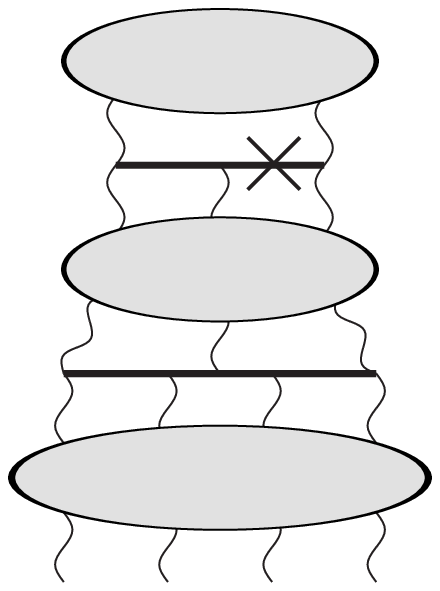}
  \end{minipage}
 \, .
\end{equation}
The $2\to3$ vertex with the jet is the same as introduced before and given in
appedix in eq. \eqref{eq:Gamma3}.
eq.(\ref{eq:12ZA3I}).
Below this vertex, a $t$-channel state of three reggeized
gluons appears which, after BKP evolution, through new effective vertices ${}_i\oc{W}{}{4}$,
turns into four reggeized gluons. These new $3 \to 4$ transition vertices
are conveniently expressed in terms of the integral (uncut) operators
$\op{\Gamma}{}{i}{4}$ which are listed in Appendix \ref{App:iGamman}
(eqs.(\ref{eq:iGammanUncut})):
\begin{subequations}
  \label{eq:iW4}
\begin{eqnarray}
  \label{eq:1W4}
  {}_1\oc{W}{}{4} &=&
  \delta^{a_1 a_2} \delta^{a_3 a_4}
  \; \op{\Gamma}{1234}{1}{4} +
  \delta^{a_1 a_3} \delta^{a_2 a_4}
  \; \op{\Gamma}{1324}{1}{4} +
  \delta^{a_1 a_4} \delta^{a_2 a_3}
  \; \op{\Gamma}{1423}{1}{4} \, , \\
  \label{eq:2W4}
  {}_2\oc{W}{}{4} &=&
  \delta^{a_1 a_2} \delta^{a_3 a_4}
  \; \op{\Gamma}{1234}{2A}{4} +
  \delta^{a_1 a_3} \delta^{a_2 a_4}
  \; \op{\Gamma}{1234}{2B}{4} +
  \delta^{a_1 a_4} \delta^{a_2 a_3}
  \; \op{\Gamma}{1243}{2B}{4} \, , \\
  \label{eq:3W4}
  {}_3\oc{W}{}{4} &=&
  \delta^{a_1 a_2} \delta^{a_3 a_4}
  \; \op{\Gamma}{1234}{3}{4} +
  \delta^{a_1 a_3} \delta^{a_2 a_4}
  \; \op{\Gamma}{1324}{3}{4} +
  \delta^{a_2 a_3} \delta^{a_1 a_4}
  \; \op{\Gamma}{2314}{3}{4} \, .
\end{eqnarray}
\end{subequations}
The $\op{\Gamma}{}{i}{4}$  have the same symmetry properties
\eqref{eq:iGammanSymProp} as their cut counterparts. Therefore, also 
the effective vertices ${}_1\oc{W}{}{4}$ are symmetric under the 
exchange of gluons on each side of the cut. Furthermore, they can be shown 
to satisfy Ward identities.
Again one may note that for the cut on the lhs ($i=1$) or on the rhs ($i=3$)
it is trivial to rewrite $\obZ{}{1}{4}^{I4}$, $\obZ{}{3}{4}^{I4}$ in terms of ${}_1\oc{W}{}{4}$
and $\os{\Gamma}{}{1}{3}$, ${}_3\oc{W}{}{4}$
and $\os{\Gamma}{}{2}{3}$, respectively.

Let us summarize our results for $\obZ{}{i}{4}$ in eqs.(\eqref{eq:BSeq4}).
For each position of the cutting line - denoted by $1=1,2,3$ - we have 
reggeizing and irreducible pieces. The irreducible pieces, for the case 
$i=1$ and $i=2$, are collected in Figs.19 and 20, resp.: 
\begin{center}
\epsfig{file=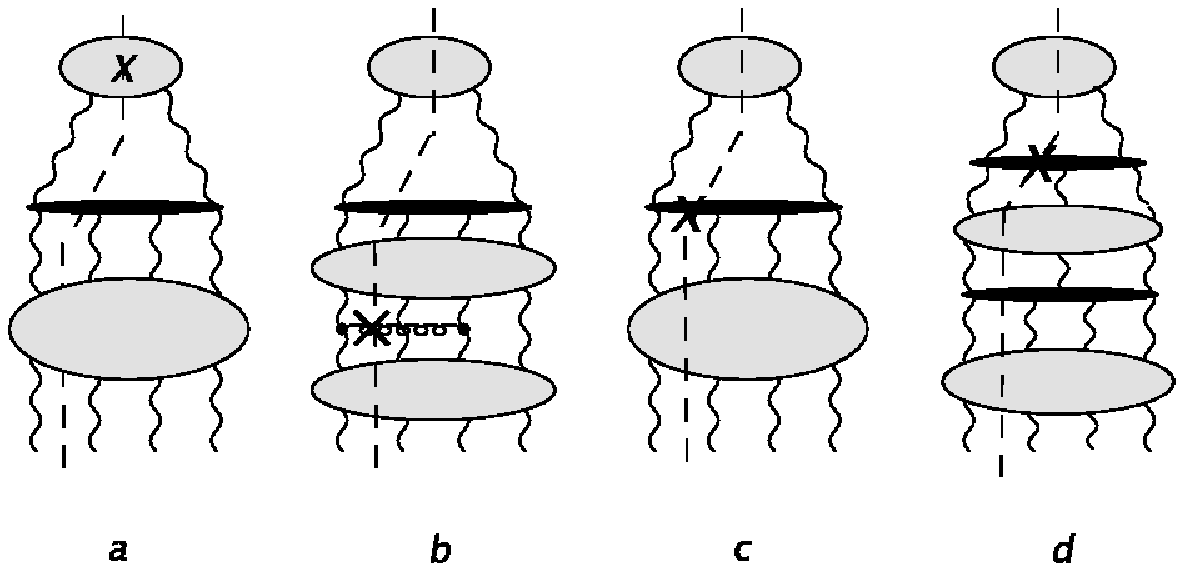,width=12cm,height=3.6cm}\\ 
Fig.19: the four pieces of the single jet inclusive cross section.\\
(a) and (b): production above and below the $2 \to 4$ transition,\\
(c) and (d): production inside the $2 \to 4$ transition.  
\end{center}
\begin{center}
\epsfig{file=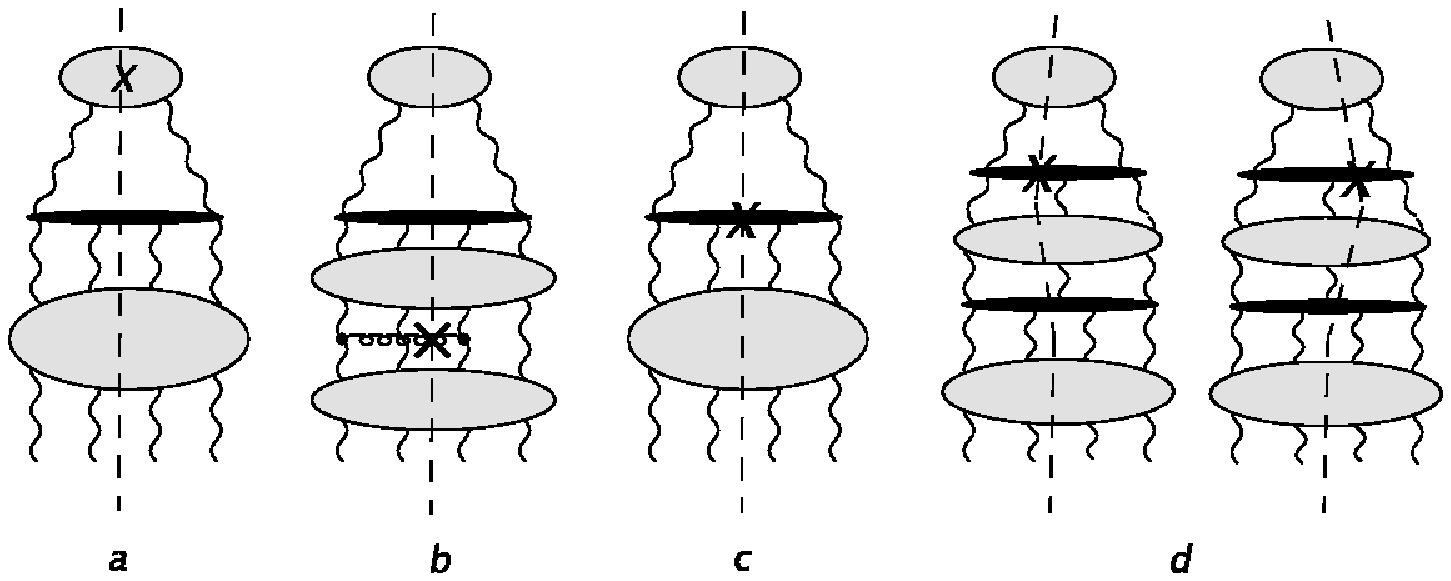,width=12cm,height=3.6cm}\\ 
Fig.20: the same as Fig.19, for the cut $i=2$.
\end{center}
They satisfy the 
Ward identities, and they are invariant under permutations of the gluons 
on both sides of the cut. 
They come in four different classes of contributions. If the jet is produced 
above or below the $2 \to 4$ transition vertex 
(groups $1$ and $2$, Figs.19 a and b),
the contributions are identical for all cuts (i.e. independent 
of $i$). The $2 \to 4$ vertex is the same as in the total cross section. 
As a result, these contributions can be added in the same way, as in the case 
of the total cross section, In particular, 
Group $2$ will cancel, due to the AKG counting rules~\cite{agk-qcd2}. 
If the jet is produced inside the $2 \to 4$ transition 
(groups $3$ and $4$, Figs.19c and d), 
the cuts $i=1,2,3$ differ from each other, and the vertices are new. 
In particular, there is a novel contribution (Fig.19d) which contains 
a $t$-channel state consisting of $3$ reggeized gluons.  

Finally, let us comment on the reggeizing pieces which do not satisfy 
Ward identities and symmetry requirements. Here we have found two groups 
which are illustrated in eqs.(\ref{Z4R1}) and 
(\ref{eq:4R2-1})-(\ref{eq:4R2-3}). As we have discussed at the end of 
section 3.1, these contributions introduce {\it higher order correlators} 
inside the target. We illustrate them in Fig.21:   
\begin{center}
\epsfig{file=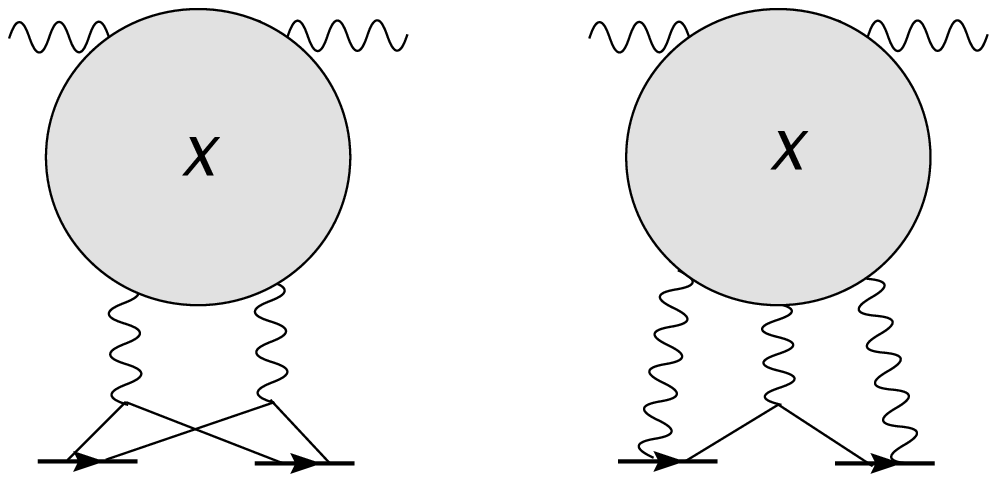,width=8cm,height=3cm}\\ 
Fig.21: inclusive jet production with higher order color correlaters
inside the two-nucleon target.   
\end{center}
A more detailed discussion will be given elsewhere, and for the rest of 
this paper we will restrict our discussion to the irreducible pieces which 
constitute the two Pomeron contribution to the inclusive cross section.  

\section{The 1-jet inclusive cross section}
\label{sec:discussion}
In the previous section we have described the computation of the triple
discontinuities of the amplitudes, with one gluon being fixed in transverse
momentum $\bp_1$ and rapidity $y_1$. Due to this gluon, the decomposition 
into reggeized and irreducible pieces has turned out to be quite 
different from the total cross section.

Let us now make use of these triple discontinuites and return to 
the inclusive cross section in eq.(\ref{sumofcuts2}). Beginning 
with the term $N_4^c(1,2|3,4;y_1,\bp_1)$, 
we use Fig.20 and attach nucleon $1$ to lines $1$ and $2$ and nucleon $2$ 
to lines $3$ and $4$. Similarly, the second term $N_4^c(3,4|1,2;y_1,\bp_1)$ is 
obtained by interchanging nucleons $1$ and $2$. In the last term, 
$N_4^c(1,3|2,4;y_1,\bp_1)$, we connect nucleon $1$ with the gluon lines 
$1$ and $4$. Because of the symmetry under the exchange of gluons 
on both sides of the cut, we do not need to distinguish 
between $N_4^c(1,3|2,4;y_1,\bp_1)$ and $N_4^c(1,4|2,3;y_1,\bp_1)$. 
For the third and fourth lines on the rhs of eq.(\ref{sumofcuts2}), 
we use Fig.19. Again, the symmetry on the rhs of the cutting line 
allows to identify, for example, $N_4^c(1,2,3|4;y_1,\bp_1)$ and 
$N_4^c(1,3,2|4;y_1,\bp_1)$.

For each of these terms, we have the four groups corresponding to the
Figs.19a-d or Figs.20a-d. As we have said before, for the first two 
groups the different cuts $i=1,2,3$ lead to the same result. Hence we 
can, in eq.(\ref{sumofcuts2}), simply sum over the phase factors. This leads, 
in the case of the first group (Fig.19a and 20a) to the usual 
AGK counting: 2 - 8 + 4 = -2. In the second group (Figs.19b and 20b) we find 
complete cancellation~\cite{agk-qcd2}: 2 - 6 + 4 = 0\footnote{Here we make use of the fact 
that the coupling of the two gluon pairs to the two nucleons also satifies the
symmetry properties: invariance under the interchange of the two 
nucleons, and - for each nucleon separately - symmetry under interchange of the
two gluons.}. In contrast to this, for the remaining contributions to the
inclusive cross section there is no simple way of summing the different cuts,
and the inclusive cross section remains of the form given in eq.(\ref{sumofcuts2}).   
For the first group (two groups (Figs.19c and 20c) we illustrate the integrand 
of eq. \eqref{sumofcuts2} in the following equation:
\begin{equation}
  \label{eq:ImjetV24}
  \begin{split}
  \Big[
    \xi_1 \xi_2^* \sum
    \begin{minipage}{2cm}
      \includegraphics[width=2cm]{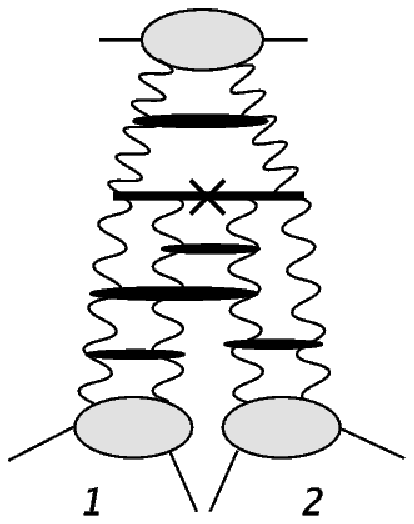}
   \end{minipage} +\xi_2 \xi_1^* 
    \sum \begin{minipage}{2cm}
      \includegraphics[width=2cm]{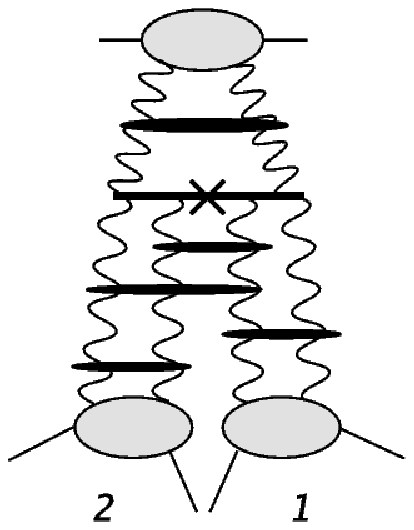}
    \end{minipage} + \hspace{3cm}
\\  + 2\, {\rm Im} \xi_1\left( (i\xi_2)^*  
          \sum \begin{minipage}{2cm}
      \includegraphics[width=2cm]{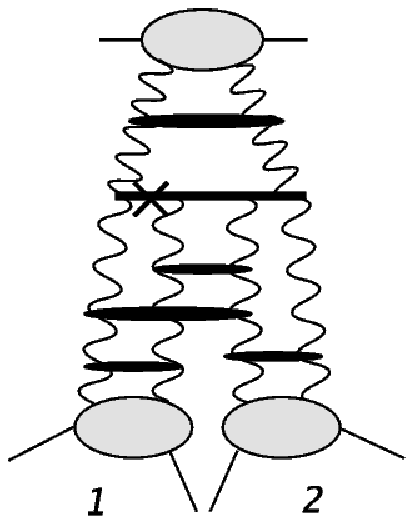}
    \end{minipage}
       +  c.c.\right)
+2\, {\rm Im} \xi_2  \left( i\xi_1 
             \sum \begin{minipage}{2cm}
      \includegraphics[width=2cm]{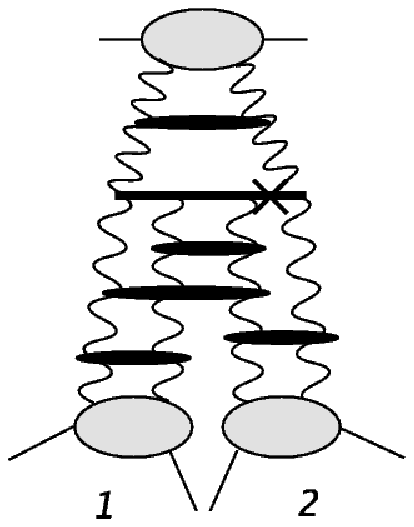}
               \end{minipage} + c.c.\right)
\\ + 4 {\rm Im} \xi_1 {\rm Im} \xi_2 
    \sum \begin{minipage}{2cm}
      \includegraphics[width=2cm]{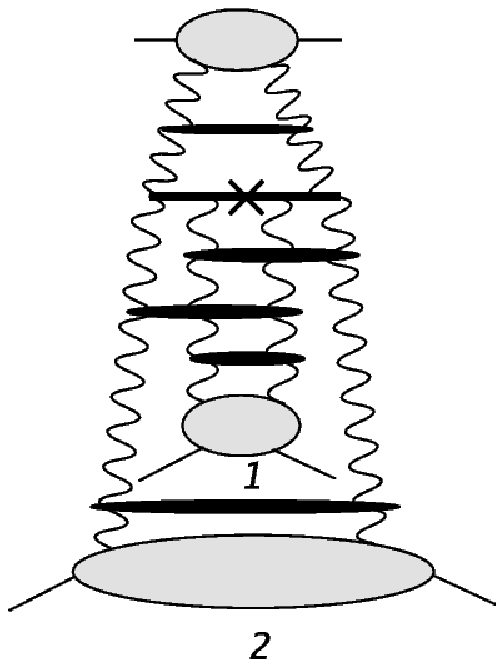} 
    \end{minipage} \Big] \hspace{3cm}
  \end{split}
\end{equation}
In this expression 
above in the first line, which corresponds to the 'diffractive
cut', one has to insert all the contributions constructed with the effective
vertex $\ocs{V}{}{2}{4}$ given in eq. \eqref{eq:2V4}, inserted in
eq. \eqref{eq:iZ4I3}.
The two contributions with complex conjugate phase factors are associated 
to the two possible ways of coupling to the two nucleons in the deuteron.
The 'single absorptive cut' contribution in the second line of eq. \eqref
{eq:ImjetV24}, is given by the sum of 4 terms, two associated with the jet
produced along the cut which goes to one nucleon and the other two when the
cut goes through the second nucleon. The two cases are constructed similarly
to the previous one employing the vertices $\ocs{V}{}{1}{4}$ and
$\ocs{V}{}{3}{4}$ given respectively in eqs. \eqref{eq:1V4} and
\eqref{eq:3V4}. 
The third line) in eq. \eqref{eq:ImjetV24} is associated
to the 'double cut' contribution, and it is built again from $\ocs{V}{}{2}{4}$.
The coupling to the nucleons selects the structure equivalent to
$N_4^c (1,3|2,4)=N_4^c (2,3|1,4)$ and is associated to a purely real phase.
Because of the symmetry of $N_4^c$ under permutations on both sides of the 
cutting line we do not need to include another term with nucleons 
$1$ and $2$ interchanged. Let us note that in our approximation we shall
choose purely imaginary BFKL pomeron phases, $\xi_{1,2}=i$.  
 
The final group (Figs.19d and 20d) is illustrated in the following 
equation: 
\begin{equation}
  \label{eq:ImjetV23}
  \begin{split}
    \Big[
    \xi_1 \xi_2^* \left( \sum 
    \begin{minipage}{2cm}
      \includegraphics[width=2cm]{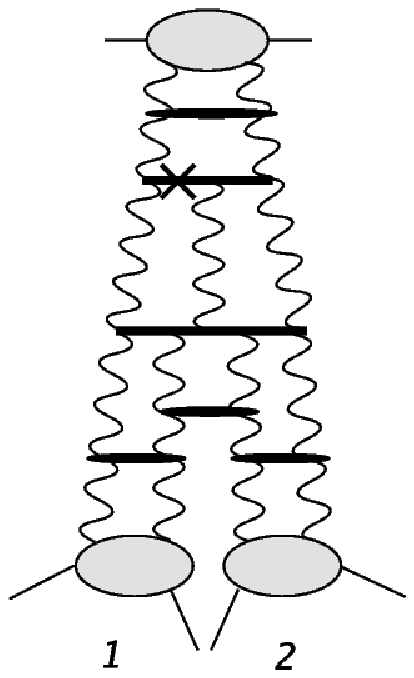}
    \end{minipage} + 
\sum \begin{minipage}{2cm}
      \includegraphics[width=2cm]{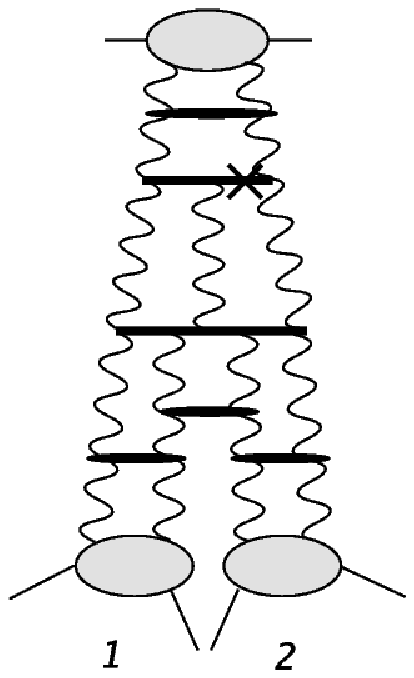}
    \end{minipage} \right)
+ c.c.  \hspace{3cm}
\\ + 2\, {\rm Im} \xi_1 \left( (i\xi_2)^* \sum
    \begin{minipage}{2cm}
      \includegraphics[width=2cm]{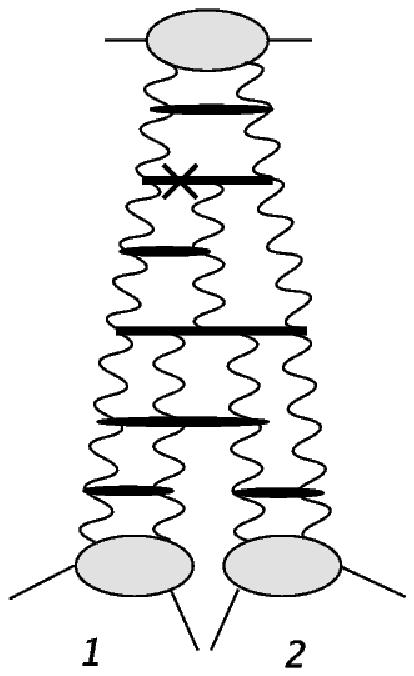}
    \end{minipage}  + c.c.\right)  
+2\, {\rm Im} \xi_2 \left( i\xi_1 \sum
    \begin{minipage}{2cm}
      \includegraphics[width=2cm]{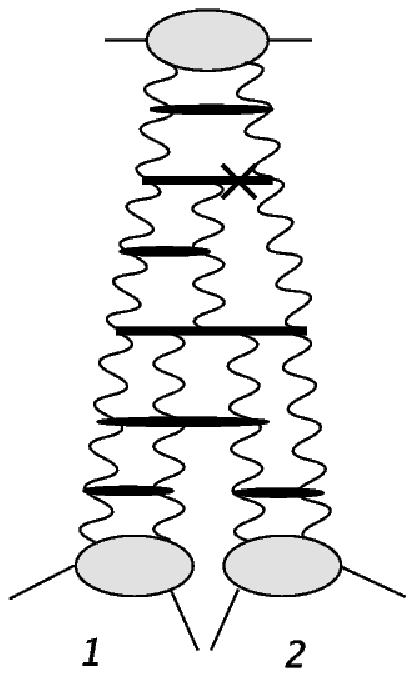}
    \end{minipage} + c.c.\right)
     + \\ 
+ 4 {\rm Im} \xi_1 {\rm Im} \xi_2 \left( \sum
    \begin{minipage}{2cm}
      \includegraphics[width=2cm]{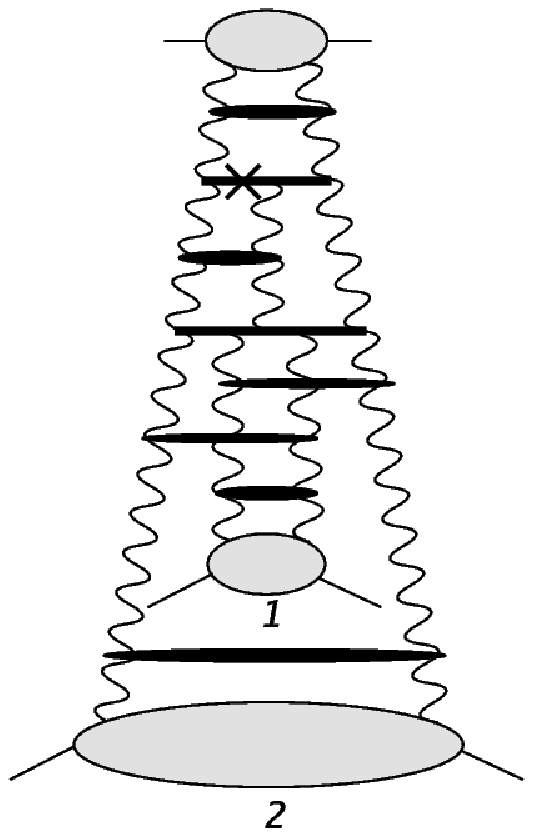} 
    \end{minipage} + \sum \begin{minipage}{2cm}
      \includegraphics[width=2cm]{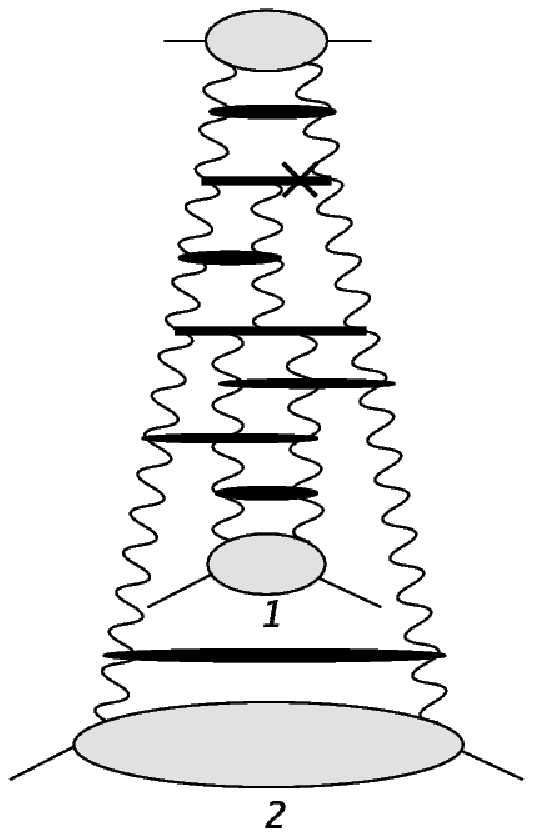} 
    \end{minipage} \right) \Big] \hspace{3cm}.
  \end{split}
\end{equation}
These terms are novel and quite peculiar since they are characterized by
the emission of a jet inside the effective vertices $\os{\Gamma}{}{}{3}$ and
$\os{\Gamma}{}{i}{3}$ which allow, in the $t$-channel, the transition from
$2$ to $3$ reggeized gluons. After rapidity evolution 
a second splitting, described by the vertices ${}_i\oc{W}{}{4}$ 
is taking place. In this effective $3 \to 4$ transition there is always one 
gluon which acts as a spectator.  Finally, the resulting
$t$-channel four gluon state, after a BKP evolution, is coupled to the 
deuteron form factor.

The 'diffractive contributions' in the first line are constructed using 
the effective vertex $\os{\Gamma}{}{}{3}$, given in eq. \eqref{eq:Gamma3} of appendix
\ref{App:iGamman}, which contains contributions from the jet emitted in the
two possible positions. The subsequent $3\to4$ transition is described by the
effective vertex ${}_2\oc{W}{}{4}$, given in eq. \eqref{eq:2W4}. One is
therefore led to use eq.\eqref{eq:2Z3I4},
which has to be integrated with the four reggeon Green's functions
and the deuteron form factor.
The 'single absorptive cut' contributions in the second line of
eq. \eqref{eq:ImjetV23} are expressed in terms of the effective vertices
$\os{\Gamma}{}{1}{3}$ and $\os{\Gamma}{}{2}{3}$ (or also using
$\os{\Gamma}{}{}{3}$ as in eq. \eqref{eq:1Z3I4}) defined in
eq. \eqref{eq:iGamma3} of appendix \ref{App:iGamman}. They contain 
contributions from the jet emitted only on the left or on the right of the
effective $2\to3$ vertex. These two cases are associated to two corresponding $3\to4$
splittings described by the effective vertices ${}_1\oc{W}{}{4}$ and
${}_3\oc{W}{}{4}$, listed in eqs. \eqref{eq:1W4} and \eqref{eq:3W4}, respectively.

Finally the 'double cut contribution' in the third line of 
eq. \eqref{eq:ImjetV23}) is, again, constructed in the same way as the diffractive case,
with the produced gluon inside the effective $2\to3$ vertex being either on the left or on
the right hand side. What distinguishes this case from the diffractive one is the 
coupling to the deuteron form factors.  

We complete this section with the large-$N_c$ limit which somewhat simplifies our results.
As the main feature, the four-gluon evolution above the two nucleons turns into two 
non-interacting BFKL Pomerons, one for each nucleon. We illustrate this 
in Fig.22:
\begin{center}
\epsfig{file=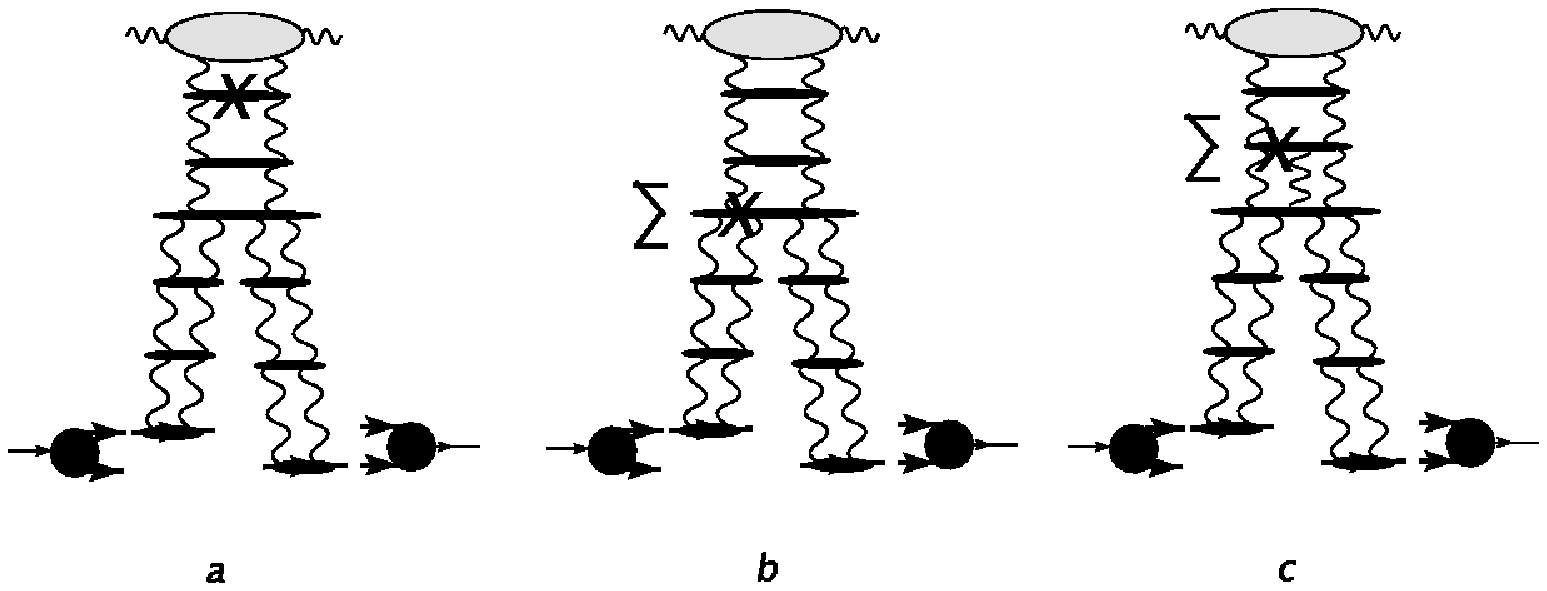,width=14cm, height=6cm}\\
Fig.22: the large-$N_c$ approximation (a) production above the $2\to4$ 
transition, (b) production inside the $2\to4$ transition, (c) production 
in the $2\to3$ transition.
\end{center} 
In the first contribution, shown in Fig.22a (which corresponds to Fig.19a and 
Fig.20a), these two Pomerons couple directly to the $2 \to 4$ vertex, 
selecting the color structure $\delta_{a_1 a_2} \delta_{a_3 a_4}$. 
This is the triple Pomeron vertex, which also appears in the nonlinear 
Balitsky Kovchegov (BK) evolution equation~\cite{BK}. In the second contributions 
illustrated in Fig.22b (corresponding to Figs.19c and 20c), 
the two Pomerons couple to the new production vertex $\ocs{V}{}{i}{4}$ 
listed in eqs.\eqref{eq:1V4} - \eqref{eq:3V4},
and the different cut positions lead to different expressions for the vertex. 
For each $i$, only one of the color structures contributes to the large-$N_c$ limit.
We expect that the rather lengthy expressions for the vertices that we have obtained 
may simplify, if we make use of the Moebius representation of the BFKL Pomerons. 
This will be discussed in a subsequent paper.  
Finally the new contribution in Fig.22c (corresponding to Figs.19d and 20d):
here the two Pomerons arrive at the effective $3 \to 4$ vertices,
${}_i\oc{W}{}{4}$, listed in eqs.\eqref{eq:1W4} 
- \eqref{eq:3W4}. Again, each cut picks one color structure, dismissing the other ones 
as subleading. In particular, there is no $N_c$ supression of this novel piece with the 
$3$ gluon contribution. Again, simpflifications of the kernels will be discussed  
elsewhere. Note that the jet production from one of the two ladders below the 
$2 \to 4$ vertex cancels because of the AGK rules.  

\section{Conclusions}
\label{sec:conclusions}

In this paper we have investigated, within the BFKL framework of pQCD,
the single-jet inclusive cross section in the scattering of a virtual photon 
on a weakly coupled nucleus (deuteron). We have identified the two-Pomeron 
exchange between the jet and the nucleus, and we have derived an analytic 
expression for the jet vertex. Invoking Regge factorization, the same 
vertex can also be used in $pp$ collisions where the jet, in rapidity, 
is close to one of the protons, but has a large rapidity separation 
from the other proton. Our analysis has been done in momentum space, 
and we stress that the results are valid for {\it finite} $N_c$.      

On the theoretical side, our analyis shows several new features.
First, the jet vertex contains a new 
structure not seen before, namely a three gluon $t$-channel state which, 
in a total cross section, would violate signature conservation and, hence,
never appears. This contribution to the jet production vertex is not 
suppressed in the limit $N_c \to \infty$, and there are no extra powers in  
$g^2$ which are not compensated by factors $\ln 1/x$. This latter statement 
simply follows from the fact that all our results are derived from production  
amplitudes which are all of the same order:
$g^2 \times g^2 \times g^2 \times \sum_k (g^2 y)^k \times g^4$ 
(where the last factor $g^4$ belongs to the coupling to the nucleons), and 
all subsequent steps amount to a re-ordering\footnote{In particular, the contributions in Fig.22c
are of the same order as those of Fig.22a: the $3 \to 4$ gluon vertex is
of the order $g^3$ (cf. eq.(A.16a)), i.e. in the transition from 3 to 4 gluons 
one gluon remains a 'spectator'.}

This last term the last term seems to be missing in 
previous studies, in particular in both ~\cite{KT} and ~\cite{B2006}, 
and we feel that it is very important to clarify this discrepancy. Whereas, at the moment, 
we feel unable to comment on ~\cite{KT}, we do see a possible reason why ~\cite{B2006} does 
not find this piece. At first sight, ~\cite{B2006} follows a strategy very 
similar to ours. It starts from discontinuities, 
computed in momentum space, and it then separates reggeizing pieces from 
nonreggeizing ones. In contrast to our strategy, however, this separation 
is done in the same way as for the total cross section, i.e. before fixing the 
momenta of the jet. In our approach, however, we do the separation of 
reggeizing and irreducible pieces only after fixing the momenta of the 
jets. As it turns out, the results for the inclusive cross section 
do depend on the order of these steps.    
Connected with these new contributions are new production vertices 
and transition vertices of reggeized gluons, which represent building 
blocks of QCD reggeon field theory. 

In order to clarify the connection of our result with those of, e.g.,
 ~\cite{KT,KL,BGV} it will be useful to first translate our results 
into configuaration space, making use of the Moebius representation 
~\cite{BLV2}, and also taking the large-$N_c$ limit. We plan to do this 
in a forthcoming work 

Returning to the further interpretation of our analysis, they also shows that, 
for the two-Pomeron exchange in the inclusive cross section formula, the 
AGK counting rules have to be used with care: if the jet is produced inside 
the $2\to3$ or the $2\to4$ transition vertex, the relative weights of the 
different cuttings across the two Pomeron exchange differ from the 
AGK counting derived for the total cross section. This supports 
the findings of ~\cite{KT,B2006}.
On the other hand, the cancellation of the rescattering corrections 
across the jet vertex remains valid and has been confirmed by our analysis.    

Another result is the appearance of the reggeizing pieces. In the 
inclusive cross section formula, reggeizing pieces belong to single BFKL 
ladder. They appear in the coupling to the nucleus and introduce 
higher order correlators between the nucleons.      

As to practical applications, the most interesting aspect, at present, 
is the search for saturation. For the total $\gamma^*\,\, nucleus$ 
cross section,
the high energy behavior (small-$x$ limit), in the large-$N_c$  
approximation, is described by the nonlinear Balitsky-Kovchegov (BK)
evolution equation, and solutions to this equation have been investigated 
in some detail. In order to derive the BK equation in momentum 
space one investigates the scattering of a virtual photon on nuclear
targets consisting of $2$, $3$,...
nucleons and separates reggeizing and nonreggeizing contributions.
For the case of $2$ nucleons, the corresponding QCD diagrams 
have been analyzed before (and summarized in this paper), and the 
validity of the BK equation is intimately connected with the dominance of
the 'fan-like' structure of the QCD ladder diagrams. In particular, 
there is no direct coupling of two Pomerons to the photon impact factor,
and the splitting of a single Pomeron into two Pomerons goes via the 
$2 \to 4$ gluon vertex which, in the large-$N_c$ limit, coincides with 
the integral kernel of the BK equation.              

As the main intention of the present paper was the generalization of this 
analysis, from the total cross section to the single inclusive cross section,
we can, again, look at the structure the leading QCD-diagrams, illustrated in 
Fig.22. The first term, Fig.22a, suggests that, below the $2\to 4$, we see the 
beginning of the same fan-like structure as in the total cross section.
That is, when generalizing our analysis to the scattering on a nucleus 
consisting of $3$ or more nucleons, we expect to see the fan structure
which sums up to the familiar nonlinear BK-equation.  
The second and the third terms (Fig.22b and c), however, do not fit 
into this pattern: the evolution below the jet vertex starts with double 
Pomeron exchange, and in the last term the new three gluon state introduces 
a new Pomeron component which survives in the large-$N_c$ limit. One might 
interpret it as a nonlocal (in rapidity) contribution to the effective 
$2 \to 4$ transition vertex. 

A comment on $k_t$ factorization might be in place. The structure of our 
large-$N_c$ cross section can be read off from Fig.22. All three 
contribtions to the inclusive cross section have in common that, 
in transverse momentum, they factorize into 
a production vertex and gluon amplitudes above and below the vertex.
In detail, however, there are some differences compared to the usual 
factorization pattern. In the first term, Fig.22a, 
we still have the usual $k_t$-factorization: momentum dependent 
amplitudes (unintegrated gluon densitities) from above and below, convoluted  
(in transverse momentum) with the gluon emission vertex. In Fig.22b,  
we still have, above the gluon emission, a single unintegrated gluon density,
whereas from below we now have two gluon amplitudes, and this leads to a 
threefold transverse momentum integration. 
In Fig.22c, the emission vertex has a single gluon density from above,   
a three gluon amplitude from below. Figs.22b and c thus 
introduce gluon correlation functions of four and three gluons, resp.
It is the three-gluon correlator which seems to be absent in previous 
studies. In order to understand the further rapidity evolution 
of Figs.22 b and c it will be necessary to study the scattering of a photon on a
three nucleon state.

Finally, one might wonder how our result would generalize in the analysis of the
equations describing the corrections to inclusive two-jet production cross
sections. This case has been considered in the framework of the
color dipole-CGC picture~\cite{JM-K}. Clearly we expect the pattern of gluon
reggeization to be broken further, leading possibly to new terms with 
even higher order gluon correlators.
This is a challenging analysis which we hope to address in the future.
\\[0.5cm] 
%
{\bf Acknowledgements:} We thank Mikhail Braun for lively and very helpful discussions.
We gratefully acknowledge the support of the Galileo Galilei Institute in 
Florence where part of this work has been done.  

\begin{appendix}

\setcounter{section}{0}
\setcounter{equation}{0}
\renewcommand{\thesection}{\Alph{section}}
\renewcommand{\theequation}{\Alph{section}.\arabic{equation}}

\section{Appendices} 
\subsection{Color identities}
\label{App:color}

The structure of the $SU(N_c)$ algebra is determined by
the structure constant $f^{abc}$ fixing the commutation
relations $[t^a,t^b] = i f^{abc} t^c$, with the generators
$t^a$ normalized such that $\tr(t^at^b) = \delta^{ab}/2$.
From the generators is possible to get the structure
constants via
\begin{equation}
  \label{eq:fa1a2a3asatrace}
  f^{a_1 a_2 a_3} =
  -2i (tr(t^{a_1} t^{a_2} t^{a_3}) - tr(t^{a_3} t^{a_2} t^{a_2})) \, ,
\end{equation}
and defining the symmetric structure constant $d^{a_1 a_2 a_3}$
through the anticommutators of the generators,
\begin{equation}
  \label{eq:anticom}
  \{t^{a_1},t^{a_2}\} = \frac{1}{N_c} \delta^{a_1 a_2} +
  d^{a_1 a_2 a_3} t^{a_3} \, ,
\end{equation}
we have
\begin{equation}
  \label{eq:dasatrace}
  d^{a_1 a_2 a_3} =
  2 (tr(t^{a_1} t^{a_2} t^{a_3}) + tr(t^{a_3} t^{a_2} t^{a_2})) \, .
\end{equation}

It turns out to be useful to define as well tensors $f^{a_1 \ldots a_n}$
and $d^{a_1 \ldots a_n}$ for $n>3$:
\begin{subequations}
  \begin{eqnarray}
    \label{eq:fnadda1an}
    f^{a_1 a_2 \ldots a_n} &=&
    -i \big( tr(t^a_1 t^{a_2} \ldots t^{a_n}) -
    tr(t^{a_n} \ldots t^{a_2} t^{a_1}) \big) \, ,\\
    d^{a_1 a_2 \ldots a_n} &=&
    tr(t^a_1 t^{a_2} \ldots t^{a_n}) +
    tr(t^{a_n} \ldots t^{a_2} t^{a_1}) \, .
  \end{eqnarray}
\end{subequations}
Both $f$ and $d$ tensors are evidently invariant under cyclic permutation,
and moreover $f^{a_1 a_2 a_3}$ is antisymmetric under the transposition
of two indices, while $d^{a_1 a_2 a_3}$ is symmetric.
\begin{subequations}
  \begin{eqnarray}
    \label{eq:fanddsymmetry}
    f^{a_1 a_2 a_3} &=& -f^{a_2 a_1 a_3} \\
    d^{a_1 a_2 a_3} &=& d^{a_2 a_1 a_3}
  \end{eqnarray}
\end{subequations}
A very useful relation is the Fierz identity,
\begin{equation}
  \label{eq:Fierz}
  (t^a)_{i_1 i_2}(t^a)_{j_1 j_2} =
  \frac{1}{2} \delta_{i_1 j_1} \delta_{i_2 j_2} -
  \frac{1}{2 N_c} \delta_{i_1 i_2} \delta_{j_1 j_2} \, .
\end{equation}
Other essential relations are the Jacobi identity,
\begin{equation}
  \label{eq:jacobi}
  f^{a_1 a_2 b} f^{b a_3 a_4} -
  f^{a_1 a_3 b} f^{b a_2 a_4} +
  f^{a_1 a_4 b} f^{b a_2 a_3} =
  0 \, ,
\end{equation}
the decomposition of $d^{a_1 a_2 a_3 a_4}$ in terms of rank three tensors,
\begin{equation}
  \label{eq:da1a2a3a4intermsoffddelta}
  d^{a_1 a_2 a_3 a_4} =
  \frac{1}{4}(d^{a_1 a_2 b} d^{b a_3 a_4} - f^{a_1 a_2 b} f^{b a_3 a_4}) +
  \frac{1}{2 N_c} \delta^{a_1 a_2} \delta^{a_3 a_4} \, ,
\end{equation}
and some contractions of various tensors
\begin{subequations}
  \begin{eqnarray}
    \label{eq:l2}
    f^{b_1 a_1 b_2} f^{b_2 a_2 b_1} &=& -N_c \delta^{a_1 a_2} \, ,\\
    \label{eq:l3}
    f^{b_1 a_1 b_2} f^{b_2 a_2 b_3} f^{b_3 a_3 b_1} &=&
    -\frac{N_c}{2} f^{a_1 a_2 a_3} \, ,\\
    \label{eq:l3d}
    d^{b_1 a_1 b_2} f^{b_2 a_2 b_3} f^{b_3 a_3 b_1} &=&
    -\frac{N_c}{2} d^{a_1 a_2 a_3} \, ,\\
    \label{eq:l4}
    f^{b_1 a_1 b_2} f^{b_2 a_2 b_3} f^{b_3 a_3 b_4} f^{b_4 a_4 b_1} &=&
    N_c~ d^{a_1 a_2 a_3 a_4} + \nonumber \\
    &&+\frac{1}{2} (\delta^{a_1 a_2} \delta^{a_3 a_4} +
    \delta^{a_1 a_3} \delta^{a_2 a_4} +
    \delta^{a_1 a_4} \delta^{a_2 a_3}) \, , \\
    \label{eq:dffa}
    d^{a_1 a_2 b_1 b_2} f^{b_1 a_3 c} f^{c a_4 b_2} &=&
    -\frac{N_c}{2} d^{a_1 a_2 a_3 a_4} -
    \frac{1}{4} \delta^{a_1 a_2} \delta^{a_3 a_4} \, ,\\
    \label{eq:dffb}
    d^{a_1 b_1 a_3 b_2} f^{b_1 a_2 c} f^{c a_4 b_2} &=&
    \frac{1}{4} \delta^{a_1 a_2} \delta^{a_3 a_4} +
    \frac{1}{4} \delta^{a_1 a_4} \delta^{a_2 a_3} \, .
  \end{eqnarray}
\end{subequations}

\subsection{The 2-to-4 effective vertex $\cV_4$}
\label{App:V4}

The integral operators ${}_i\Gamma_n$ are given in terms of
the infrared safe $\cG$ function (first introduced in \cite{Bartels:1994jj}
in the forward direction and later generalized and investigated in
\cite{Braun:1997nu,Vacca:1998kc}). Its action on a two gluon function
$\phi$ is given by
\begin{equation}
  \label{eq:Gfunction}
  \cG \phi_2 = 
  \cK_3 \phi + \frac{g}{N_c} \bigg(
  \oo{2} \ov{\phi}{(1 \cdot) 3}{} +
  \oo{2} \ov{\phi}{1 (\cdot 3)}{} -
  \oo{(12)} \ov{\phi}{\cdot 3}{} -
  \oo{(23)} \ov{\phi}{1 \cdot}{}
  \bigg) \, .
\end{equation}
This object is nothing but a regularized version of the two-to-three
operator $\cK_3$, being the trajectories in \eqref{eq:Gfunction}
the precise subtraction terms necessary to get rid of the divergences.
Note that when the transverse momentum $\bk_2$ of the central
leg is put to zero, $\cG$ reduces to the singlet version of the BFKL operator
$\cH$; we indicate $\bk_2 = \bzero$ putting a small circle $\circ$
in its position:
\begin{equation}
  \label{eq:GtoH}
  \delta^{a_1 a_2} \oc{G}{1 \circ 2}{} \phi = -\frac{N_c}{g}~
  \oc{H}{12}{2} \phi \, ,
\end{equation}
with $\phi$ a two gluon color neutral function. The vertex $\cV_4$ introduced
in \eqref{eq:D4I} is then defined by
\begin{equation}
  \label{eq:Vcal4}
  \cV_4 = \delta^{a_1 a_2} \delta^{a_3 a_4}~ \overset{1234}{V} +
  \delta^{a_1 a_3} \delta^{a_2 a_4}~ \overset{1324}{V} +
  \delta^{a_1 a_4} \delta^{a_2 a_3}~ \overset{1423}{V} \, ,
\end{equation}
where the operator $V$, which is M\"obius
invariant~\cite{Bartels:1995kf,Braun:1997nu},
is defined as
\begin{eqnarray}
  \overset{1234}{V} \phi &=& ~~\frac{g}{2} \bigg(
  \oc{G}{1(23)4}{} \; +
  \oc{G}{1(24)3}{} \; +
  \oc{G}{2(13)4}{} \; +
  \oc{G}{2(14)3}{} \; +
  \oc{G}{(12)\circ(34)}{} + \nonumber \\&&\qquad\qquad
  -\oc{G}{(12)34}{} \; -
  \oc{G}{(12)43}{} \; -
  \oc{G}{12(34)}{} \; -
  \oc{G}{21(34)}{} \;
  \bigg) \overset{\cdot \cdot}{\phi}
  \, . \nonumber
\end{eqnarray}

\subsection{Definitions of the operators ${}_i\slashed{\Gamma}_n$
and ${}_i\Gamma_n$}
\label{App:iGamman}

 The effective vertices $\os{\Gamma}{}{i}{3}$ describing the
transition 2-to-3 reggeized gluons with associated jet production
are conveniently expressed in term of an auxiliary operator
$\os{\Gamma}{}{}{3}$ defined as
\begin{equation}
  \label{eq:Gamma3}
  \os{\Gamma}{}{}{3} \phi = \frac{1}{2} \bigg(
  g\obs{K}{(13)2}{}{2} -
  \obs{K}{132}{1}{3} -
  \obs{K}{213}{2}{3} \bigg) \overset{\cdot \cdot}{\phi}
  -\frac{g}{2}
  \obs{K}{13}{}{2} \overset{(\cdot \cdot) 2}{\phi} \, .
\end{equation}
In terms of $\os{\Gamma}{}{}{3}$ we have
\begin{subequations}
  \label{eq:iGamma3}
\begin{eqnarray}
  \label{eq:1Gamma3}
  \os{\Gamma}{}{1}{3} \phi &=& \frac{1}{2} \bigg(
  \os{\Gamma}{123}{}{3} - \os{\Gamma}{132}{}{3}
  \bigg)\, , \\
  \label{eq:2Gamma3}
  \os{\Gamma}{}{2}{3} \phi &=& \frac{1}{2} \bigg(
  \os{\Gamma}{123}{}{3} - \os{\Gamma}{213}{}{3}
  \bigg) \, .
\end{eqnarray}
\end{subequations}

In the case of the transitions 2-to-4 there are four different vertices,
one each for the cuts 1 and 3 and two for the cut 2. They are
\begin{subequations}
  \label{eq:iGamma4}
\begin{eqnarray}
  \os{\Gamma}{}{1}{4} \phi &=& ~~\frac{1}{4} \bigg(
  \obs{K}{2134}{2}{4} + \obs{K}{2314}{2}{4} +
  g\obs{K}{1(23)4}{1}{3} - g\obs{K}{(12)34}{1}{3} - g\obs{K}{23(14)}{2}{3}
  \bigg) \overset{\cdot \cdot}{\phi} + \nonumber \\
  \label{eq:1Gamma4}
  &&+\frac{g}{4} \obs{K}{134}{1}{3} \overset{(\cdot \cdot) 2}{\phi} +
  \frac{g}{4} \obs{K}{132}{1}{3} \overset{(\cdot \cdot) 4}{\phi} +
  (3 \leftrightarrow 4) + \\
  &&+\frac{1}{4} \bigg(
  g^2\obs{K}{(12)(34)}{}{2} + g^2\obs{K}{(134)2}{}{2} - g\obs{K}{12(34)}{1}{3} -
  g\obs{K}{21(34)}{2}{3} + g\obs{K}{1(34)2}{1}{3} -  g\obs{K}{(34)12}{2}{3}
  \bigg) \overset{\cdot \cdot}{\phi} + \nonumber \\
  &&-\frac{g^2}{4} \obs{K}{1(34)}{}{2}  \overset{(\cdot \cdot) 2}{\phi}
  -\frac{g^2}{4} \obs{K}{12}{}{2}  \overset{(\cdot \cdot) (34)}{\phi}
  \, , \nonumber \\
  \os{\Gamma}{}{2A}{4} \phi &=& ~~\frac{1}{4} \bigg(
  \obs{K}{1234}{2}{4} + \obs{K}{1324}{2}{4}
  - g\obs{K}{(23)14}{1}{3} - g\obs{K}{14(23)}{2}{3}
  + g^2\obs{K}{(13)(24)}{}{2}
  \bigg) \overset{\cdot \cdot}{\phi} + \nonumber \\ &&
  + \frac{g}{4} \obs{K}{234}{1}{3} \overset{1 (\cdot \cdot)}{\phi}
  + \frac{g}{4} \obs{K}{123}{2}{3} \overset{(\cdot \cdot) 4}{\phi}
  \nonumber \\ &&
  \label{eq:2AGamma4}
  + (1 \leftrightarrow 2) + ( 3 \leftrightarrow 4)
  + (1 \leftrightarrow 2, 3 \leftrightarrow 4) + \\ &&
  + \frac{g}{4} \bigg(
  \obs{K}{3(12)4}{2}{3} - \obs{K}{(12)34}{1}{3}
  \bigg) \overset{\cdot \cdot}{\phi}
  - \frac{g^2}{4} \obs{K}{(123)4}{}{2} \overset{\cdot \cdot}{\phi}
  - \frac{g^2}{4} \obs{K}{(12)3}{}{2} \overset{(\cdot \cdot) 4}{\phi}
  + ( 3 \leftrightarrow 4) + \nonumber \\ &&
  + \frac{g}{4} \bigg(
  \obs{K}{1(34)2}{1}{3} - \obs{K}{12(34)}{2}{3}
  \bigg) \overset{\cdot \cdot}{\phi}
  -\frac{g^2}{4}  \obs{K}{1(234)}{}{2} \overset{\cdot \cdot}{\phi}
  - \frac{g^2}{4} \obs{K}{2(34)}{}{2} \overset{1 (\cdot \cdot)}{\phi}
  + ( 1 \leftrightarrow 2) + \nonumber \\ &&
  + \frac{g^2}{2} \obs{K}{(12)(34)}{}{2} \overset{\cdot \cdot}{\phi}
  \, , \nonumber
\end{eqnarray}
\begin{eqnarray}
  \os{\Gamma}{}{2B}{4} \phi &=& ~~\frac{1}{4} \bigg(
  \obs{K}{1234}{2}{4} + \obs{K}{1324}{2}{4}
  + g\obs{K}{1(34)2}{1}{3} + g\obs{K}{3(12)4}{2}{3}
  - g\obs{K}{13(24)}{1}{3} - g\obs{K}{(13)24}{2}{3} + \nonumber \\ &&
  - g\obs{K}{(13)24}{1}{3} - g\obs{K}{13(24)}{2}{3}
  + g^2\obs{K}{(13)(24)}{}{2}
  \bigg) \overset{\cdot \cdot}{\phi} + \nonumber \\ &&
  + \frac{g}{4} \obs{K}{234}{1}{3} \overset{1 (\cdot \cdot)}{\phi}
  + \frac{g}{4} \obs{K}{123}{2}{3} \overset{(\cdot \cdot) 4}{\phi}
  - \frac{g^2}{4} \obs{K}{13}{}{2} \overset{(\cdot \cdot) (24)}{\phi}
  \nonumber \\&&
  \label{eq:2BGamma4}
  + (1 \leftrightarrow 2, 3 \leftrightarrow 4) + \\&&
  + \frac{g}{4} \bigg(
  \obs{K}{3(12)4}{2}{3} - \obs{K}{(12)34}{1}{3}
  \bigg) \overset{\cdot \cdot}{\phi}
  - \frac{g^2}{4} \obs{K}{(123)4}{}{2} \overset{\cdot \cdot}{\phi}
  - \frac{g^2}{4} \obs{K}{(12)3}{}{2} \overset{(\cdot \cdot) 4}{\phi}
  + ( 3 \leftrightarrow 4) + \nonumber \\&&
  + \frac{g}{4} \bigg(
  \obs{K}{1(34)2}{1}{3} - \obs{K}{12(34)}{2}{3}
  \bigg) \overset{\cdot \cdot}{\phi}
  - \frac{g^2}{4} \obs{K}{1(234)}{}{2} \overset{\cdot \cdot}{\phi}
  - \frac{g^2}{4} \obs{K}{2(34)}{}{2} \overset{1 (\cdot \cdot)}{\phi}
  + ( 1 \leftrightarrow 2) + \nonumber \\&&
  + \frac{g^2}{2} \obs{K}{(12)(34)}{}{2} \overset{\cdot \cdot}{\phi}
  \, , \nonumber \\
  \os{\Gamma}{}{3}{4} \phi &=& ~~\frac{1}{4} \bigg(
  \obs{K}{1243}{2}{4} + \obs{K}{1423}{2}{4} +
  g\obs{K}{1(23)4}{2}{3} - g\obs{K}{12(34)}{2}{3} - g\obs{K}{(14)23}{1}{3}
  \bigg) \overset{\cdot \cdot}{\phi} + \nonumber \\&&
  \label{eq:3Gamma4}
  +\frac{g}{4} \obs{K}{124}{2}{3} \overset{3(\cdot \cdot)}{\phi} +
  \frac{g}{4} \obs{K}{324}{2}{3} \overset{1(\cdot \cdot)}{\phi} +
  (1 \leftrightarrow 2) + \\&&
  +\frac{1}{4} \bigg(
  g^2\obs{K}{(12)(34)}{}{2} + g^2\obs{K}{3(124)}{}{2} - g\obs{K}{(12)34}{2}{3} -
  g\obs{K}{(12)43}{1}{3} + g\obs{K}{3(12)4}{2}{3} -  g\obs{K}{34(12)}{1}{3}
  \bigg) \overset{\cdot \cdot}{\phi} + \nonumber \\&&
  -\frac{g^2}{4} \obs{K}{(12)4}{}{2}  \overset{3(\cdot \cdot)}{\phi}
  -\frac{g^2}{4} \obs{K}{34}{}{2}  \overset{(12) (\cdot \cdot)}{\phi} \, .
  \nonumber
\end{eqnarray}
\end{subequations}

The integral operators ${}_i\Gamma_n$ are given in terms of
the infrared safe $\cG$ function defined in \eqref{eq:Gfunction}.
Analougusly to \eqref{eq:iGamma4}, there are two different operators
for the central cut:
\begin{subequations}
\label{eq:iGammanUncut}
\begin{eqnarray}
  \op{\Gamma}{}{1}{4} \phi &=& \frac{1}{4} \bigg(
  + \oc{G}{234}{} \; \overset{1 \cdot \cdot}{\phi}
  - \oc{G}{432}{} \; \overset{1 \cdot \cdot}{\phi}
  + \oc{G}{134}{} \; \overset{\cdot 2 \cdot}{\phi}
  - \oc{G}{134}{} \; \overset{\cdot \cdot 2}{\phi}
  + \oc{G}{132}{} \; \overset{\cdot 4 \cdot}{\phi}
  - \oc{G}{132}{} \; \overset{\cdot \cdot 4}{\phi}
  - (3 \leftrightarrow 4) \bigg) + \nonumber \\ &&
  \label{eq:1GammanUncut}
  + \frac{g}{4} \bigg(
  - \oc{G}{2 \circ (34)}{} \; \overset{1 \cdot \cdot}{\phi}
  + \oc{G}{(34) \circ 2}{} \; \overset{1 \cdot \cdot}{\phi}
  - \oc{G}{1 \circ (34)}{} \; \overset{\cdot 2 \cdot}{\phi}
  + \oc{G}{1 \circ (34)}{} \; \overset{\cdot \cdot 2}{\phi}
  - \oc{G}{1 \circ 2}{} \; \overset{\cdot (34) \cdot}{\phi}
  + \oc{G}{1 \circ 2}{} \; \overset{\cdot \cdot (34)}{\phi}
  \bigg) \, , \qquad\quad \\[10pt]
  \op{\Gamma}{}{2A}{4} \phi &=& \frac{1}{4} \bigg(
    \oc{G}{123}{} \; \overset{\cdot \cdot 4}{\phi}
  + \oc{G}{234}{} \; \overset{1 \cdot \cdot}{\phi}
  + \oc{G}{124}{} \; \overset{\cdot 3 \cdot}{\phi}
  + \oc{G}{134}{} \; \overset{\cdot 2 \cdot}{\phi}
  - \oc{G}{132}{} \; \overset{\cdot \cdot 4}{\phi}
  - \oc{G}{234}{} \; \overset{1 \cdot \cdot}{\phi} + \nonumber \\ &&\qquad
  + (1 \leftrightarrow 2) + (3 \leftrightarrow 4)
  + (1 \leftrightarrow 2,3 \leftrightarrow 4) \bigg) + \nonumber \\ &&
  + \frac{g}{4} \bigg(
    \oc{G}{1 \circ 2}{} \; \overset{\cdot \cdot (34)}{\phi}
  - \oc{G}{2 \circ (34)}{} \; \overset{1 \cdot \cdot}{\phi}
  - \oc{G}{1 \circ (34)}{} \; \overset{\cdot 2 \cdot}{\phi}
  - (1 \leftrightarrow 2) \bigg) + \\ &&
  + \frac{g}{4} \bigg(
  - \oc{G}{3 \circ 4}{} \; \overset{(12) \cdot \cdot}{\phi}
  - \oc{G}{(12) \circ 3}{} \; \overset{\cdot \cdot 4}{\phi}
  - \oc{G}{(12) \circ 4}{} \; \overset{\cdot 3 \cdot}{\phi}
  + (3 \leftrightarrow 4)
  \bigg) \, , \nonumber
\end{eqnarray}
\begin{eqnarray}
  \op{\Gamma}{}{2B}{4} \phi &=& \frac{1}{4} \bigg(
  - \oc{G}{213}{} \; \overset{\cdot \cdot 4}{\phi}
  - \oc{G}{134}{} \; \overset{2 \cdot \cdot}{\phi}
  + \oc{G}{214}{} \; \overset{\cdot 3 \cdot}{\phi}
  + \oc{G}{234}{} \; \overset{\cdot 1 \cdot}{\phi}
  - \oc{G}{231}{} \; \overset{\cdot \cdot 4}{\phi}
  - \oc{G}{314}{} \; \overset{2 \cdot \cdot}{\phi} + \\ && \qquad
  + g\oc{G}{2 \circ (13)}{} \; \overset{ \cdot \cdot 4}{\phi}
  + g\oc{G}{(13) \circ 4}{} \; \overset{2 \cdot \cdot}{\phi}
  - g\oc{G}{2 \circ 4}{} \; \overset{\cdot (13) \cdot}{\phi}
  + (1 \leftrightarrow 2,3 \leftrightarrow 4)
  \bigg) \, , \nonumber \\[10pt]
  \op{\Gamma}{}{3}{4} \phi &=& \frac{1}{4} \bigg(
    \oc{G}{123}{} \; \overset{\cdot \cdot 4}{\phi}
  - \oc{G}{321}{} \; \overset{\cdot \cdot 4}{\phi}
  + \oc{G}{124}{} \; \overset{\cdot 3 \cdot}{\phi}
  - \oc{G}{124}{} \; \overset{3 \cdot \cdot}{\phi}
  + \oc{G}{324}{} \; \overset{\cdot 1 \cdot}{\phi}
  - \oc{G}{324}{} \; \overset{1 \cdot \cdot}{\phi}
  + (1 \leftrightarrow 2) \bigg)+ \\
  \label{eq:3GammanUncut}
  &&
  +  \frac{g}{4} \bigg(
  - \oc{G}{(12) \circ 3}{} \; \overset{\cdot \cdot 4}{\phi}
  + \oc{G}{3 \circ (12)}{} \; \overset{\cdot \cdot 4}{\phi}
  - \oc{G}{(12) \circ 4}{} \; \overset{\cdot 3 \cdot}{\phi}
  + \oc{G}{(12) \circ 4}{} \; \overset{3 \cdot \cdot}{\phi}
  - \oc{G}{3 \circ 4}{} \; \overset{\cdot (12) \cdot}{\phi}
  + \oc{G}{3 \circ 4}{} \; \overset{(12) \cdot \cdot}{\phi}
  \bigg) \, . \nonumber
\end{eqnarray}
\end{subequations}

The Ward identities fulfilled by all these operators (cut and uncut)
can be verified directly from these expressions. Moreover, thanks to the
properties of the function $\cG$\cite{Braun:1997nu},
these operators define M\"obius (conformal) invariant objects. 

\end{appendix}



\begin{thebibliography}{100}

\bibitem{B2000} M.Braun, Phys.\ Lett.\ {\bf B 483} 105 (2000)

\bibitem{KT} Yu.V.Kovchegov, K.Tuchin, Phys.\ Rev.\ {\bf D 65} 074026 (2002) 

\bibitem{KL} M.Kovner, M.Lublinsky, JHEP {\bf 0611} 083 (2006)

\bibitem{BGV} G.-P.Blaizot, F.Gelis, R.Venugopalan, Nucl.\ Phys. {\bf A 743} 57 (2004)

\bibitem{B2005} M.Braun, EPJ {\bf C 42} 169 (2005)

\bibitem{B2006} M.Braun, EPJ {\bf C 48} 501 (2006)

\bibitem{Abramovsky:1973fm}
  V.~A.~Abramovsky, V.~N.~Gribov and O.~V.~Kancheli,
  Yad.\ Fiz.\  {\bf 18} (1973) 595
  [Sov.\ J.\ Nucl.\ Phys.\  {\bf 18} (1974) 308].

\bibitem{agk-qcd1}
  J.~Bartels and M.~G.~Ryskin,
  Z.\ Phys.\  C {\bf 76} (1997) 241
  [arXiv:hep-ph/9612226];\\

\bibitem{agk-qcd2}
  J.~Bartels, M.~Salvadore and G.~P.~Vacca,
  Eur.\ Phys.\ J.\  C {\bf 42} (2005) 53
  [arXiv:hep-ph/0503049].

\bibitem{Bartels:1994jj}
  J.~Bartels and M.~Wusthoff,
  Z.\ Phys.\ C {\bf 66} (1995) 157.

\bibitem{Kuraev:1976ge}
  E.~A.~Kuraev, L.~N.~Lipatov and V.~S.~Fadin,
  Sov.\ Phys.\ JETP {\bf 44} (1976) 443
  [Zh.\ Eksp.\ Teor.\ Fiz.\  {\bf 71} (1976) 840].

\bibitem{Kuraev:1977fs}
  E.~A.~Kuraev, L.~N.~Lipatov and V.~S.~Fadin,
  Sov.\ Phys.\ JETP {\bf 45} (1977) 199
  [Zh.\ Eksp.\ Teor.\ Fiz.\  {\bf 72} (1977) 377].

\bibitem{Balitsky:1978ic}
  I.~I.~Balitsky and L.~N.~Lipatov,
  Sov.\ J.\ Nucl.\ Phys.\  {\bf 28} (1978) 822
  [Yad.\ Fiz.\  {\bf 28} (1978) 1597].

\bibitem{Bartels:1978fc}
  J.~Bartels,
  Nucl.\ Phys.\ B {\bf 151} (1979) 293.

\bibitem{Bartels:1980pe}
  J.~Bartels,
  Nucl.\ Phys.\ B {\bf 175} (1980) 365.

\bibitem{Kwiecinski:1980wb}
  J.~Kwiecinski and M.~Praszalowicz,
  Phys.\ Lett.\ B {\bf 94} (1980) 413.

\bibitem{Jaroszewicz:1980mq}
  T.~Jaroszewicz,
  Acta Phys.\ Polon.\ B {\bf 11} (1980) 965.

\bibitem{Lipatov:1985uk}
  L.~N.~Lipatov,
  Sov.\ Phys.\ JETP {\bf 63} (1986) 904
  [Zh.\ Eksp.\ Teor.\ Fiz.\  {\bf 90} (1986) 1536].

\bibitem{Bartels:2005ji}
  J.~Bartels, L.~N.~Lipatov, M.~Salvadore and G.~P.~Vacca,
  Nucl.\ Phys.\  B {\bf 726}, 53 (2005)
  [arXiv:hep-ph/0506235].

\bibitem{Mueller:1993rr}
  A.~H.~Mueller,
  Nucl.\ Phys.\  B {\bf 415}, 373 (1994).

\bibitem{Nikolaev:1994uu}
  N.~N.~Nikolaev and B.~G.~Zakharov,
  Phys.\ Lett.\  B {\bf 327}, 149 (1994)
  [arXiv:hep-ph/9402209].

\bibitem{Bartels:2004ef}
  J.~Bartels, L.~N.~Lipatov and G.~P.~Vacca,
  Nucl.\ Phys.\  B {\bf 706}, 391 (2005)
  [arXiv:hep-ph/0404110].

\bibitem{integrab}  L.N. Lipatov, {\it Phys. Lett.} {\bf B309} (1993) 394;\\
  L.~N.~Lipatov,
  arXiv:hep-th/9311037;
%
\bibitem{JW}    R.A. Janik and J. Wosiek, Phys. Rev. Lett. {\bf 82}
                (1999) 1092.
%
\bibitem{BLV}
  J.~Bartels, L.~N.~Lipatov and G.~P.~Vacca,
  Phys.\ Lett.\  B {\bf 477}, 178 (2000)
  [arXiv:hep-ph/9912423].
%
\bibitem{BBCV}
  J.~Bartels, M.~A.~Braun, D.~Colferai and G.~P.~Vacca,
  Eur.\ Phys.\ J.\  C {\bf 20}, 323 (2001)
  [arXiv:hep-ph/0102221]

\bibitem{Bartels:2003zu}
  J.~Bartels, M.~A.~Braun and G.~P.~Vacca,
  Eur.\ Phys.\ J.\  C {\bf 33}, 511 (2004)
  [arXiv:hep-ph/0304160].
%
\bibitem{dkkm}
  S.~E.~Derkachov, G.~P.~Korchemsky, J.~Kotanski and A.~N.~Manashov,
  Nucl.\ Phys.\  B {\bf 645}, 237 (2002)
  [arXiv:hep-th/0204124].
%
\bibitem{dv-lip2}
  H.~J.~de Vega and L.~N.~Lipatov,
  Phys.\ Rev.\  D {\bf 66}, 074013 (2002)
  [arXiv:hep-ph/0204245].

\bibitem{Vacca:2000bk}
  G.~P.~Vacca,
  Phys.\ Lett.\  B {\bf 489} (2000) 337
  [arXiv:hep-ph/0007067].

\bibitem{Lotter:1996vk}
  H.~Lotter,
  arXiv:hep-ph/9705288.

\bibitem{Iafelice:2007dc}
  P.~L.~Iafelice and G.~P.~Vacca,
  Eur.\ Phys.\ J.\  C {\bf 52} (2007) 581
  [arXiv:0709.0655 [hep-th]].

\bibitem{Bartels:1999aw}
  J.~Bartels and C.~Ewerz,
  JHEP {\bf 9909} (1999) 026
  [arXiv:hep-ph/9908454].

\bibitem{MSthesis}
  M.~Salvadore, Aspects of Multipartonic Interactions in Small-x QCD,
  PhD thesis, University of Bologna, Bologna, Italy (2006).

\bibitem{BK}  I.I. Balitsky, {\it Nucl. Phys.}\ {\bf B463} (1996) 99,
{\it Phys. Rev.} {\bf D60} (1999) 014020; Y.V. Kovchegov, {\it Phys. Rev.}
{\bf D60} (1999) 034008, {\it Phys. Rev.} {\bf D61} (2000) 074018.

\bibitem{Braun:1997nu}
  M.~A.~Braun and G.~P.~Vacca,
  Eur.\ Phys.\ J.\ C {\bf 6} (1999) 147
  [arXiv:hep-ph/9711486].

\bibitem{Vacca:1998kc}
  G.~P.~Vacca,
  arXiv:hep-ph/9803283.

\bibitem{Bartels:1995kf}
  J.~Bartels, L.~N.~Lipatov and M.~Wusthoff,
  Nucl.\ Phys.\  {\bf B 464} (1996) 298
  [arXiv:hep-ph/9509303].

\bibitem{BLV2} J.Bartels, L.N.Lipatov, G.P.Vacca, 
         Nucl.\ Phys.\ {\bf B706} 391-410,2005.\\

\bibitem{JM-K}
  J.~Jalilian-Marian and Y.~V.~Kovchegov,
  Phys.\ Rev.\  D {\bf 70} (2004) 114017
  [Erratum-ibid.\  D {\bf 71} (2005) 079901]
  [arXiv:hep-ph/0405266].

      

\end{thebibliography}
\end{document}